\documentclass[useAMS,usenatbib,usegraphicx]{mn2e}
\include{epsf}
\usepackage{fixltx2e}
\title{Cosmological Simulations using GCMHD+}
\author[Barnes et. al]
{David J. Barnes, Daisuke Kawata, Kinwah Wu
\\
Mullard Space Science Laboratory, University College London, Holmbury St. Mary, Dorking, Surrey, RH5 6NT
}
\date{Accepted .
      Received ;
      in original form }

\pagerange{\pageref{firstpage}--\pageref{lastpage}}
\pubyear{2011}

\begin{document}

\maketitle

\label{firstpage}

\begin{abstract}
Radio observations of galaxy clusters show that the intra cluster medium is permeated by $\rm{\mu G}$ magnetic fields. The origin and evolution of these cosmological magnetic fields is currently not well understood and so their impact on the dynamics of structure formation is not known. Numerical simulations are required to gain a greater understanding and produce predictions for the next generation of radio telescopes.  We present the galactic chemodynamics smoothed particle magnetohydrodynamic (SPMHD) code  (GCMHD+), which is an MHD implementation for the cosmological smoothed particle hydrodynamic code GCD+. The results of 1, 2 and 3 dimensional tests are presented and the performance of the code is shown relative to the ATHENA grid code. GCMHD+ shows good agreement with the reference solutions produced by ATHENA. The code is then used to simulate the formation of a galaxy cluster with a simple primordial magnetic field embedded in the gas. A homogeneous seed field of $10^{-11} \rm{G}$ is amplified by a factor of $10^3$ during the formation of the cluster. The results show good agreement with the profiles found in other magnetic cluster simulations of similar resolution.
 
\vspace{2mm}

\vspace{0.1cm}
\end{abstract}

\section{Introduction}

Cosmological magnetic fields are thought to be ubiquitous through out the universe and have been detected by radio observations on scales as large as galaxy clusters. On the galactic cluster scale, magnetic fields have been detected via Faraday rotation measurements of distant quasars/AGN, by the observation of radio galaxies within the cluster and by the detection of diffuse synchrotron emission from radio haloes \citep[e.g.][]{obs94,obs02,obs04,obs11,obs01}. The current observations find magnetic fields of $\rm{\mu G}$ strength permeates the intra-cluster medium (ICM) of most galaxy clusters. Beyond the cluster scale, measurements of the magnetic field are far less certain. Cosmological fields can be generated via Weibel's instability \citep{sch03,med06}, Biermann's battery \citep{bbm50,sub94,kul97,gne00}, structure formation \citep{har70,ici06}, galactic winds \citep{bec96,ber06,zfe09}, relativistic charged particles \citep{min11} and various processes in the early universe \citep{the02}. Current observations of the structural detail of these fields is limited, but the next generation of radio telescopes, such as the Square Kilometer Array (SKA) \citep[e.g.][]{ska06,ska08}, will produce a wealth of observational data on the strength and structure of cosmological magnetic fields. To compare these observations with our knowledge of cosmological magnetic fields we require numerical simulations. The predictions made by simulations allow for a comparison of the theory with the observation. This will produce a more detailed understanding of cosmological magnetic fields.

The evolution of primordial magnetic fields in a cosmological setting has been studied using both smoothed particle magnetohydrodynamics \citep[SPMHD, e.g.][]{cp09,niMHD11} and adaptive mesh refinement codes \citep[e.g.][]{pmc08,pmc11}. The different techniques produce good agreement on the predicted strength and profile of a magnetic field in a galaxy cluster. They show that the initial structure of the field has little influence on the final field. The central cluster field strength predicted by these simulations agrees with the values inferred from observation. A few simulations have followed the creation of a cosmological magnetic field from either galactic wind pollution of the ICM \citep{zfe09}, AGN pollution of the cluster \citep{zfe10} or the Biermann battery mechanism during structure formation \citep{zfe98} to generate a field from zero field initial conditions. The magnetic field strength can also be predicted a posteriori using the velocity fields from a purely hydrodynamical cosmological simulations \citep{zfe08}. These simulations produce different field strengths than the values found from following the evolution of a primordial magnetic field, especially in filament structures. Further simulations are required to examine the different origins of a magnetic field in the ICM and to test the validity of the predictions made. 

We present the implementation of magnetohydrodynamics (MHD) for the existing galactic chemodynamics code (GCD+) \citep{gcd03,gcd09,gcd11}. GCD+ is a three-dimensional tree N-body/Smoothed Particle Hydrodynamics (SPH) code which incorporates self-gravity, hydrodynamics, radiative cooling, star formation, supernova feedback and metal enrichment.  The following MHD implementation is fully compatible with all of the original features of the code and allows for a complete cosmological simulation to be run. We discuss the choice of method to ensure tensile instability is avoided to an appropriate level. We also discuss the effect of an applied time dependant dissipation scheme \citep{tdd97} for the magnetic field and the effect of the Balsara switch \citep{cp95} on the test simulations, including its need in a cosmological simulation. For the tests shown in this paper we ignore the additional processes and concentrate on the effect of introducing MHD to the simulations and the accuracy of the non-radiative solutions produced. As such, radiative cooling, star formation, supernovae feedback and metal enrichment are switched off in all results presented.

In Section 2 we present the numerical implementation of the MHD equations used in GCMHD+. Section 3 shows the performance of the code in various test simulations. Section 4 shows the results of a cosmological simulation for the formation of a galaxy cluster with a simple primordial magnetic field embedded in the gas. Our conclusions are then given in Section 5.

\section{MHD Implementation}

The hydrodynamic components in the code are adopted from \citet{gcd11}. In this section we summarise the addition of the MHD implementation and the updated parameters for the artificial viscosity for MHD simulations. We note that while our scheme does not ensure the $\nabla\cdot \textbf{B}=0$ constraint, we present results in Section 3.5 and 4 which show that it is well satisfied for all simulations.

\subsection{Hydrodynamic Parameters}
Artificial viscosity (A.V.) is used in hydrodynamic codes to smooth discontinuities, such as shocks, in the velocity field of the simulation, thus allowing the code to capture the physics correctly. Applying a constant level of A.V. to the simulation causes smoothing when it is not required. \citet{tdd97} suggest an A.V. switch where the level of A.V. is allowed to vary in space and time for particles in the simulation. The minimum level of A.V. in a simulation is set by the parameter $\alpha^{AV}_{min}$ and for a purely hydrodynamic simulation this is set to  $\alpha^{AV}_{min} = 0.1$\citep[e.g.][]{cp07}. While running our MHD test suite for the MHD implementation we found some post shock oscillation in the density field, hence for MHD simulations we increased the minimum A.V. to $\alpha^{AV}_{min} = 0.6$.

We also implement artificial thermal conductivity (A.C.) following \citet{cp07}. We allow the thermal conductivity parameter, $\alpha^C$, to vary with time for each particle. The parameter varies such that $0 \leq \alpha^C \leq 2$ via
\begin{equation}
 \frac{d\alpha_i^C}{dt}=-\frac{\alpha_i^C}{\tau_i}+S_i^C,
\end{equation}
where $S_i^C=0.05h_i|\nabla^2u_i|/\sqrt{u_i}$ is the source term. Unless explicitly stated, A.C. is turned on in all simulations.

The Balsara switch \citep{cp95} reduces the A.V. when the code detects a shear flow. This prevents the artificial viscosity from becoming the dominant force and generating spurious forces in a shear flow. While running the MHD test suite with the code, especially shocktube test 5A in Section 3.1, it was found that this switch was causing the velocity to being captured poorly. We then removed the switch from the code and found that it produced a significant improvement in the code's ability to produce the velocity solution, with minimal negative effects on other tests.

\subsection{Induction Equation}
The magnetic field is evolved via the induction equation which, neglecting any form of dissipation and enforcing the $\nabla\cdot \textbf{B}=0$ constraint, takes the form 
\begin{equation}\label{eq:indc}
 \frac{1}{c}\frac{d\textbf{B}}{dt}=(\textbf{B}\cdot\nabla)\textbf{v}-\textbf{B}(\nabla\cdot \textbf{v}).
\end{equation}
This is the standard form of the induction equation and it is the correct choice as it is unaffected by magnetic monopoles. We then convert this to SPH components and the equation takes the form of:
\begin{equation}
\frac{1}{c}\frac{dB_i^k}{dt}=\frac{1}{\Omega_i\rho_i}\left[\displaystyle\sum_{j=1}^N m_j(v_{ij}^kB_i^l-B_i^kv_{ij}^l) \frac{\partial W_i}{\partial u}\frac{r_{ij}^l}{|\textbf{r}_{ij}|}\right],
\end{equation}
where $B^k$ are the components of the magnetic field and we sum over the $l$ components, $\rho_i$ is the density at particle $i$, $m_j$ is the mass of particle $j$, $v_{ij}$ is the velocity between the two particles, $r_{ij}$ is the distance between the two particles and $1/\Omega_i$ is the factor correcting for the use of variable particle smoothing lengths \citep{cp04b}. The discrete form of equation (\ref{eq:indc}) no longer enforces the $\nabla\cdot \textbf{B}=0$ constraint. The magnetic field is allowed to act back on the fluid via a Lorentz force term in the momentum equation.

\subsection{Lorentz Force}
The conservative form of the magnetic stress tensor, derived by \citet{cp85}, is given by
\begin{equation}
 M_i^{kl} =\frac{1}{4\pi}\left(B_i^kB_i^l-\frac{1}{2}|\textbf{B}_i|^2\delta^{kl}\right).
\end{equation}
The form of the stress tensor ensures conservation of momentum at shocks. This generates a force that produces an additional term in the momentum equation, which in component form is
\begin{eqnarray}
 \left(\frac{dv_i^k}{dt}\right)^{(\rm{mag})} &=& \frac{1}{4\pi}\displaystyle\sum^N_{j=1}m_j\left(\frac{M^{kl}_i}{\Omega_i\rho_i^2}\frac{\partial W_i}{\partial u}\frac{r^l_{ij}}{|\textbf{r}_{ij}|}\right. \nonumber \\
&&\left. + \frac{M^{kl}_j}{\Omega_j\rho_j^2}\frac{\partial W_j}{\partial u}\frac{r^l_{ij}}{|\textbf{r}_{ij}|}\right).
\end{eqnarray}
When the magnetic field dominates the gas pressure, this exactly momentum conserving form of the force becomes unstable. The magnetic stress can become negative and this leads to the tensile instability. The negative stress, or an attractive force, between particles can cause them to clump together and the simulation becomes highly unstable. This requires a stable MHD code to have an additional term in the momentum equation to suppress the possibility of clumping, which takes the form of an instability correction.

\subsection{Tensile Instability Correction}

The Tensile Instability is a well known problem for SPMHD and several methods have been proposed to suppress it. \citet{cp00} suggested the addition of an anti-clumping term to the momentum equation which prevents the occurrence of the tensile instability. At short distances a steepened kernel ensures that this additional term becomes significant and the particles repel each other. This was found to be effective for one and two dimensional simulations. However, for 3D simulations with variable smoothing lengths this method is no longer effective at preventing clumping.

 \citet{ic01} suggested directly subtracting any non-vanishing monopole terms from the momentum equation. This can be implemented by an additional term in the momentum equation, which takes the form
\begin{eqnarray}
 \left(\frac{dv_i^k}{dt}\right)^{(\rm{corr})} &=& \frac{\hat{\beta}}{4\pi}B_i^k\displaystyle\sum^N_{j=1}m_j\left(\frac{B_i^l}{\Omega_i\rho_i^2}\frac{\partial W_i}{\partial u}\frac{r^l_{ij}}{|\textbf{r}_{ij}|} \right. \nonumber \\
&& \left.+ \frac{B^l_j}{\Omega_j\rho_j^2}\frac{\partial W_j}{\partial u}\frac{r^l_{ij}}{|\textbf{r}_{ij}|}\right),
\end{eqnarray}
where the parameter $\hat{\beta}$ controls the level of non-vanishing divergence subtracted. In principle the the addition of this term to the momentum equation breaks the momentum conservation of the formulation, but it does not seem to cause any major effects in our test simulations. \citet{ic01} used a value of $\hat{\beta}=1$ and \citet{cp09} found from testing that $\hat{\beta}=1$ produced no harm to their results. \citet{ic04} suggested that stability can be achieved with $\hat{\beta}< 1$ and that $\hat{\beta} =0.5$ should be used to minimise any non-conservative contribution. After testing we find that $\hat{\beta}=0.5$ produces the optimal results for our code and that it removes the tensile instability effectively (see Section 3.5).

\subsection{Artificial Magnetic Dissipation}
When no magnetic field is present a good estimate of the speed at which a signal can be sent from one particle to another is given by
\begin{equation}
 v_{ij}^{sig} = \frac{1}{2}(c_i+c_j)-\beta \textbf{v}_{ij}\cdot\hat{\textbf{e}}_{ij}.
\end{equation}
where $v_{ij}^{sig}$ is the signal velocity between the particles $i$ and $j$, $c_i$ is the sound speed at particle $i$ and $v_{ij}$ is the velocity between the particles. In the presence of a magnetic field a variety of MHD waves can propagate. The simplest generalization of the signal velocity is to replace the sound speed with the fastest magnetic wave \citep{cp04a}. The sound speed is then given by
\begin{eqnarray}
 v_i &=& \frac{1}{\sqrt{2}}\left[\left(c_i^2 + \frac{\textbf{B}_i^2}{4\pi\rho_i}\right) \right. \nonumber \\
&& \left. \times \sqrt{\left(c_i^2 + \frac{\textbf{B}_i^2}{4\pi\rho_i}\right)^2-4\frac{c_i^2(\textbf{B}_i\cdot\textbf{r}_{ij}/|\textbf{r}_{ij}|)^2}{4\pi\rho_i}}\right]^{0.5}.
\end{eqnarray}
In order to treat MHD shock fronts correctly a dissipation term is required to resolve any steep gradients in the magnetic field. We implement an artificial magnetic dissipation analogous to an artificial viscosity, based on the change of the total magnetic field, following \citet{cp04a}. This artificial dissipation is included via an additional term in the induction equation, which takes the form
\begin{eqnarray}
 \frac{1}{c}\left(\frac{d\textbf{B}_i}{dt}\right)^{(\rm{dis})} &=& \frac{\rho_i\alpha^B}{2}\displaystyle\sum^N_{j=1}\frac{m_jv_{ij}^{sig}}{\hat{\rho}_{ij}^2} \nonumber \\
&& \times(\textbf{B}_i-\textbf{B}_j)\frac{\textbf{r}_{ij}}{|\textbf{r}_{ij}|}\cdot\nabla_iW_i.
\end{eqnarray}
The strength of the dissipation is controlled by the parameter $\alpha^B$. The dissipation will lead to a generation of entropy according to
\begin{eqnarray}
 \left(\frac{da_i}{dt}\right)^{(\rm{dis})} &=& -\frac{\gamma-1}{\rho_i^{\gamma-1}}\frac{\alpha^B}{16\pi} \nonumber \\
&& \times \displaystyle\sum^N_{j=1}\frac{m_jv_{ij}^{sig}}{\hat{\rho}_{ij}^2}(\textbf{B}_i-\textbf{B}_j)^2\frac{\textbf{r}_{ij}}{|\textbf{r}_{ij}|}\cdot\nabla_i\bar{W}_{ij},
\end{eqnarray}
where $a = (P/\rho^{\gamma})$ \citep{ent02}. It was found that the artificial dissipation significantly reduced the noise, however with a constant value of $\alpha^B$ it also lead to a smoothing out of acute features. \citet{cp05} proposed that this could be avoided by making $\alpha^B$ independent and time varying for each particle. We allowed $\alpha^B$ to vary by integrating an equation very similar to the one for time-dependant viscosity,
\begin{equation}
 \frac{d\alpha^B}{dt}=-\frac{(\alpha_B-\alpha^B_{min})}{\tau} + S,
\end{equation}
where $\alpha_B^{min}$ is the minimum level of dissipation applied to each particle in the simulation and the source term, $S$, was chosen to be
\begin{equation}
 S=\rm{max}\left(\frac{|\nabla\times\textbf{B}|}{\sqrt{4\pi\rho}},\frac{|\nabla\cdot\textbf{B}|}{\sqrt{4\pi\rho}}\right).
\end{equation}
The parameter $\tau$ controls how fast the dissipation decays. This combined with the signal velocity defines the distance from the shock for the artificial dissipation to return to the minimum value. It is defined by
\begin{equation}
 \tau=\frac{h_i}{Cv^{sig}},
\end{equation}
where $h_i$ is the smoothing length for particle i and a value of $0.2$ was chosen for the constant $C$, i.e. 5 smoothing lengths. In order to conserve momentum the average value $\bar{\alpha}^B = \frac{1}{2}(\alpha^B_i+\alpha^B_j)$ is used in all simulations. Tests indicate that the use of a particle variable dissipation parameter greatly improves the code's shock capturing capabilities without significant smoothing of sharp features. 

\begin{table*}
\begin{flushleft}
 \begin{tabular}[t]{ccccccccccc}
 \hline
  Test & $N_L$ & $\rho_L$ & $V_L$ & $B_L$ & $P_L$ & $N_R$ & $\rho_R$ & $V_R$ & $B_R$ & $P_R$ \\
 \hline
 1A & 540 & 1.00 & (10.0,0.0,0.0) & (5.0,5.0,0.0)/$(4\pi)^{0.5}$ & 20.0 & 540 & 1.00 & ($-$10.0,0.0,0.0) & (5.0,5.0,0.0)/$(4\pi)^{0.5}$ & 1.00 \\
 1B & 1000 & 1.00 & (0.0,0.0,0.0) & (3.0,5.0,0.0)/$(4\pi)^{0.5}$ & 1.00 & 100 & 0.10 & (0.0,0.0,0.0) & (3.0,2.0,0.0)/$(4\pi)^{0.5}$ & 10.0 \\
 2A & 540 & 1.08 & (1.2,0.01,0.5) & (2.0,3.6,2.0)/$(4\pi)^{0.5}$ & 0.95 & 500 & 1.0 & (0.0,0.0,0.0) & (2.0,4.0,2.0)/$(4\pi)^{0.5}$ & 1.00 \\
 2B & 1000 & 1.00 & (0.0,0.0,0.0) & (3.0,6.0,0.0)/$(4\pi)^{0.5}$ & 1.00 & 100 & 0.10 & (0.0,2.0,1.0) & (3.0,1.0,0.0)/$(4\pi)^{0.5}$ & 10.0 \\
 3A & 550 & 0.10 & (50.0,0.0,0.0) & $-$(0.0,1.0,2.0)/$(4\pi)^{0.5}$ & 0.40 & 550 & 0.10 & (0.0,0.0,0.0) & (0.0,1.0,2.0)/$(4\pi)^{0.5}$ & 0.20 \\
 3B & 550 & 1.00 & ($-$1.0,0.0,0.0) & (0.0,1.0,0.0) & 1.00 & 550 & 1.00 & (1.0,0.0,0.0) & (0.0,1.0,0.0) & 1.00 \\
 4A & 1000 & 1.00 & (0.0,0.0,0.0) & (1.0,1.0,0.0) & 1.00 & 200 & 0.20 & (0.0,0.0,0.0) & (1.0,0.0,0.0) & 0.10 \\
 4B & 400 & 0.40 & ($-$0.6699,0.9826,0.0) & (1.3,0.0025,0.0) & 0.5247 & 1000 & 1.000 & (0.0,0.0,0.0) & (1.3,1.0,0.0) & 1.00 \\
 4C & 650 & 0.65 & (0.667,$-$0.257,0.0) & (0.75,0.55,0.0) & 0.50 & 1000 & 1.000 & (0.4,$-$0.94,0.0) & (0.75,0.0,0.0) & 0.75 \\
 4D & 1000 & 1.00 & (0.0,0.0,0.0) & (0.7,0.0,0.0) & 1.00 & 300 & 0.300 & (0.0,0.0,1.0) & (0.7,1.0,0.0) & 0.20 \\
 5A & 960 & 1.00 & (0.0,0.0,0.0) & (0.75,1.0,0.0) & 1.00 & 120 & 0.125 & (0.0,0.0,0.0) & (0.75,$-$1.0,0.0) & 0.10 \\
\hline
 \end{tabular}
 \end{flushleft}
 \caption{Summary of the initial conditions for all 1D test simulations, where $N$ is the number of particles, $\rho$ is the density, $V$ is the 3D velocity structure, $B$ is the 3D magnetic field structure and $P$ is the pressure. $L$ and $R$ denote the left and right halves of the simulation.}
\label{tab:1Dtests}
\end{table*}

\begin{figure*}\label{fig:SHTB1A}
\begin{flushleft}
\includegraphics[width=18cm,keepaspectratio=true]{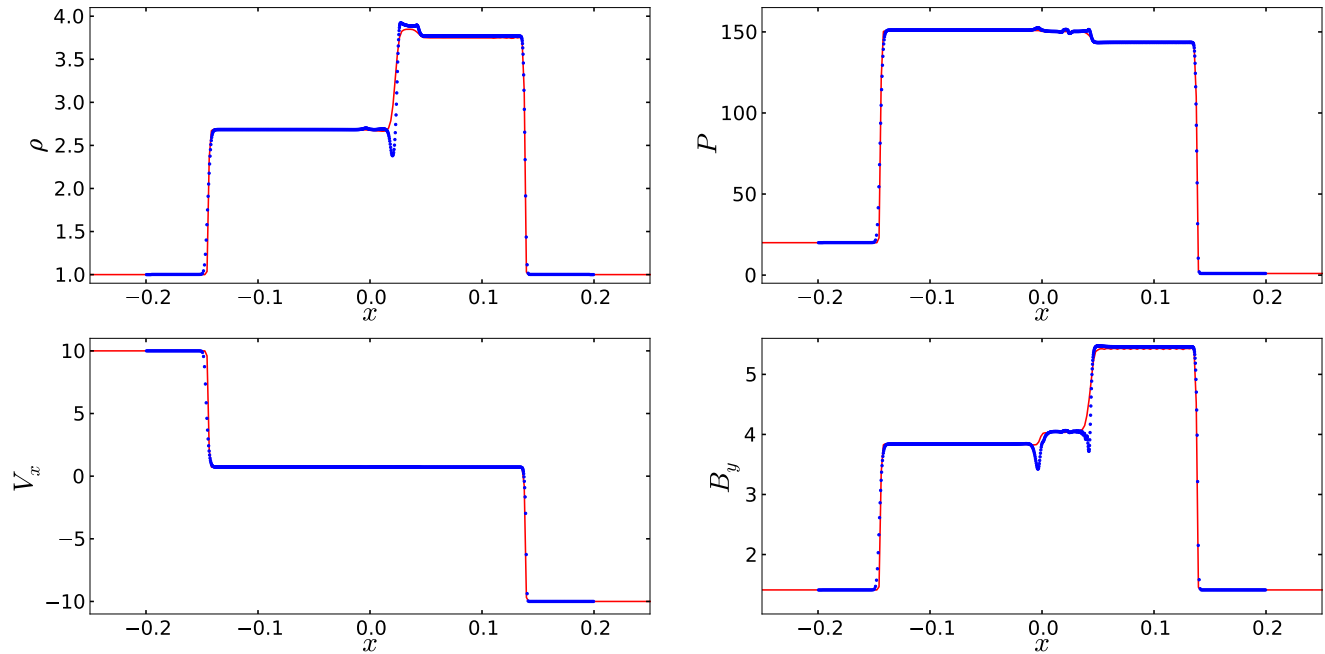}
\caption{Result for the shocktube test 1A. The ATHENA reference solution is shown by the red line and the blue points show the result produce by the code. They are in general in good agreement except for a small amount of deviation in the density and magnetic profiles.}
\end{flushleft}
\end{figure*}

\section{Test Simulations}

Having outlined the additions made to produce the GCMHD+ code, we run a series of test simulations which have become the standard set to show the performance of a numerical MHD scheme. We ran a large number of series of test simulations with different parameters associated to the MHD implementation, such as $\alpha^B_{min}$ and $\hat{\beta}$. This lead to our best parameter set for the suite of test simulations. This set is: $\alpha^B_{min}=0.05$, $\alpha^B_{max}=1.0$, $\hat{\beta}=0.5$, $\alpha^{AV}_{min}=0.6$, artificial conductivity on and the Balsara switch turned off (see Section 3.5). However, as demonstrated in Section 4 for the cosmological simulation we require $\alpha^B_{min}=0.0$. Therefore, we set $\alpha^B_{min}=0.0$ as a fiducial case. In this section we present the test simulation suite results with this fiducial set of parameters ( $\alpha^B_{min}=0.0$, $\alpha^B_{max}=1.0$, $\hat{\beta}=0.5$, $\alpha^{AV}_{min}=0.6$, artificial conductivity on and the Balsara switch turned off) and demonstrate that $\alpha^B_{min}=0.0$ leads to a satisfactory result for all the test simulations. Due to the complex nature of MHD interactions no analytical solution exists for many of the tests presented below and so the performance of the code is compared to a reference solution. This is provided by the publicly available ATHENA MHD mesh code \citep{rs08}.

\subsection{Magnetic Shock Tubes}

\begin{figure*}\label{fig:SHTB1B}
\begin{flushleft}
\includegraphics[width=18cm,keepaspectratio=true]{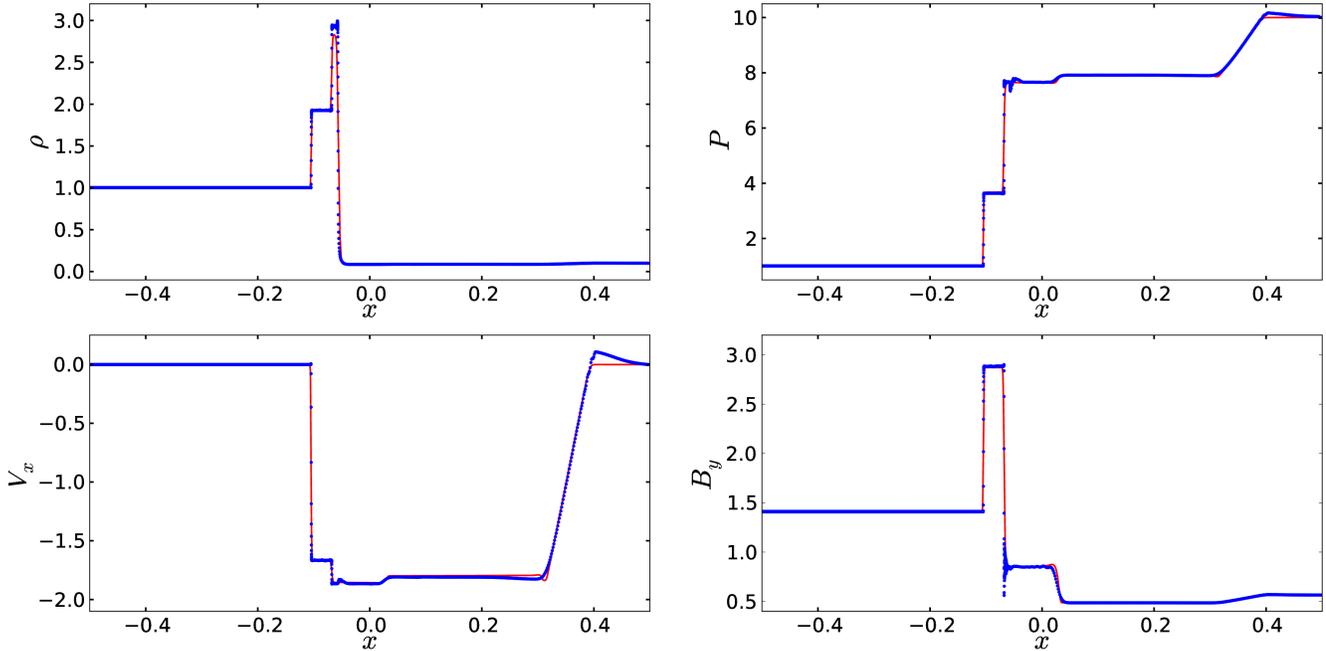}
\caption{Result for test 1B. The code shows good agreement with the reference solution in the density and pressure profiles. The velocity profile is not captured exactly in the low density region. There is a small amount of noise present in the magnetic field solution.}
\end{flushleft}
\end{figure*}

\begin{figure*}\label{fig:SHTB2A}
\begin{flushleft}
\includegraphics[width=18cm,keepaspectratio=true]{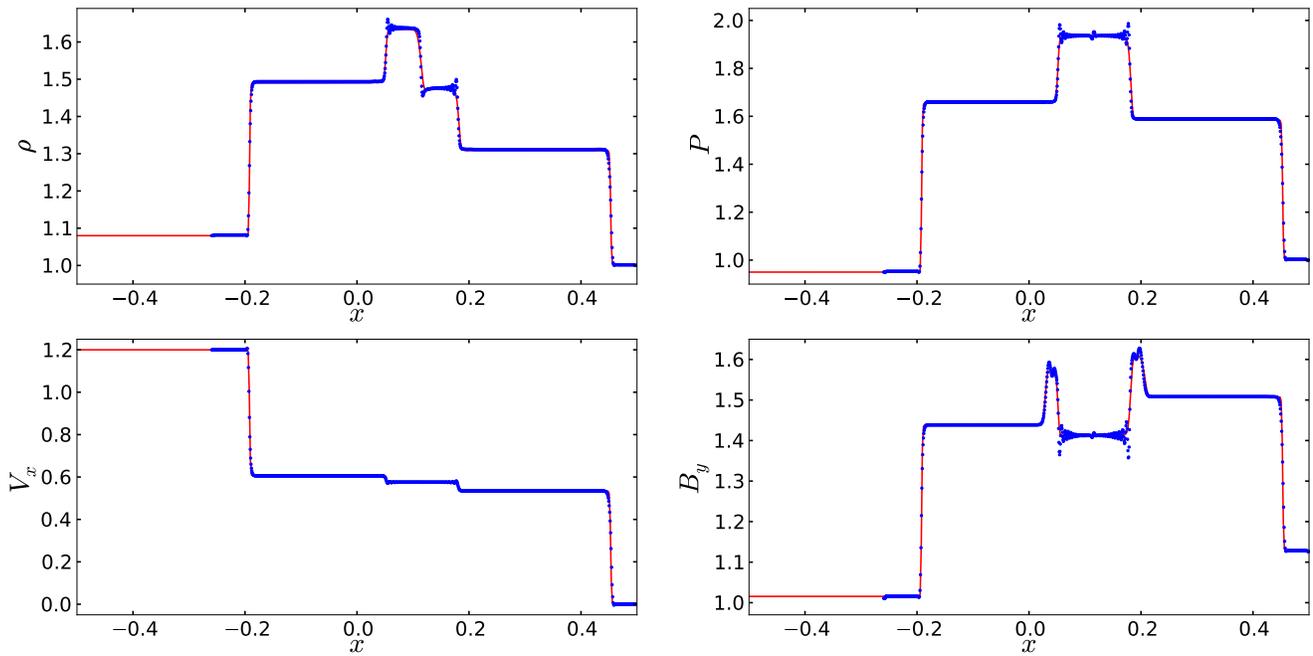}
\caption{Result for shocktube test 2A. The two solutions show good general agreement except for a small amount of noise present in the density, pressure and magnetic field.}
\end{flushleft}
\end{figure*}

\begin{figure*}\label{fig:SHTB2B}
\begin{flushleft}
\includegraphics[width=18cm,keepaspectratio=true]{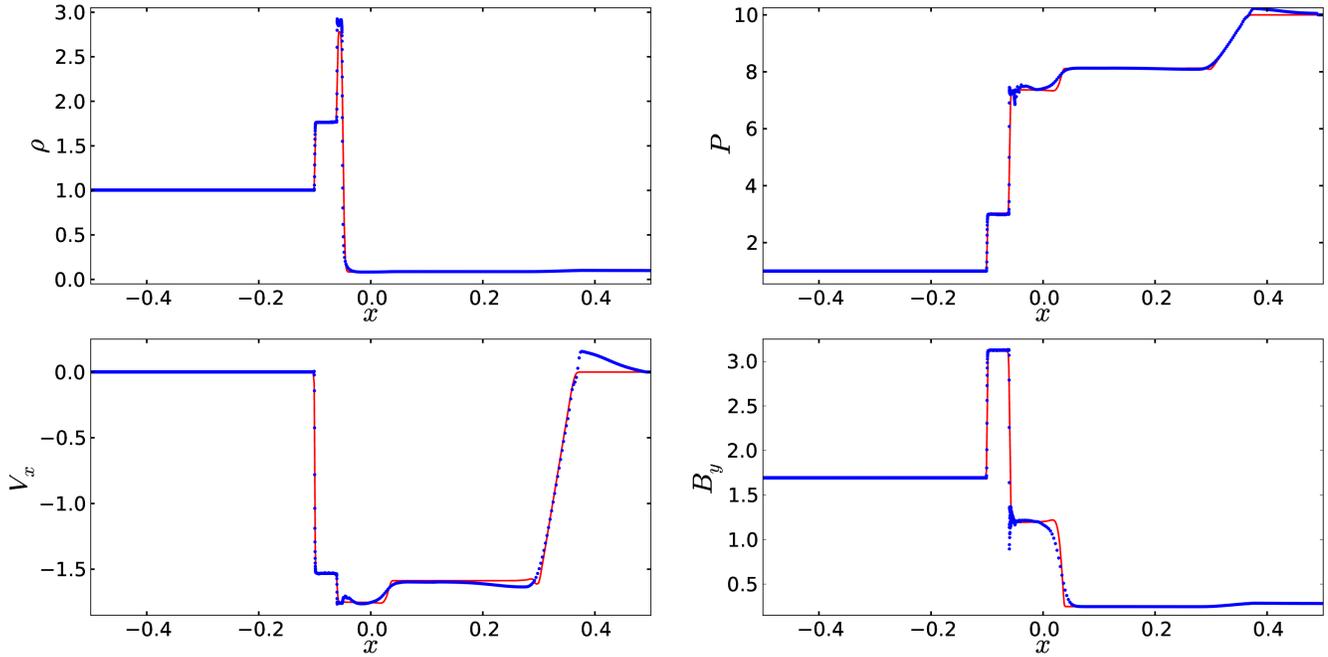}
\caption{Result for shocktube test 2B. There is a small amount of noise in the magnetic field profile and the code doesn't quite capture the velocity exactly.}
\end{flushleft}
\end{figure*}

\begin{figure*}\label{fig:SHTB3A}
\begin{flushleft}
\includegraphics[width=18cm,keepaspectratio=true]{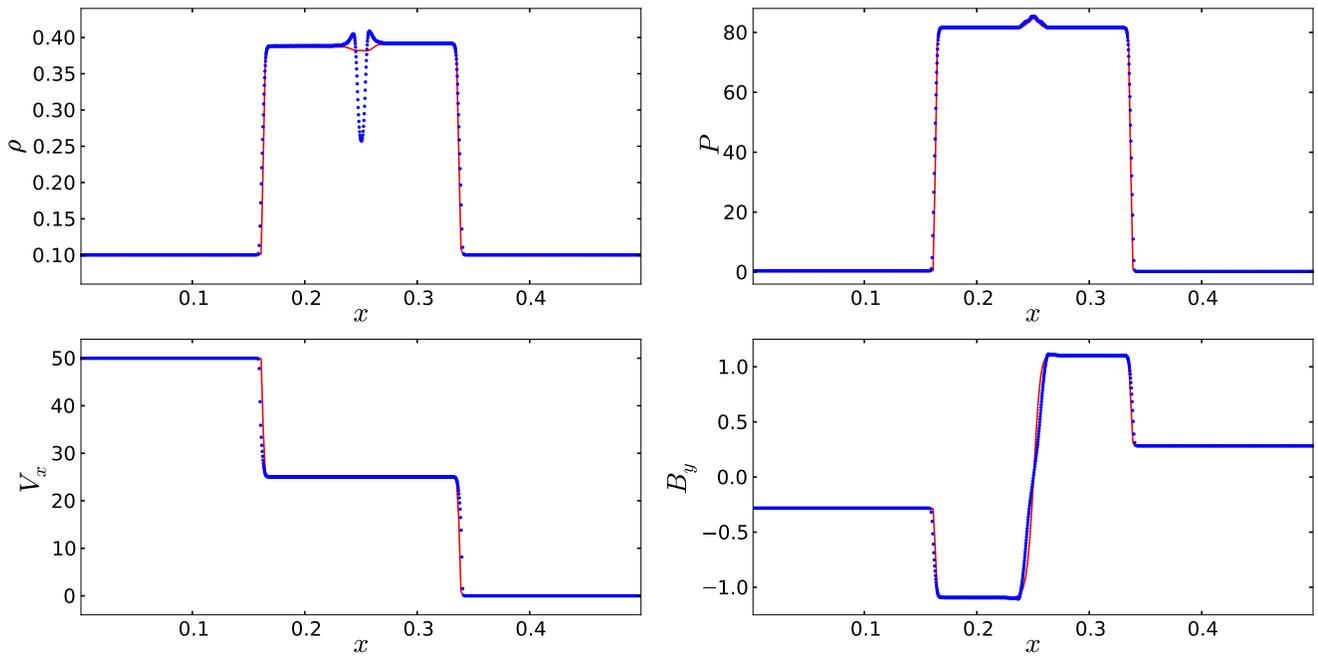}
\caption{Result for the shocktube test 3A. The pressure, velocity and magnetic field are well captured by the code when compared to the reference solution. Due to the difficult hydrodynamic set up the density is not well captured for the transition.}
\end{flushleft}
\end{figure*}

\begin{figure*}\label{fig:SHTB3B}
\begin{flushleft}
\includegraphics[width=18cm,keepaspectratio=true]{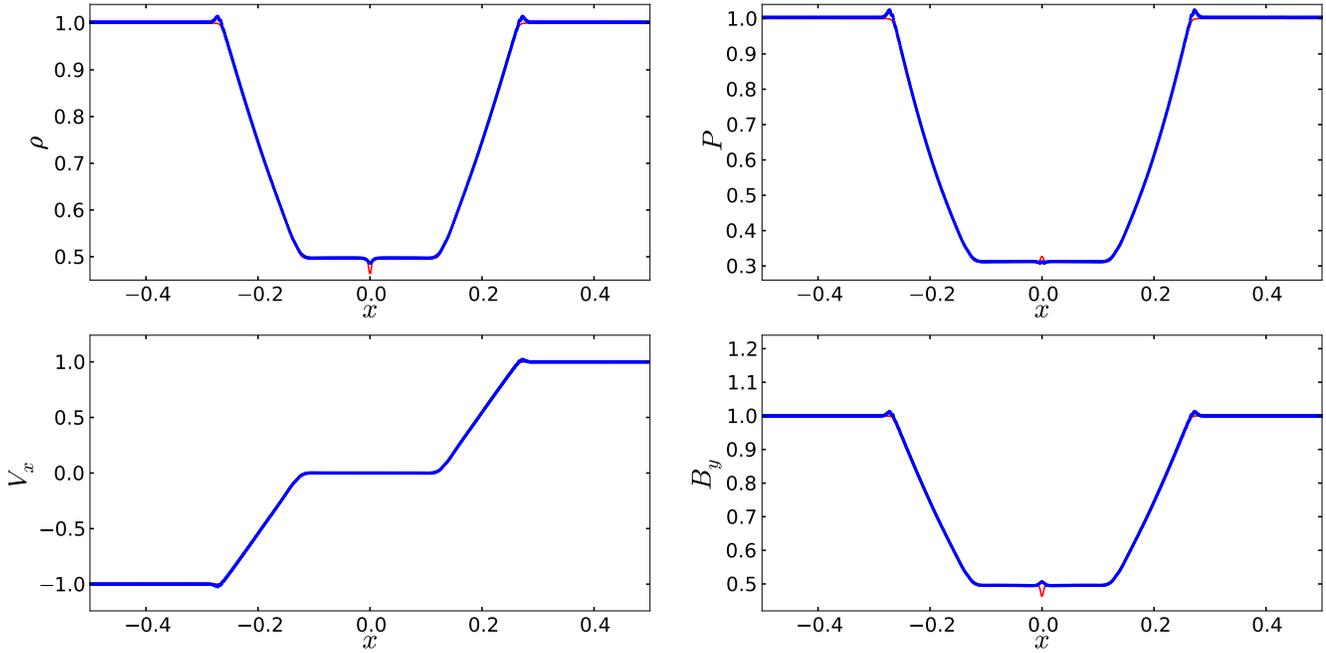}
\caption{Result for shocktube test 3B. The code captures all of the variables with very little noise. There is a discrepancy between the code and the reference at $x=0.0$ for the magnetic field.}
\end{flushleft}
\end{figure*}

\begin{figure*}\label{fig:SHTB4A}
\begin{flushleft}
\includegraphics[width=18cm,keepaspectratio=true]{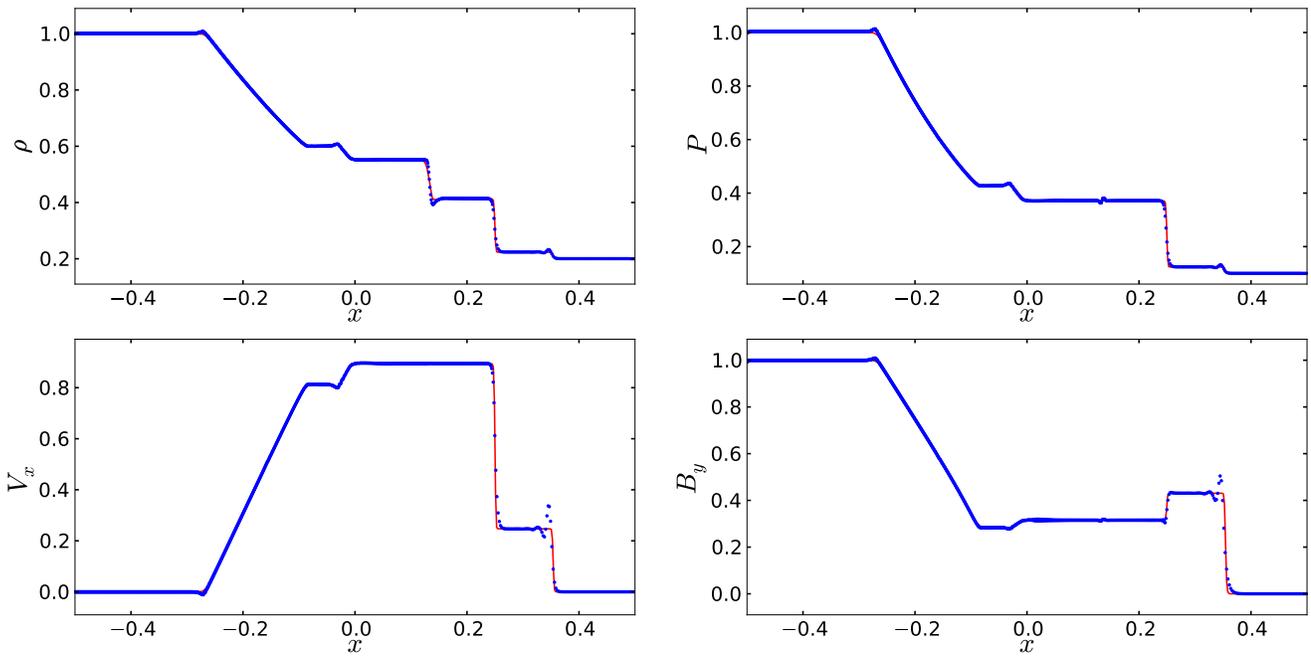}
\caption{Result for shocktube test 4A. The density, pressure, velocity and magnetic field profiles all agree with the solution from ATHENA. There is a small amount of noise in the velocity and magnetic field profile.}
\end{flushleft}
\end{figure*}

\begin{figure*}\label{fig:SHTB4B}
\begin{flushleft}
\includegraphics[width=18cm,keepaspectratio=true]{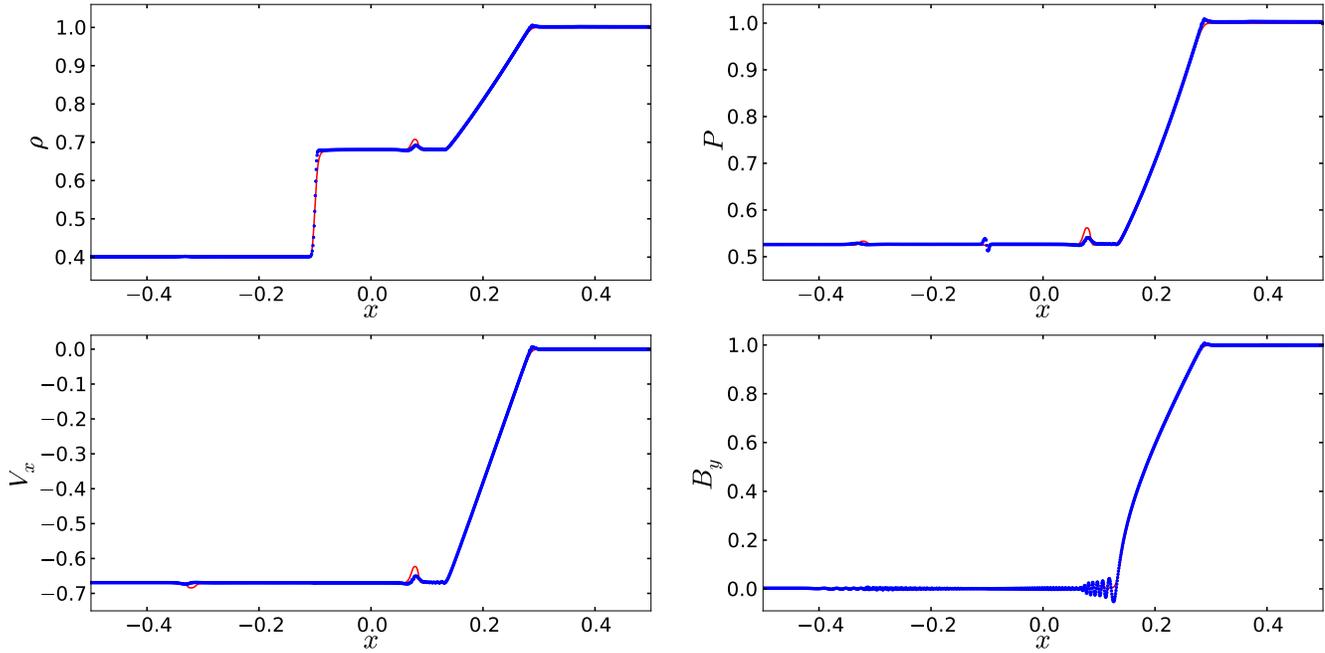}
\caption{Result for shocktube test 4B. The result produced by the code shows good agreement with the reference solution for all of the parameters. The is some noise where the magnetic field switches off.}
\end{flushleft}
\end{figure*}

\begin{figure*}\label{fig:SHTB4C}
\begin{flushleft}
\includegraphics[width=18cm,keepaspectratio=true]{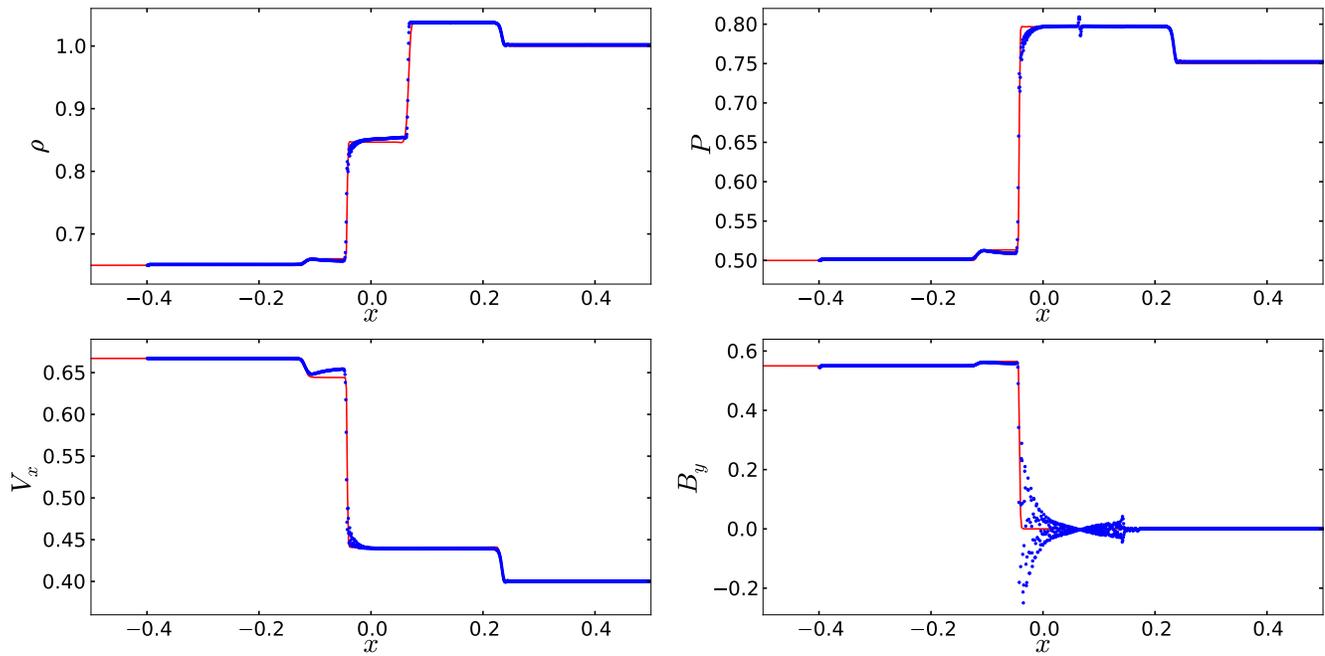}
\caption{Result for shocktube test 4C. The density, pressure and velocity agree with the reference solution. The magnetic field shows good agreement except at the shock where there is significant noise present. This can be reduced by using the test optimized parameters.}
\end{flushleft}
\end{figure*}

\begin{figure*}\label{fig:SHTB4D}
\begin{flushleft}
\includegraphics[width=18cm,keepaspectratio=true]{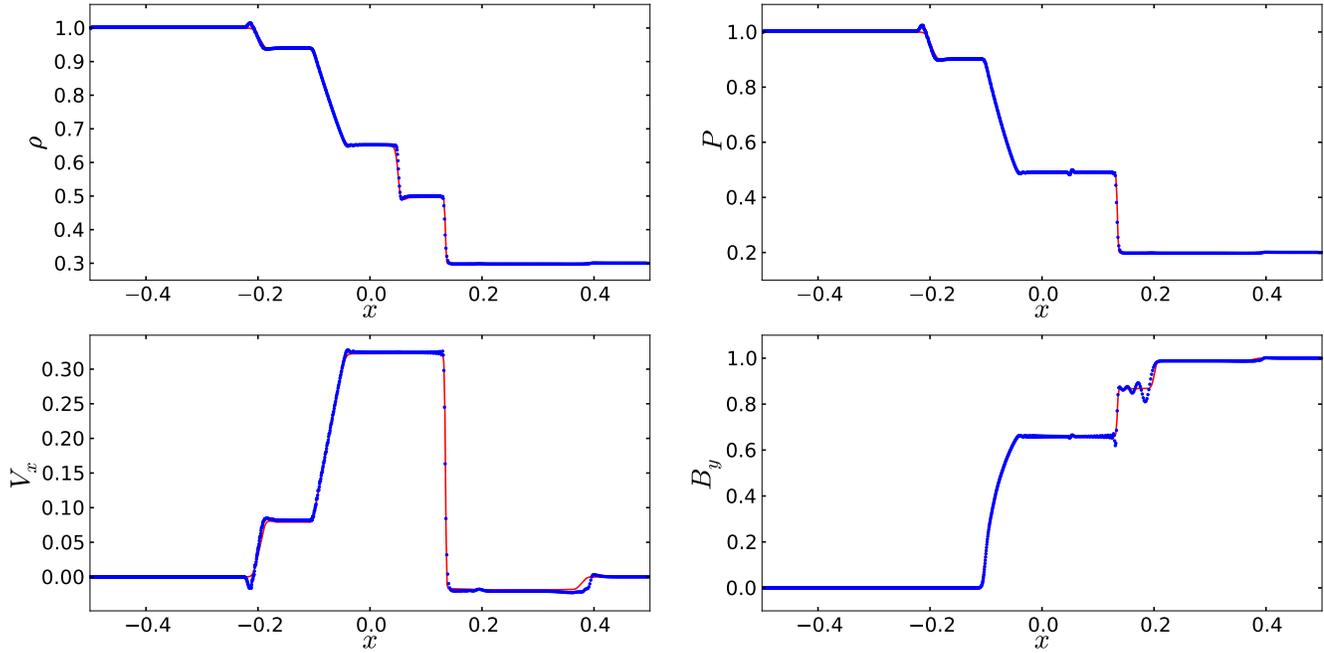}
\caption{Result for shocktube 4D. The density, pressure, velocity and magnetic field all show good agreement with the reference solution. There is a small amount of noise present in the magnetic field solution.}
\end{flushleft}
\end{figure*}

\begin{figure*}\label{fig:SHTB5A}
\begin{flushleft}
\includegraphics[width=18cm,keepaspectratio=true]{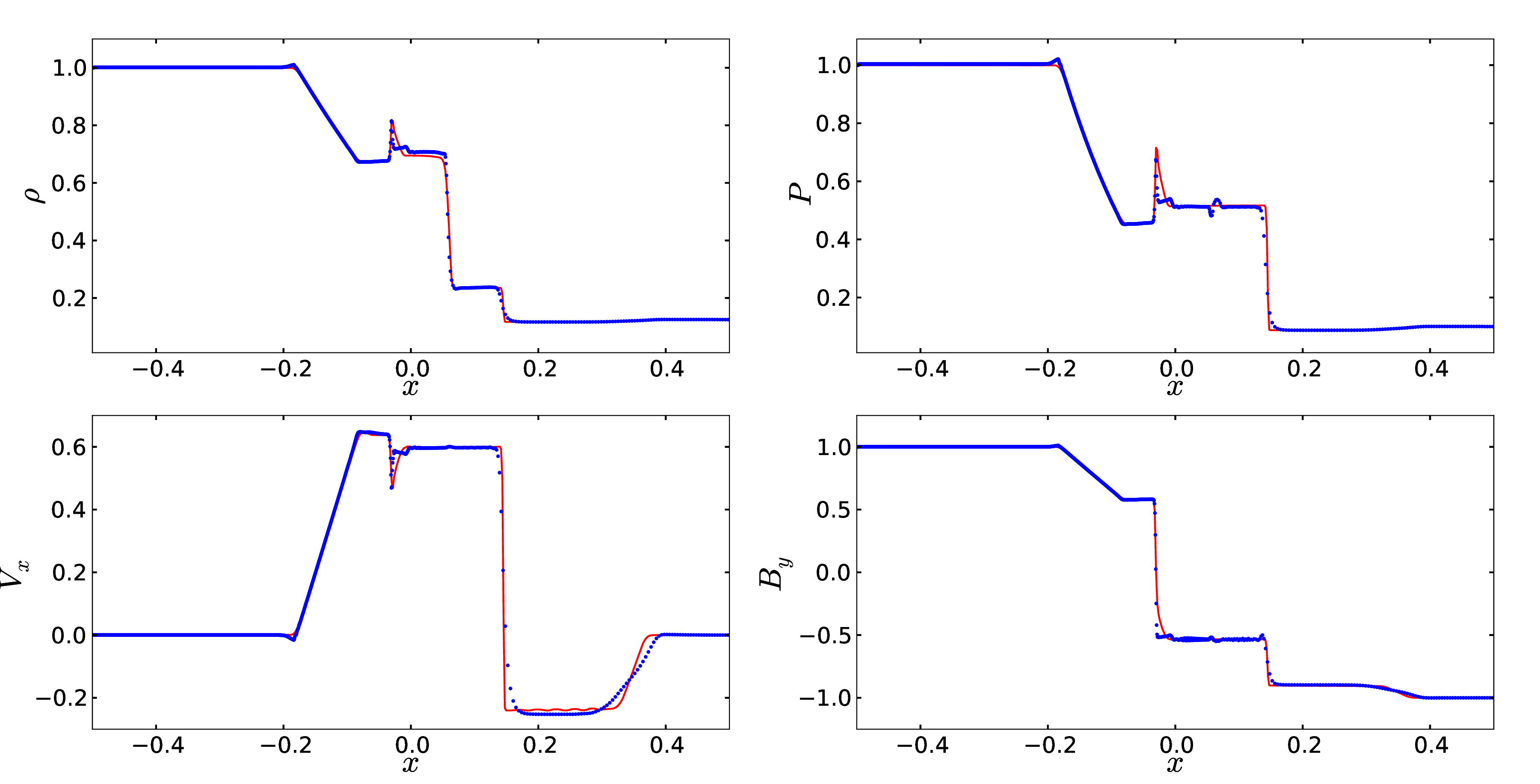}
\caption{Result for shocktube 5A. The code produces a solution which agrees with the reference for the density, pressure and magnetic field with very little noise present. The velocity is poorly captured in the low density region.}
\end{flushleft}
\end{figure*}

Magnetic shocktubes are the standard test of any numerical MHD scheme. The addition of MHD allows for more complex solutions to the Riemann problem due to the presence of slow and fast Alfv\'{e}n waves. Sharp magnetic features present in the simulation will be smoothed by the artificial dissipation and the full effects of any applied regularization scheme must be explored. This means that a full set of shock tube problems, as presented in \citet{tr95}, are required to rigorously test the effects of the applied artificial magnetic dissipation scheme and to establish the best parameters for 1D solutions produced by the code.

The set up of all the one dimensional shocktube simulations can be found in Table \ref{tab:1Dtests}. While the particle position is only allowed to vary in the $x$ direction, the particle velocity and the magnetic field are allowed to vary in 3 dimensions. As no analytic solution is known, the code is compared against a reference solution obtained using the ATHENA code. The resolution of the ATHENA simulation depended on the shock tube being run, but the parameters of each test were left unchanged from the originals provided.

Figs. 1-11 show the density, pressure, $V_x$ and $B_y$ outputs for all the shocktube tests. The code captures the majority of the features precisely. For the tests, 1A and 1B, the code captures all three stages, including the intermediate phase, for both the strong and weak shock accurately. For the density, at $x=0.02$, and magnetic field, at $x=0.00$ and $x=0.02$, of 1A and the velocity, at $x=0.39$, and magnetic field, at $x=-0.14$, of 1B there are small deviations from the reference solutions. The code's ability to capture the velocity in test 1B improves as the resolution increases.

The second set of tests, 2A and 2B, have a 3D velocity and magnetic field structure. All the features are well captured by the code and only the transition in the magnetic field of test 2B, between the intermediate and lower state, at $x=0.05$, shows any visible deviation from the reference. In both tests there is also a small amount of oscillation visible in the magnetic field at the shock boundaries. Test 2A shows some oscillation in the density and pressure as well.

The third tests, 3A and 3B, show the code's ability to handle magneto-sonic features. The pressure, velocity and magnetic field profiles produced by the code for test 3A agree well with the reference solution. The density profile in test 3A is poorly captured. There is a large dip in the density compared to the reference solution. This is due to the hard hydrodynamical conditions for the test, where some of the particles are colliding at high speed with stationary ones, producing a wall heating error. This is effectively a wall for the high speed particles and due to the restriction of only being allowed to travel in one spatial dimension they crash in to this wall and oscillate back and forth. This causes the resulting dip in the density at the wall. This problem is not experienced by the mesh code and so it produces the correct solution. The velocity profile for test 3B agrees with the solution produced by ATHENA. For the density, pressure and magnetic field solutions are well captured with a small deviation at $x=0.0$. This code shows smaller peaks compared to the ATHENA result. However, the result from \citet{tr95} for test 3B is more comparable to the solution produced by GCMHD+. We therefore believe the inconsistencies between the results could be due to an artifact in the ATHENA solution.

\begin{figure}
\centering
\includegraphics[width=\hsize,height=7.72cm]{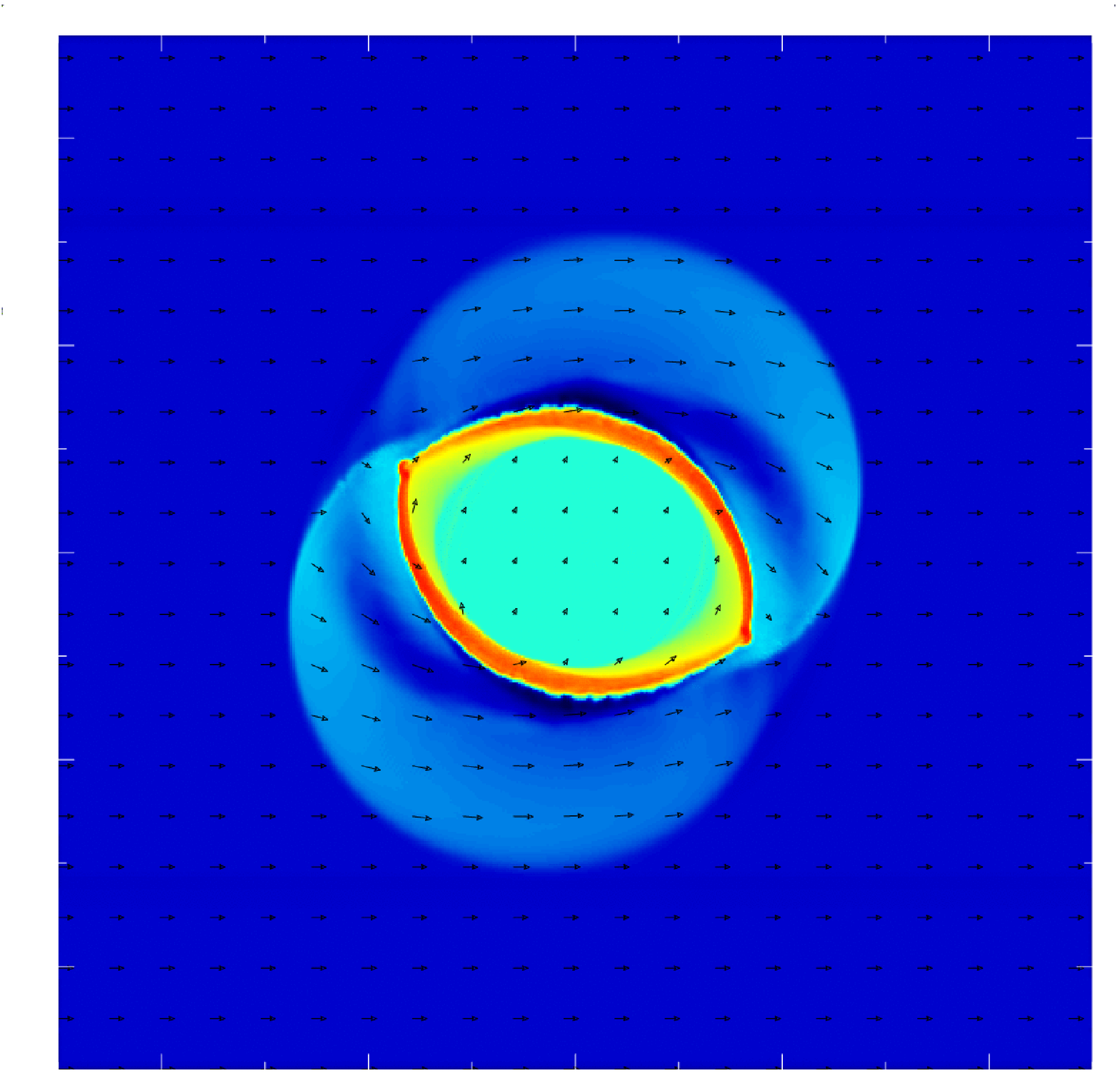}
\includegraphics[width=\hsize,height=7.72cm]{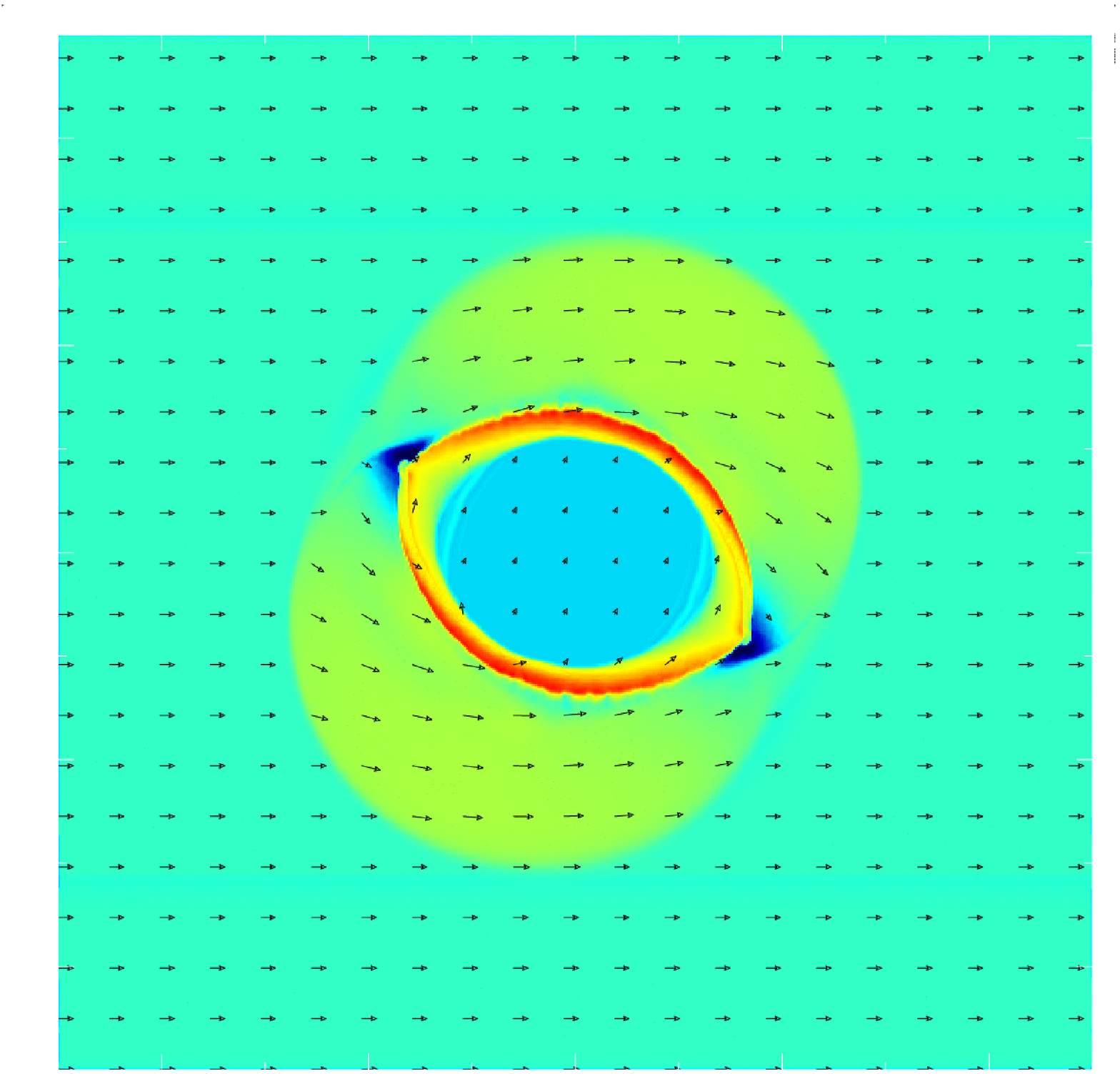}
\caption{Plots of the density (upper) and the magnetic pressure (lower) for the fast rotor test at $t=0.1$. The arrows show the size and direction of the magnetic field. The shape and position of the main features agree with other rotor test solutions published in the literature.}
\label{fig:Rotor2D}
\end{figure}

The fourth set of tests deal with features caused by the magnetic field turning on and off behind fast and slow moving rarefactions and shocks. The code produces density, pressure and velocity profiles which show good agreement with the reference solutions. The magnetic field is well captured, with a small amount of oscillation, in 4A, 4B and 4D. The magnetic field profile found in test 4C does agree with the reference solution, but there is a very large amount of oscillation present in this test at the shock front. 

The final test, 5A, is the commonly used shock tube test of \citet{tr88}. This test involves a shock and rarefaction moving together. The solution produced by the code captures all of the features very well for the density, pressure and magnetic field, with a small amount of oscillation visible in all of the profiles. The velocity is well captured in the high density region, but it is poorly captured in the low density region. Overall, the code captures all of the features present in the tests and produces very similar results to the reference solutions provided by ATHENA. 

\subsection{Rotor test}

The fast rotor test has been used many times to check the validity of solutions produced an MHD code \citep{frt00}. The 2D test is set up with a dense rotating disc embedded in a low density, static ambient medium. A uniform magnetic field is applied across the entire simulation. A constant field is set in the $x$ direction with a strength of $B_x = 2.5/\sqrt{\pi}$. The rotating disc has a radius $r_0 =0.1$, initial density of $\rho = 10$ and pressure of $P = 1$. The rotational velocity is given by $v_x = 2(y-0.5)/r_0$, $v_y = 2(x-0.5)/r_0$ and $v_z=0$. The ambient medium has a density $\rho = 1$ and pressure $P = 1$.

\begin{figure}
\includegraphics[width=\hsize,height=6.00cm]{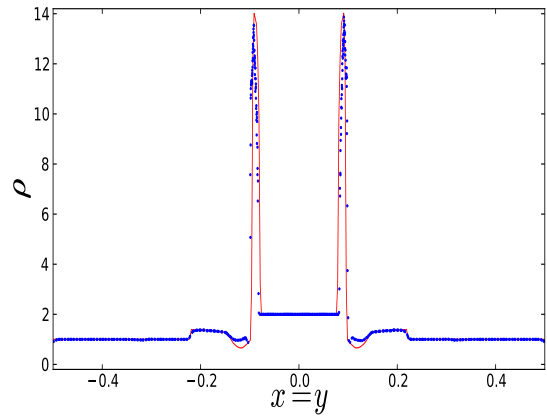}
\caption{A cut through fast rotor along $x=y$ at $t=0.1$ showing how the density (blue points) compares to result found using ATHENA (red line). The code captures the majority of the density features very well. There is a small difference between the ATHENA solution and the code either side of the main density peaks due to smoothing.}
\label{fig:RotorCUT}
\end{figure}

One way to produce the density contrast is to use particles of different mass for the disc and the background medium. However, this produces spurious unwanted effects. Instead we apply the same mass to all the particles and put ten times more particles in the disc region. First, the ambient medium is laid down in a regular hexagonal lattice. Then a second lattice is placed with smaller particle spacing in a region $2r_0 \times 2r_0$ centred at $(0,0)$. The larger separation particles for which $r<r_0$ and smaller separation particles with  $r>r_0$ are removed. This ensures that all particles in the simulation have the same mass. A hexagonal lattice is used instead a standard square lattice to reduce any discontinuities between the disc and the background. The lattice results in a set up of $400 \times 460$ particles in the background. This results in a total of 236626 particles in the simulation.

\begin{figure*}
\includegraphics[width=5.84cm,keepaspectratio=true]{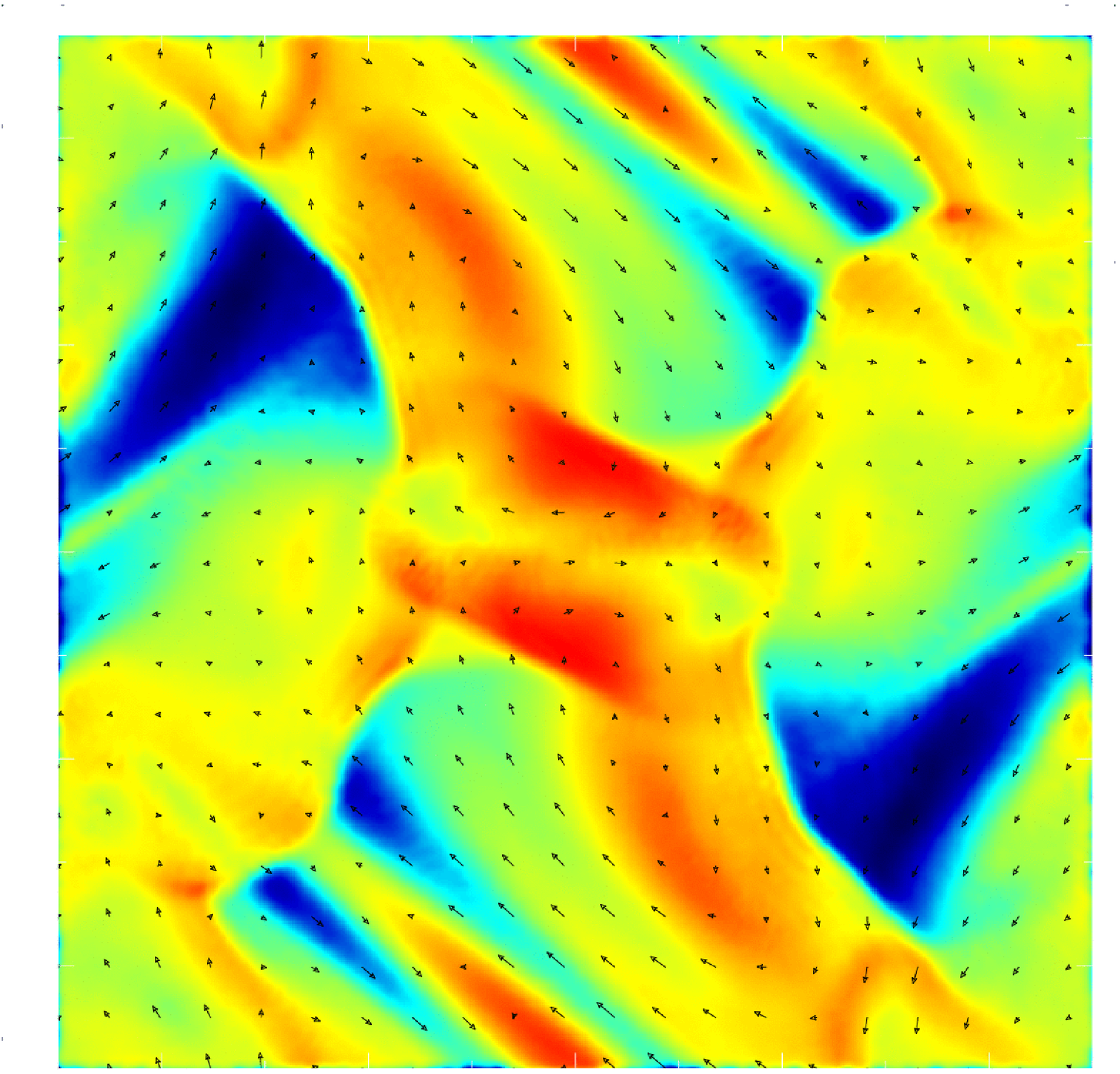}
\includegraphics[width=5.84cm,keepaspectratio=true]{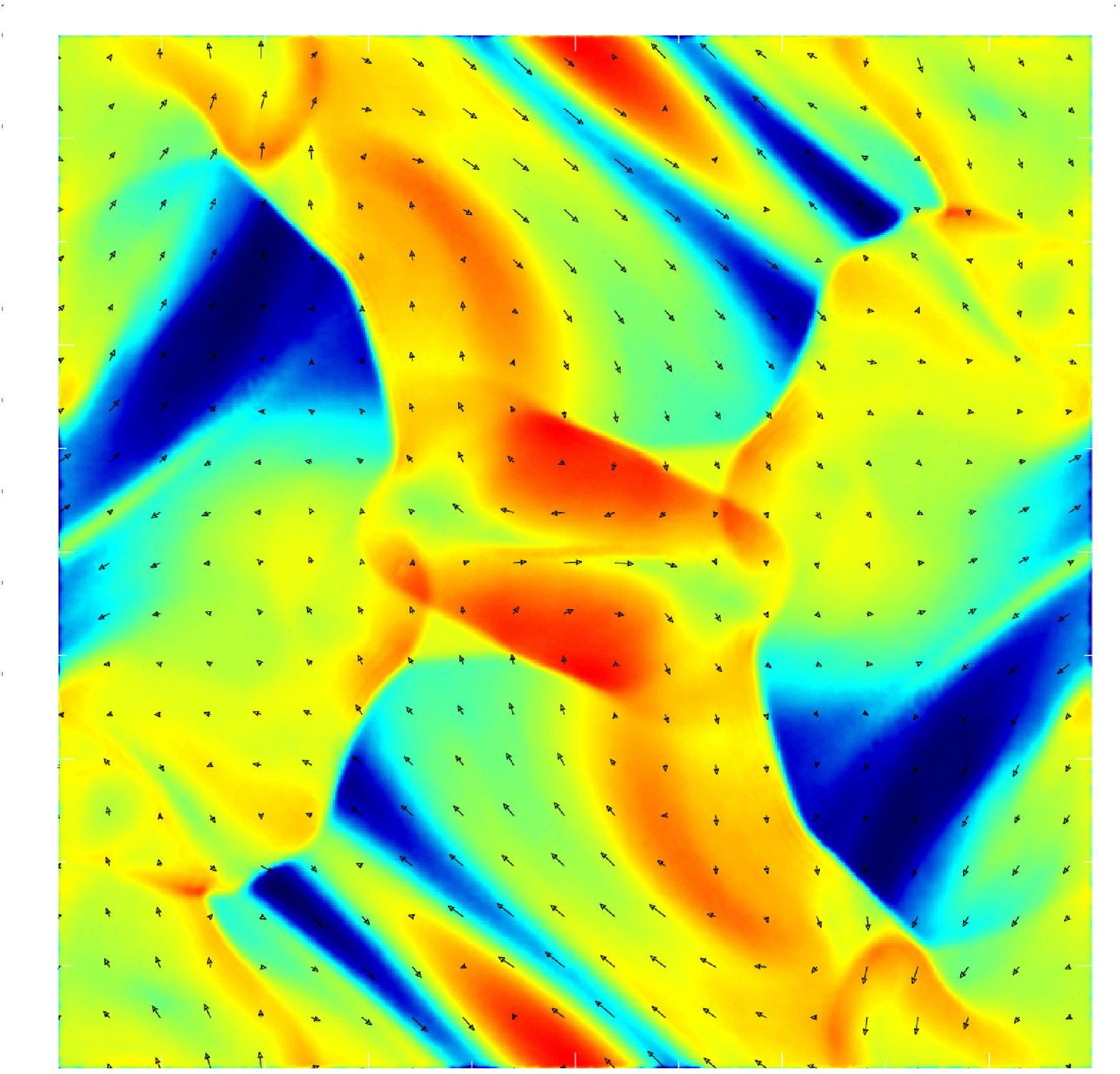}
\includegraphics[width=5.84cm,keepaspectratio=true]{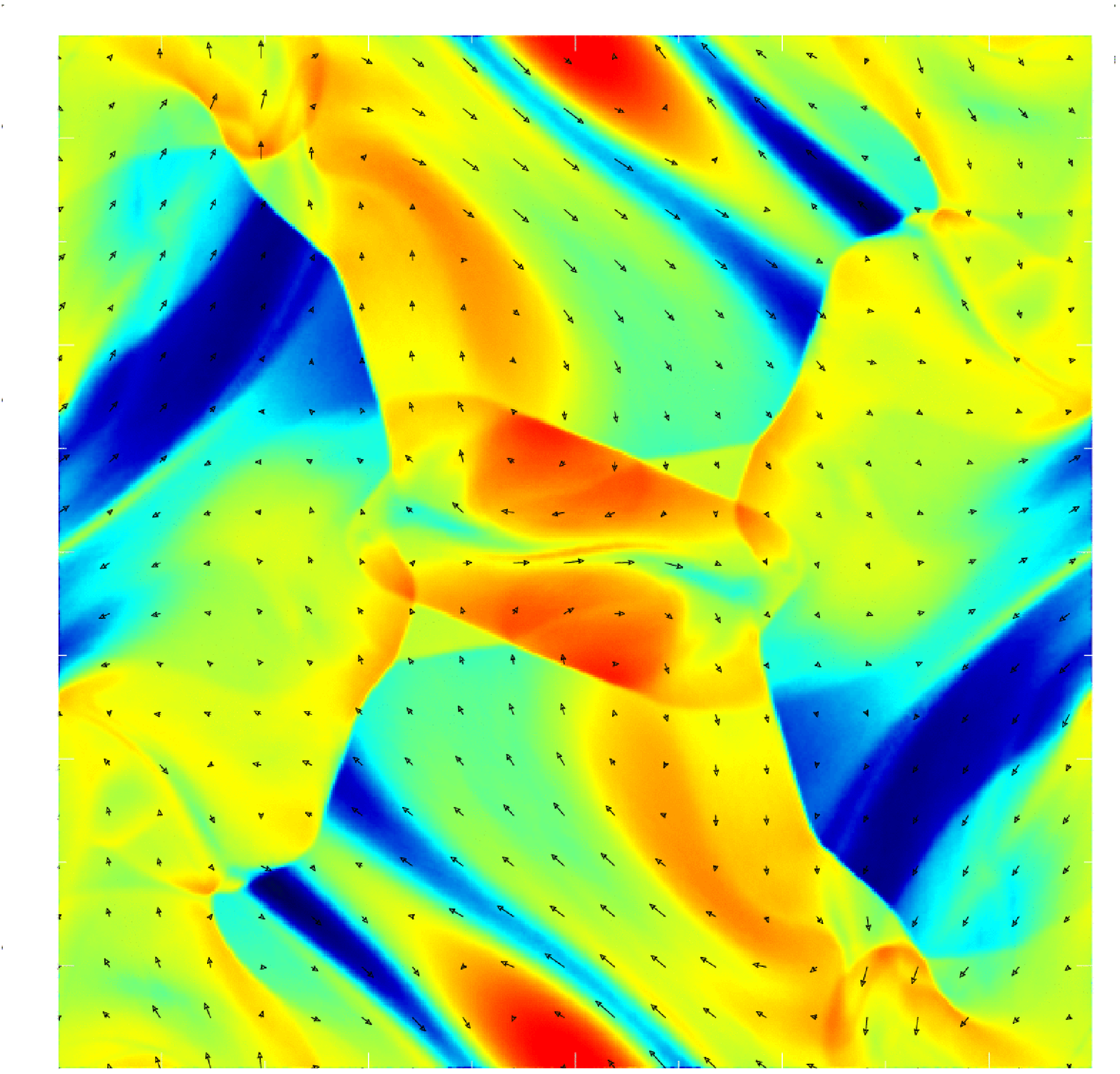}
\includegraphics[width=5.84cm,keepaspectratio=true]{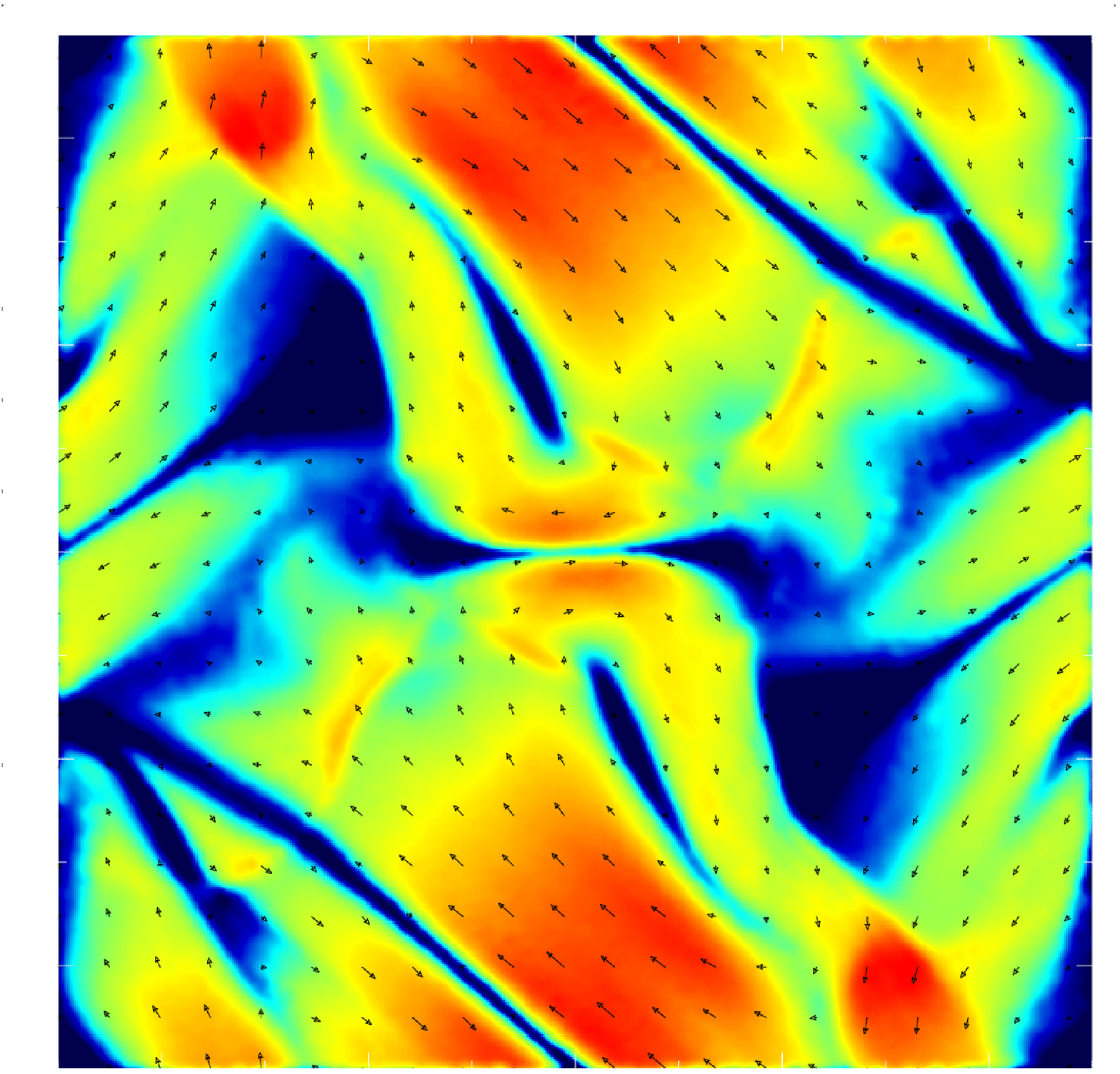}
\includegraphics[width=5.84cm,keepaspectratio=true]{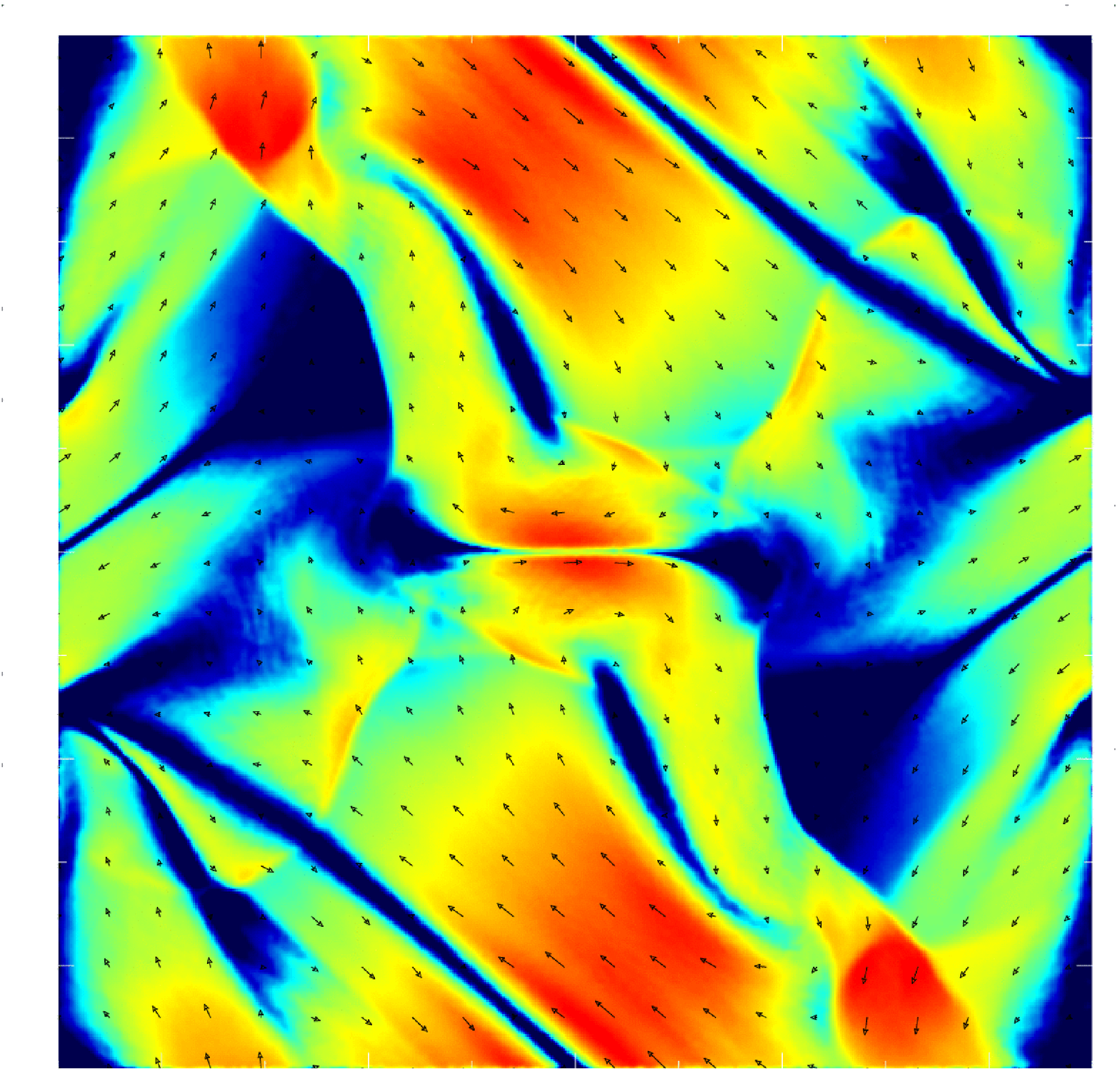}
\includegraphics[width=5.84cm,keepaspectratio=true]{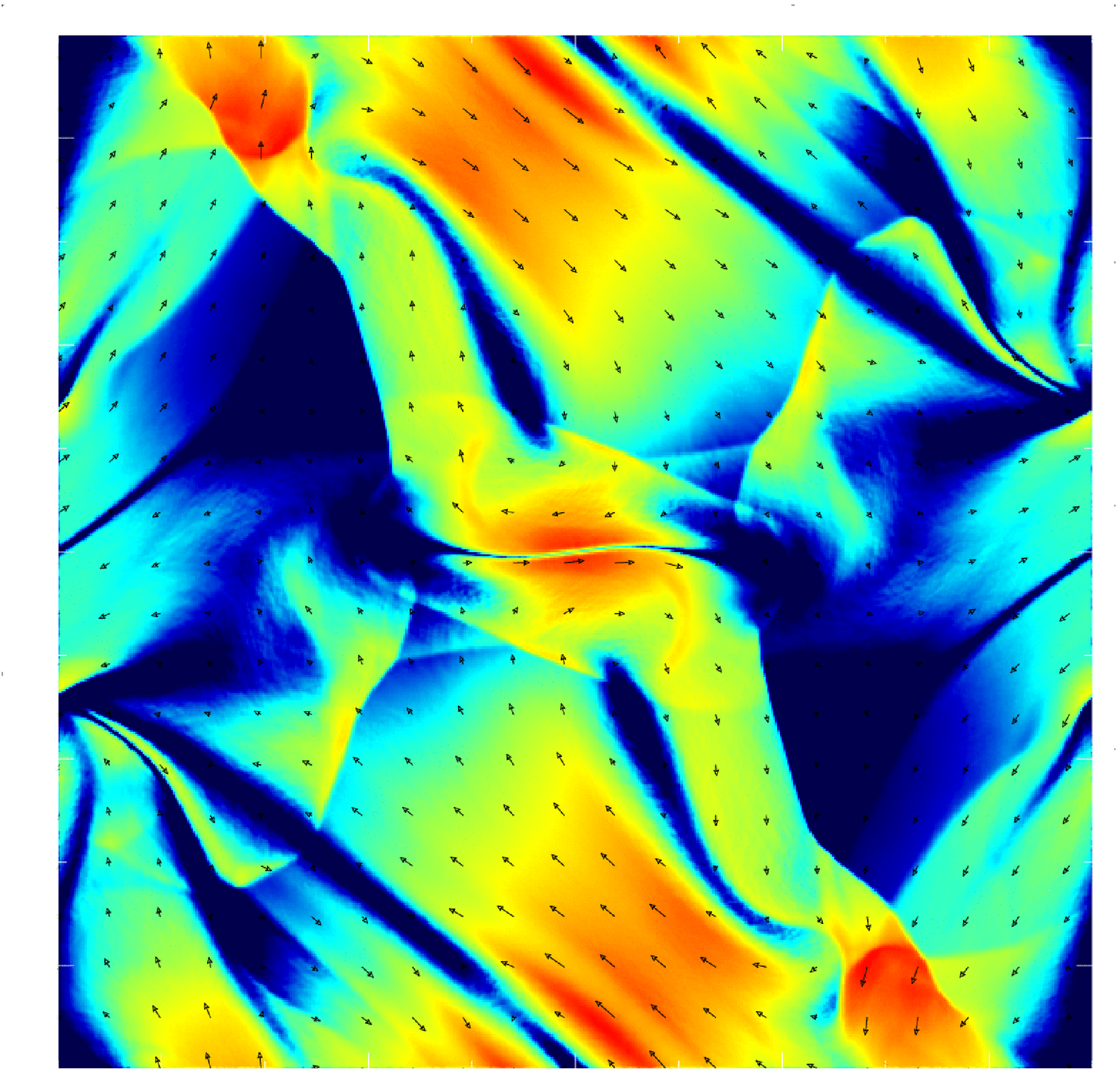}
\caption{The density (upper row) and magnetic pressure (lower row) distributions for the Orszag-Tang vortex at $t=0.5$ for three resolutions: $128 \times 146$ (left), $256 \times 294$ (centre) and $512 \times 590$ (right). The magnetic field direction and strength is shown by the arrows. The initial velocity field and periodic boundaries lead to complex interactions between the magnetic field and shock present in the simulation. The code produces results very similar to others presented in the literature.}
\label{fig:Orszag2D}
\end{figure*}

The simulation is evolved until $t=0.1$ and the result can be qualitatively seen in Fig. \ref{fig:Rotor2D}. The material contained in the disc is thrown out into the surrounding medium, but is contained by the magnetic field. The quantitative comparison between the code and the ATHENA simulation can be seen in Fig. \ref{fig:RotorCUT}. The ATHENA simulation was run with 400 x 400 cells. The result agrees well with the solution provided by ATHENA and shows very little smoothing of features by the time dependant artificial dissipation. There is a small deviation between the two solutions either side of the main density peaks. The difference between the solutions is caused by the code smoothing out the density change across the edge of the shock front. 

\subsection{Orszag-Tang Vortex}

The compressible Orszag-Tang vortex was developed from a test problem in \citet{tr79} and is a common test of MHD implementations. It shows the code's ability to handle the interaction between different classes of shock waves and the transition to MHD turbulence. Using $\gamma = 5/3$, a magnetic to thermal pressure ratio of $10/3$ and an average Mach number of unity, a uniform medium with periodic boundaries is set out in a hexagonal lattice with $P = 5/3B_0^2$ and $\rho = \gamma P/v_0$, where $B_0 = 1/\sqrt{4\pi}$ and $v_0 = 1.0$. The magnetic field is initially set as $B_x=-B_0\sin(2\pi y)$, $B_y=B_0\sin(4\pi x)$ and $B_z=0$. The gas is also given an initial velocity of $v_x = -v_0\sin(2\pi y)$, $v_y = v_0\sin(2\pi x)$ and $v_z = 0$. We performed the simulation for 3 different resolutions: $128 \times 146$, $256 \times 294$ and $512 \times 590$.

\begin{figure}
\centering
\includegraphics[width=\hsize,height=6cm]{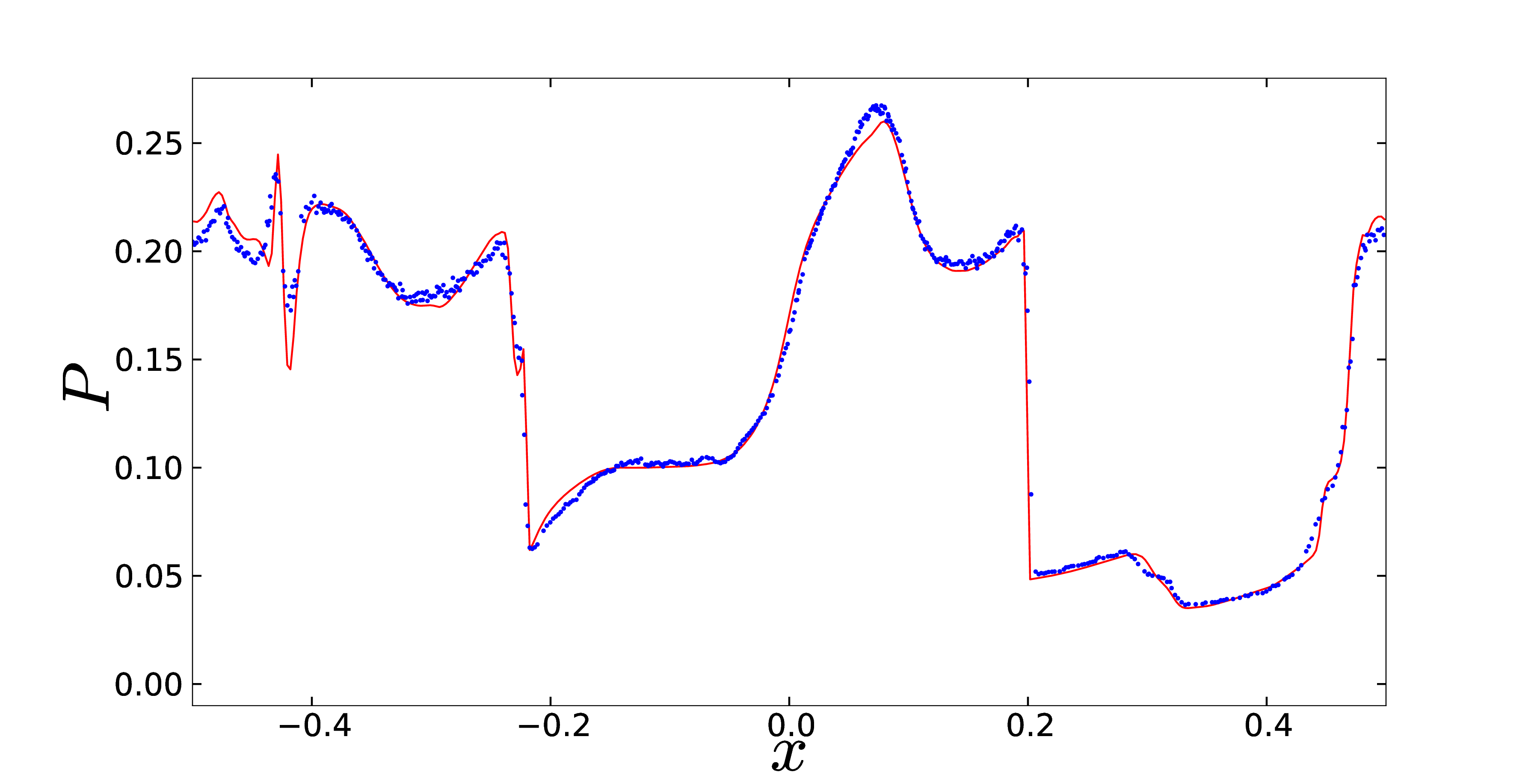}
\includegraphics[width=\hsize,height=6cm]{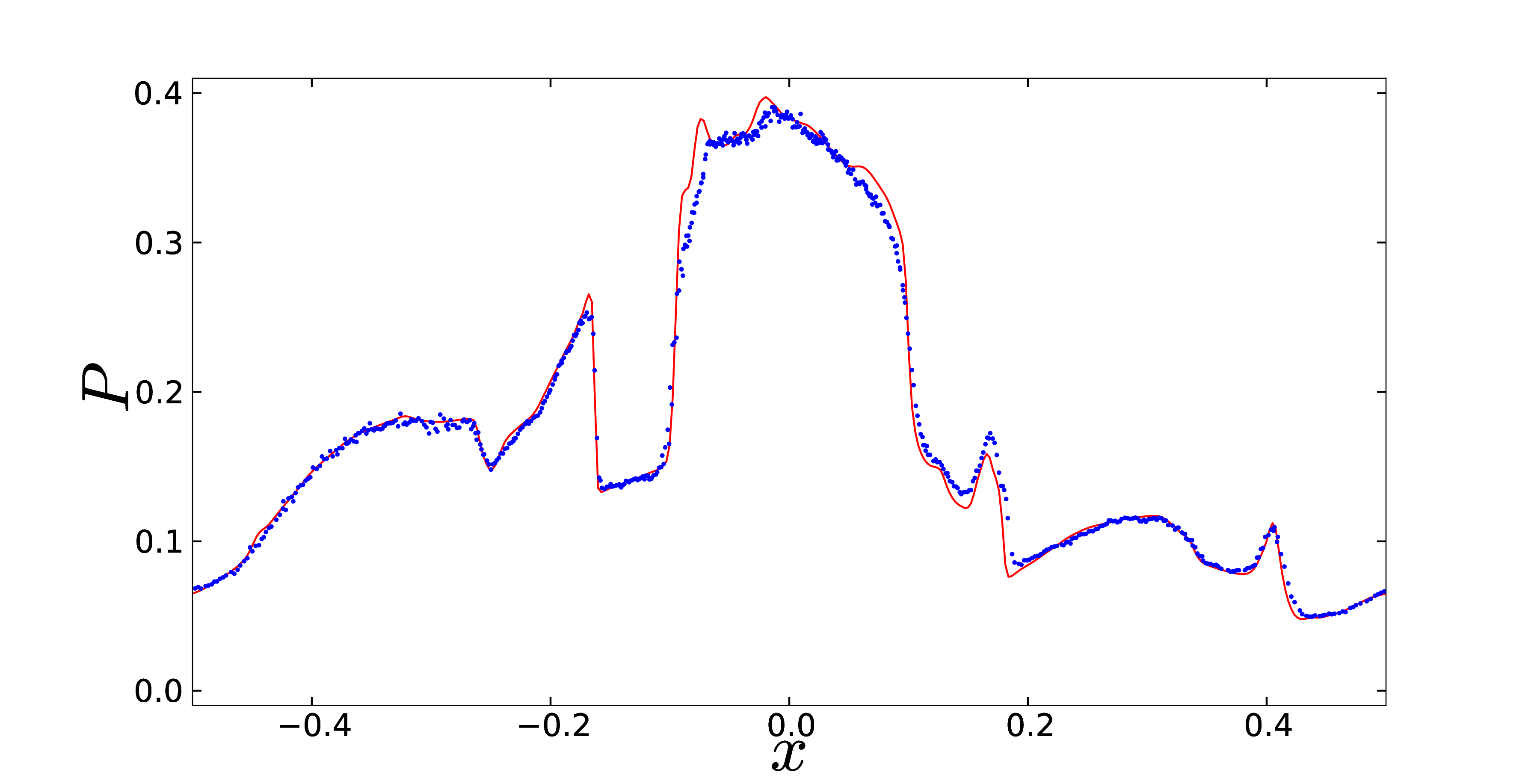}
\caption{The pressure along 1D cuts at $y=0.3125$ (upper) and $y=0.4277$ (lower) in the Orszag-Tang vortex. These two cuts have been chosen to allow comparison with other Orszag-Tang solutions presented in the literature. The ATHENA solution is plotted in red, while the solution produced by the code is shown via the blue dots. There is good agreement between the solutions.}
\label{fig:OrszagCUT}
\end{figure}

The simulation was evolved until $t=0.5$ and the results are shown in Fig. \ref{fig:Orszag2D}. The results show good agreement with those published in the literatures, such as \citet{tr95}. As the resolution increases the complex interplay between the four shock fronts and the magnetic field becomes clearer, with more small scale features easily visible in the $512 \times 590$ run compared to the $128 \times 146$ run. 

For a quantitative comparison of the code's ability, we measure the variation of the pressure as a function of $x$ at a fixed $y$ and compare the results to the solution produced by ATHENA. The results are seen in Fig. \ref{fig:OrszagCUT}. There is good agreement between the two solutions. A small amount of smoothing can be seen in the upper plot at $x=-0.45$ where the ATHENA solution produces very sharp features.

\subsection{MHD Point-like Explosion}
The MHD blast wave test is identical to the Sedov test commonly used for testing pure-hydrodynamic codes but with a magnetic field added to the simulation. The uniform field is initially added in the $x$ direction, such that $B=[B_0,0,0]$, where $B_0=3$. The medium is a constant density, $\rho=1$, with a hot point source embedded in it. The simulation box runs from $-0.5$ to $0.5$ in the $x$, $y$ and $z$ directions. The $100^3$ particles are set out on a cubic grid ensuring that a particle occupies the position at $(0,0,0)$. The central particle is then given an energy $100$ times the energy of the ambient medium. We do not apply the smoothed central high energy sphere used in some literature \citep[e.g.][]{cp09}. Instead we let the energy spread from a single high energy particle. This is a much harder initial condition than the smoothed case. Due to the strength of the field, the magnetic pressure plays an important role in the evolution of the shock. 

\begin{figure}
 \centering
 \includegraphics[width=\hsize,keepaspectratio=true]{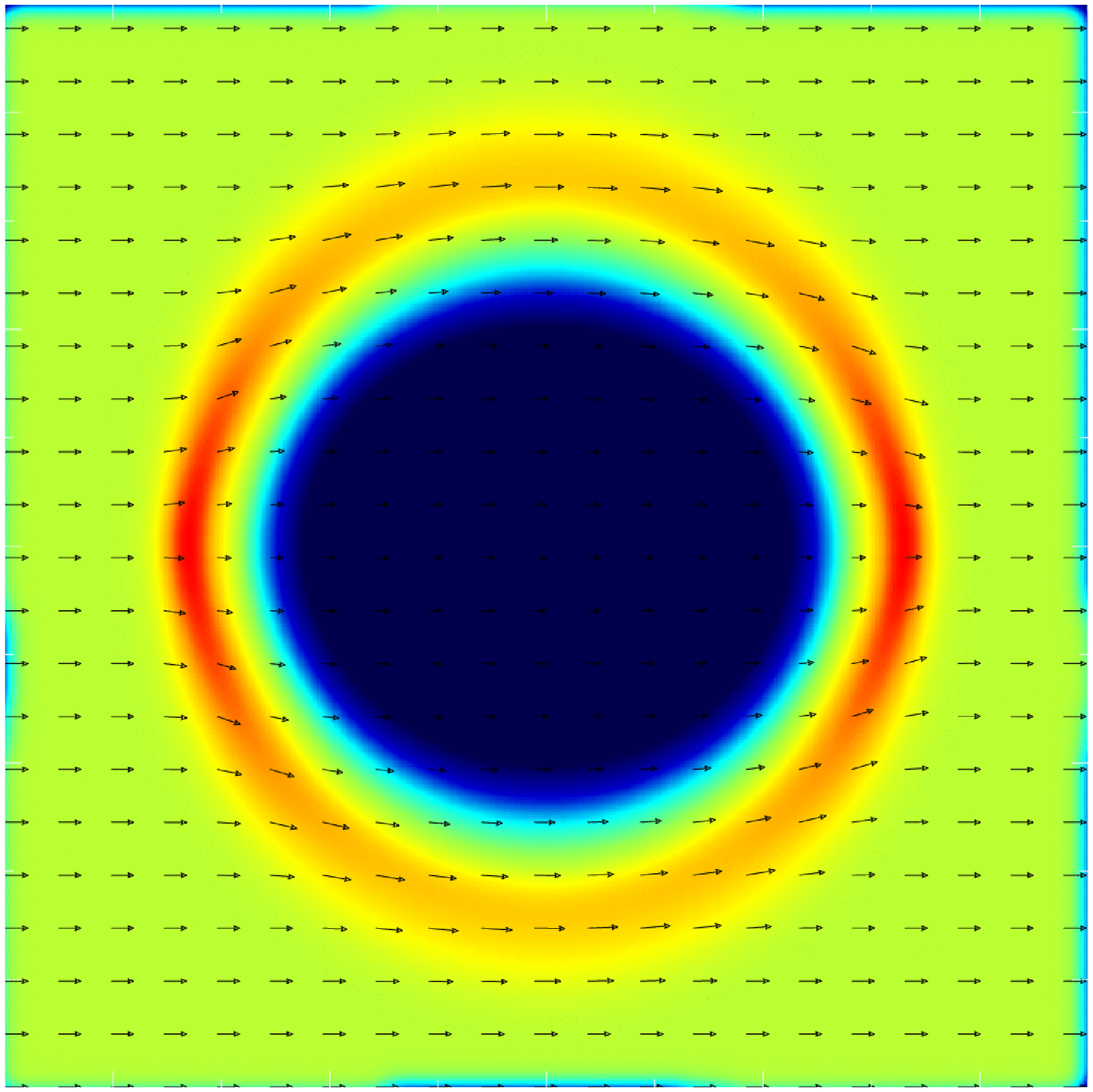}
 \includegraphics[width=\hsize,keepaspectratio=true]{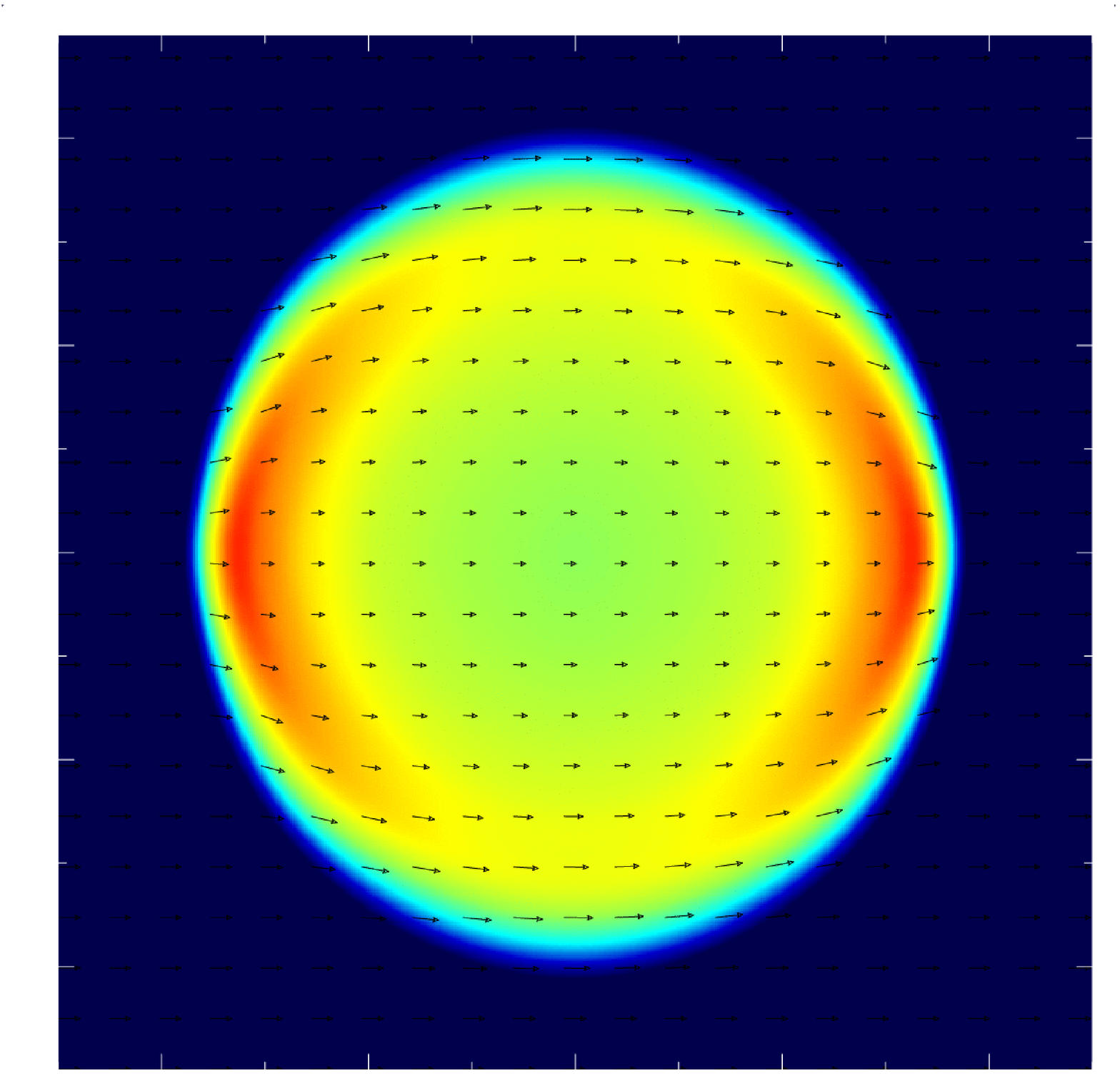}
 \caption{Plot of the density (top) and pressure (bottom) for the 3D magnetic blast wave test. The arrows indicate the strength and direction of the magnetic field in the x-y plane. The solution produced by the code agrees well with other solutions published}
 \label{fig:blastMHD2D}
\end{figure}

As Fig. \ref{fig:blastMHD2D} shows, particles moving in a direction perpendicular to the orientation of the field are constrained by the magnetic tension force, which prevents the density at the shock front from increasing to the levels seen in the $x$ direction. It is also noticeable that the field lines are not significantly bent by the shock front. The solution produced by the code agrees well with other 3D MHD blast wave results shown in the literature \citep[e.g.][]{tr01,cp07}.

\subsection{Numerical Parameters}

\begin{figure}
 \centering
 \includegraphics[width=\hsize,height=6.00cm]{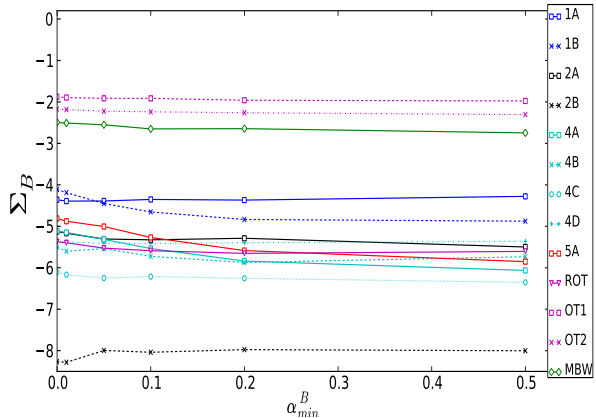}
 \caption{Mean divergence error as a function of the minimum applied magnetic dissipation, $\alpha^B_{min}$. The error is averaged across all particles in the simulation. The labels 1A to 5A represent the 1D shock tube tests in Section 3.1, ROT is the 2D rotor test in Section 3.2, OT1 and OT2 are the Orszag-Tang vortex at a resolution of $128\times146$ and $256\times294$ respectively (Section 3.3) and MBW is the MHD Point-like explosion test (Section3.4).}
 \label{fig:alphapara}
\end{figure}

The simulations shown in previous sections use our best compromised set of numerical parameters that can produce satisfactory results for all of the tests, including the cosmological simulation. The value of the magnetic dissipation is a compromise between minimising the smoothing of sharp features and reducing oscillations in the magnetic field. This can be seen in the cosmological simulation, where the magnetic field is highly dependant on the applied dissipation, and other simulations which show the presence of oscillations in the solution, such as the magnetic field for test 4C in Section 3.1. In order to get more of a quantitative measure of the quality of the solution the mean of all $ \nabla\cdot\textbf{B}\not= 0$ errors, $\Sigma_B$, in the simulation volume can be calculated using
\begin{equation}
 \Sigma_B = \left< \rm{log}_{10}\left(\frac{\it{h}\nabla\cdot \textbf{B}}{|\textbf{B}|}\right)\right>_V.
\end{equation}
The numerical parameters can then be varied and the error can be plotted as a function of the parameter. Fig. \ref{fig:alphapara} shows the mean error of the simulation as a function of the minimum applied artificial dissipation, $\alpha^B_{min}$ for all of the test simulations shown in Sections 3.1-3.4. Fig. \ref{fig:alphapara} shows that applying a small amount of dissipation to the simulation improves the divergence error, or the noise, of the solution. However, it also shows that increasing $\alpha^B_{min}$ beyond $\alpha_{min}^B=0.1$ does not lead to a further reduction in the noise, even in multi-dimensional tests. Any small amount of applied dissipation will lead to some smoothing of features and so the best range for $\alpha^B$ for the non-cosmological test simulations was to allow it to vary between $0.05$ and $1.0$ for each particle. However, due to the dependence of the cosmological magnetic field on the level of applied dissipation the value of $\alpha^B$ was allowed to vary between $0.0$ and $1.0$, and therefore we chose $\alpha^B_{min}=0.0$. Fig \ref{fig:alphapara} demonstrates that this choice of $\alpha^B_{min}$ provides a reasonably low level of error.

\begin{figure}
 \centering
 \includegraphics[width=\hsize,height=6.00cm]{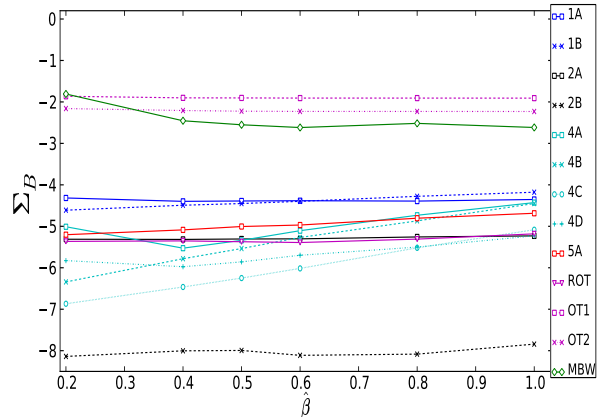}
 \caption{Mean divergence error as a function of $\hat{\beta}$. A value of $\hat{\beta}<0.2$ does not fully remove the tensile instability and the code does not properly run. The labels have the same meaning as Fig. \ref{fig:alphapara}}
 \label{fig:betapara}
\end{figure}

The second important parameter for GCMHD+ is $\hat{\beta}$, which controls the level of $\nabla\cdot \textbf{B}$ force subtraction. In \citet{ic04} it was suggested that the value of $\hat{\beta}$ could be less than one and stability still ensured. In \citet{cp09} they argue that it is unclear as to whether the conclusions by B\o rve et al. hold in 3D. We ran all of our test simulations and measured the value of $\Sigma_B$ as a function of $\hat{\beta}$. Fig. \ref{fig:betapara} shows that increasing $\hat{\beta}$ has a different effect depending on the dimensions of the simulation. Note that if we apply $\hat{\beta}<0.2$ the code collapses and cannot solve the test problems. For 1D tests, an increase in $\hat{\beta}$ generally leads to an increase in the average error. However, for higher dimension tests an increase leads to a reduction in the error, but with a reduced return for a value of $0.5$ or more. From the result shown in Fig. \ref{fig:betapara} and to minimise the violation of momentum conservation, we chose a value of $\hat{\beta}=0.5$.

\section{Cosmological simulation}
The Santa Barbara Cluster simulation was used as the base for the cosmological simulation. Since \citet{sbc99} this has become a standard test for cosmological hydrodynamic codes. The initial conditions are set up to create a galaxy cluster. It was shown in \citet{sbc99} that different codes and numerical techniques produce generally consistent results.

\begin{table}
 \centering
 \begin{tabular}[b]{ccc}
 Simulation & $\alpha^B_{min}$ & Artificial Conductivity \\
 \hline
 C1 & 0.0 & ON \\
 C2 & 0.005 & ON \\
 C3 & 0.05 & ON \\
 C4 & 0.0 & OFF \\
 C5 & 0.005 & OFF \\
 C6 & 0.05 & OFF \\
 \end{tabular}
 \caption{Properties of the cluster simulations. All other parameters were kept constant to allow the effect of dissipation and thermal conductivity to be seen.}
\label{tab:clusprop}
\end{table}

The simulation assumes a classical, flat CDM cosmology. The simulation allows a spherical region of $32h^{-1}$ to expand with the Hubble flow. The test uses open boundary conditions. The gas particles have a mass of $8.67 \times 10^8 M_{\odot}$ and the dark matter particles have a mass of $7.80 \times 10^9 M_{\odot}$. (This corresponds to a slightly lower resolution than the $1\times$ simulation of \citet{cp09}). To this  a homogeneous magnetic field of $10^{-11}G$ is initially applied. The simulation is started at a redshift of $z=20$ and evolved to the current epoch. This produces a final cluster of mass $1.16 \times 10^{15} M_{\odot}$. The cluster simulation was run varying the  value of $\alpha^B_{min}$ to test the effects of the dissipation scheme on the development of the magnetic field. The effect of artificial thermal conductivity (A.C.) on the magnetic field was also tested by running the simulation with and without it. The values for the different runs can be seen in Table \ref{tab:clusprop}.

\begin{figure}
\begin{flushleft}
 \includegraphics[width=\hsize,height=5.94cm]{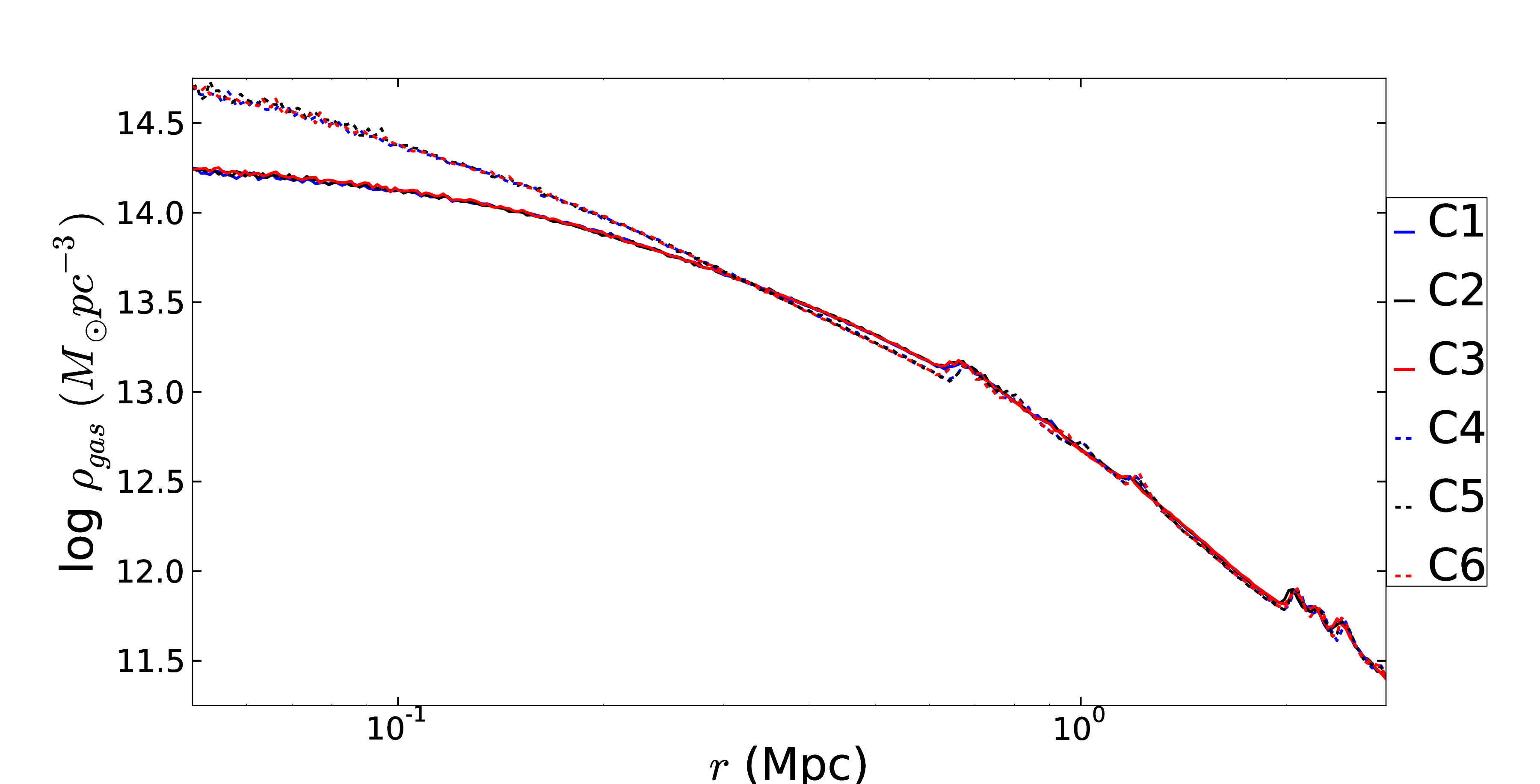}
 \includegraphics[width=\hsize,height=5.94cm]{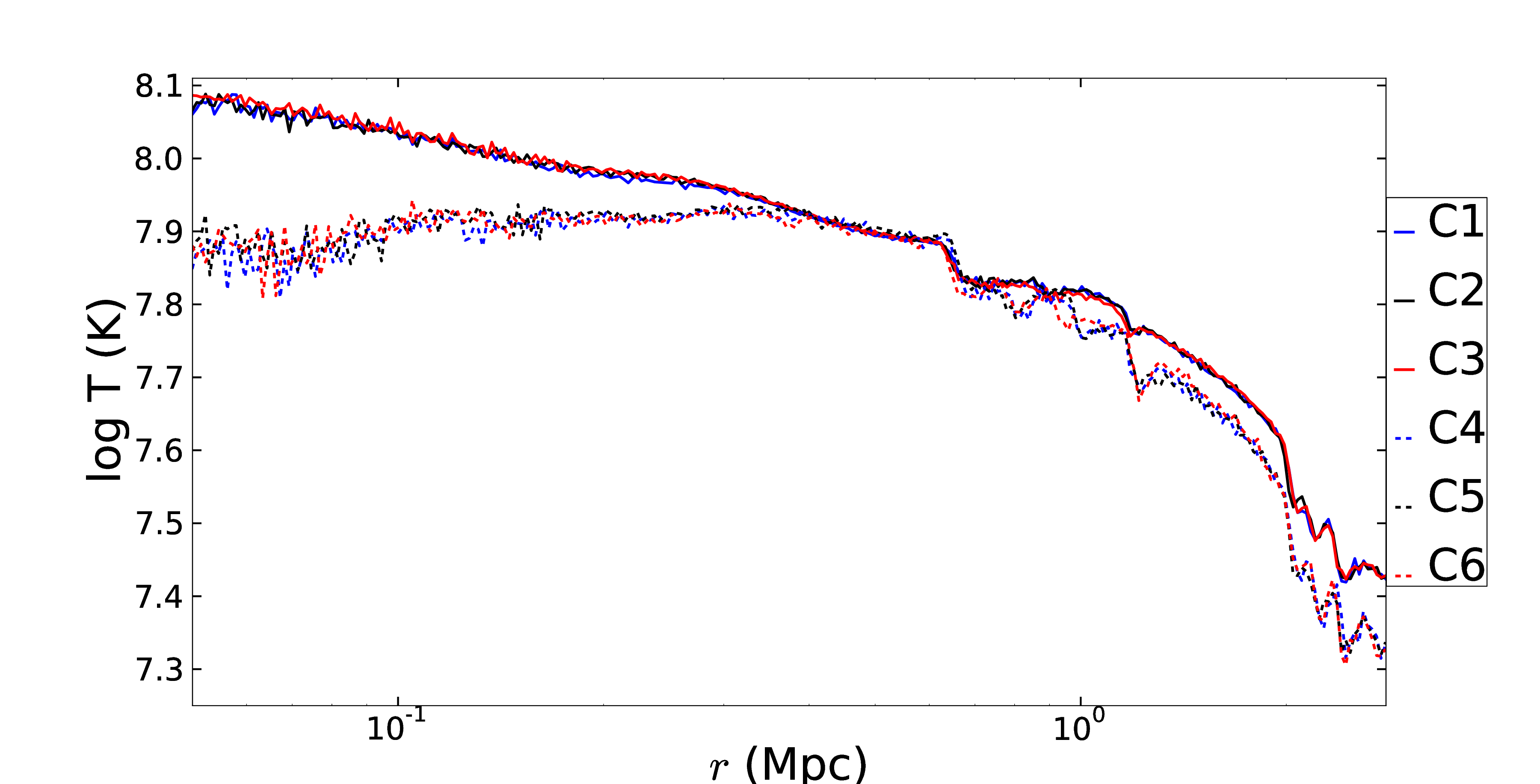}
 \includegraphics[width=\hsize,height=5.94cm]{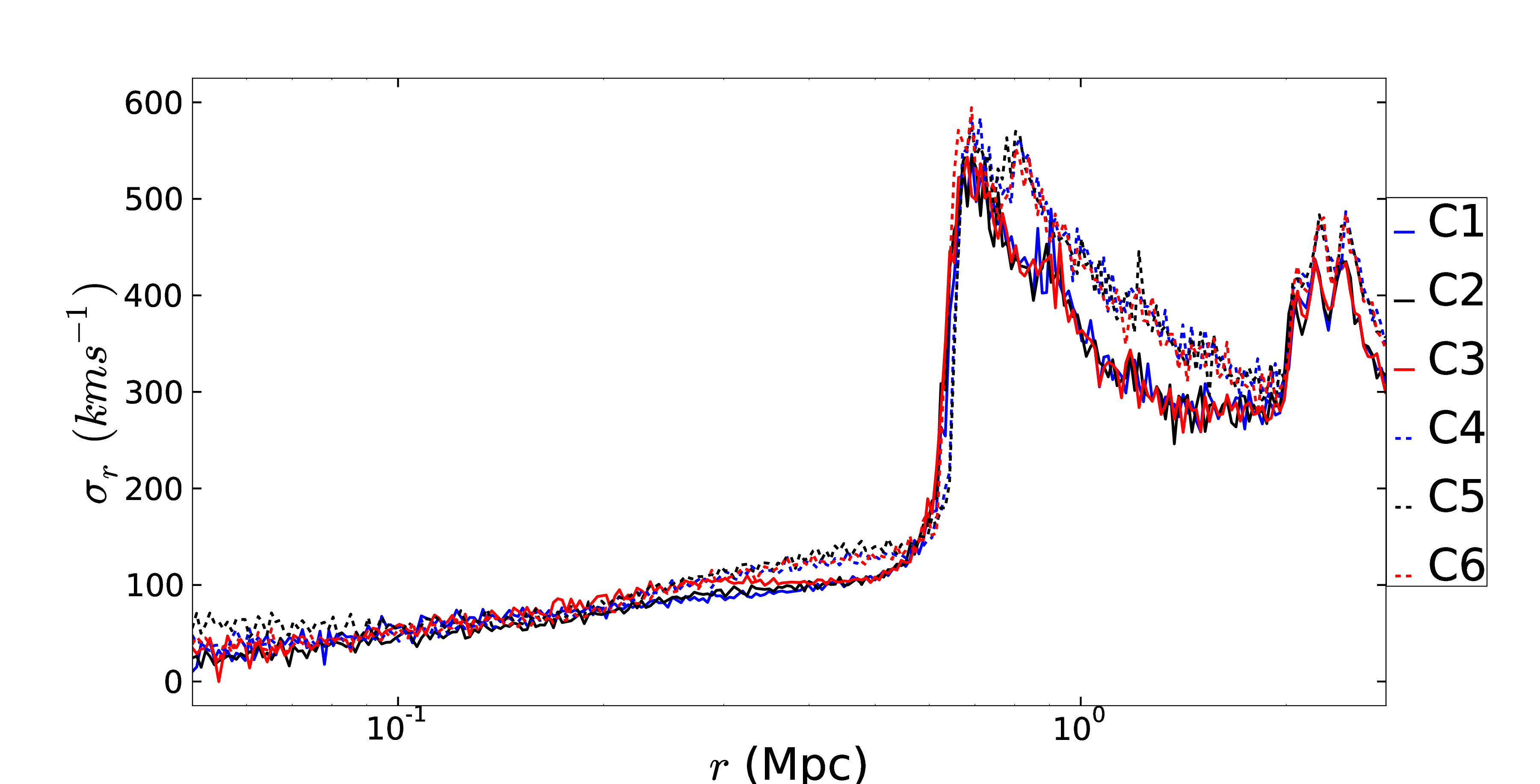}
\end{flushleft}
\caption{Radial profiles of the density (top), temperature (middle) and velocity dispersion (bottom) for all of the simulated clusters at $z=0.0$. These agree well with each other and with the results in the literature.}
\label{fig:CLUShydroprofiles}
\end{figure}

\begin{figure}
 \includegraphics[width=\hsize,height=5.931cm]{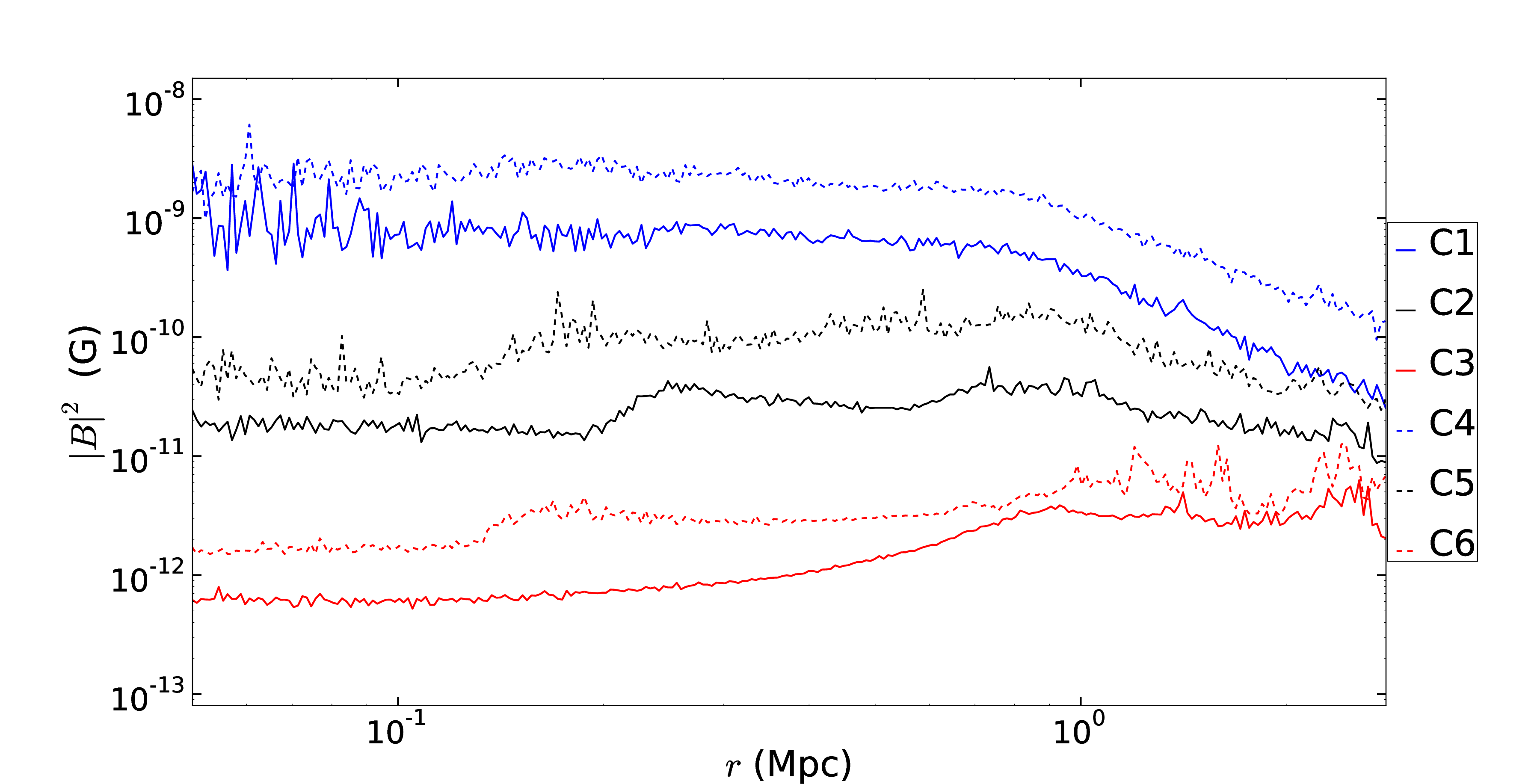}
 \includegraphics[width=\hsize,height=5.931cm]{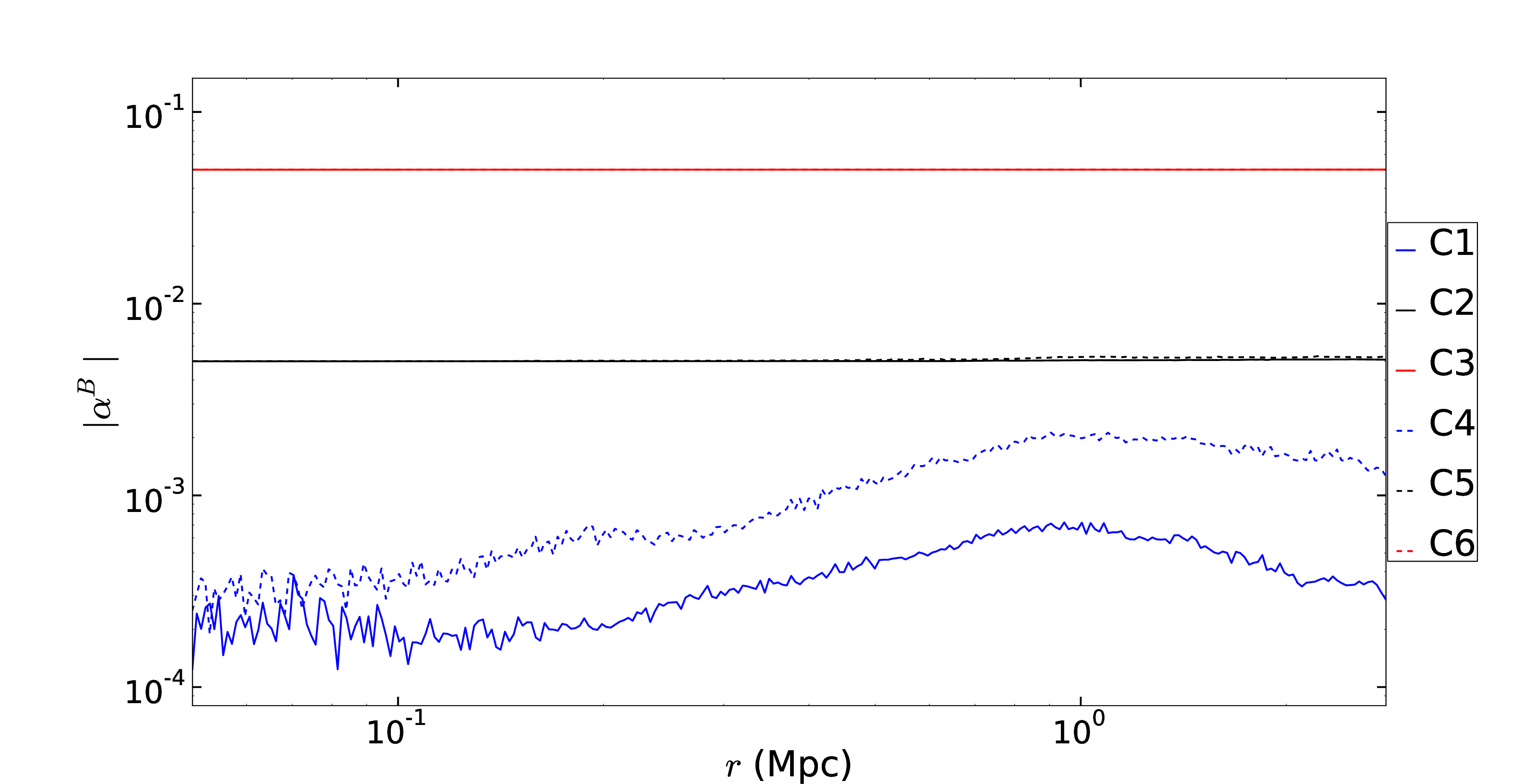}
 \includegraphics[width=\hsize,height=5.931cm]{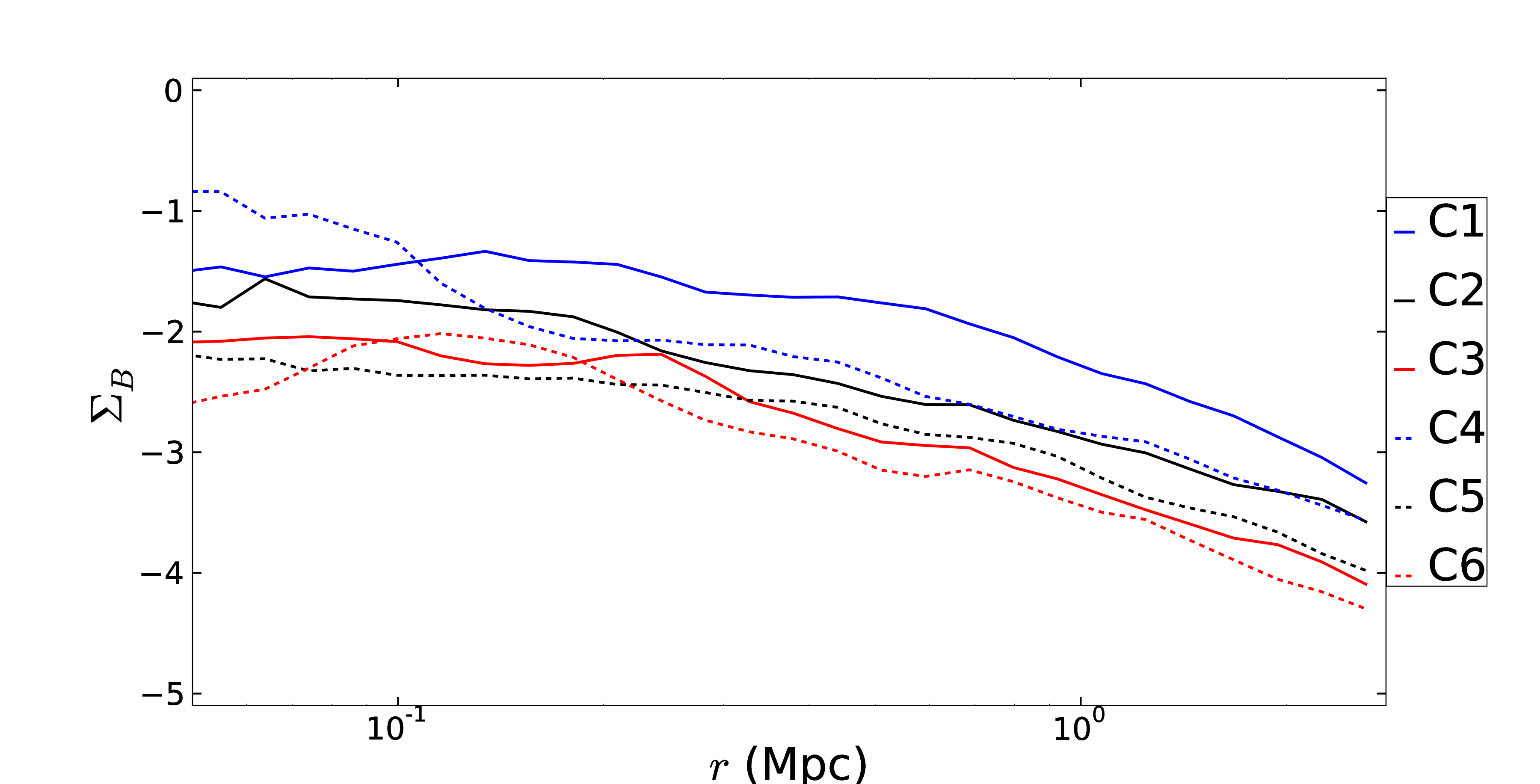}
 \caption{Radial profiles of the magnetic field, mean dissipation ($|\alpha^B|$) and mean $\nabla\cdot \textbf{B}$ error for all of the simulated clusters at $z=0.0$. The magnetic field shows that enforcing a minimum limit threshold of dissipation causes the field to decay too quickly. The middle plot shows that the simulation will settle at a lower value of dissipation, and a higher level of magnetic field, when $\alpha^B_{min}=0.0$ is applied. The bottom plot shows that $\nabla\cdot \textbf{B}$ error is kept at a satisfactory level for all of the simulated clusters.}
\label{fig:CLUSMHDprofiles}
\end{figure}

Fig. \ref{fig:CLUShydroprofiles} shows the hydrodynamic properties of the simulated clusters. The radial profiles of the density, temperature and velocity dispersion for the simulated clusters with artificial conductivity agree well with each other. The clusters without artificial conductivity show a higher core density, lower core temperature compared with the clusters that have artificial conductivity \citep{gcd11}, but their profiles agree between models with different values of $\alpha^B_{min}$. This result is expected as the magnetic energy density is very much less than the kinetic energy and so any magnetic field present should not influence the development of the cluster. The profiles of the clusters which use artificial conductivity agree well with the mesh code results presented in \citet{sbc99}, with a core density peaking at $10^{14} M_{\odot}~\rm{pc}^{-3}$ and temperature of $10^8K$, which is discussed in \citet{gcd11}.

The magnetic field of the simulated cluster can be seen in the upper plot of Fig. \ref{fig:CLUSMHDprofiles}. Clearly the final profile and strength of the magnetic field depends on the minimum value of dissipation enforced on the simulation. Clusters with higher levels of minimum dissipation have a lower strength magnetic field through the entire volume, and the strength of the field in the core is reduced by 4 orders of magnitude compared to the case with $\alpha^B_{min}=0.0$ (simulation C1). This strong dependence on the level of dissipation was also seen in results presented in \citet{cp09}. The clusters simulated without artificial conductivity show a similar radial profile for their magnetic field when compared with the cluster simulated with A.C. and the same minimum dissipation, but the strength of the field is greater. We found that A.C. lead to simulations with a much smoother density and temperature distribution \citep[see][]{gcd11}. Simulations without A.C. have more small scale structure in their density field, which amplifies the magnetic field and keeps it at a higher level compared to the models with A.C.

The middle plot of Fig. \ref{fig:CLUSMHDprofiles} shows the radial profile of $|\alpha^B|$, i.e. the average level of dissipation applied. It shows that enforcing even a very small level of minimum dissipation is too high for a cluster simulation. The radial profile for C1 shows that when $\alpha^B_{min}=0.0$ is applied, the code produces a dissipation of roughly $10^{-4}$. The average value is $\alpha^B = 3.35\times 10^{-4}$. The cluster core has a lower level of dissipation compared to the edges of the cluster, where a small amount of material is still falling in. The effect of these minor mergers can be seen in the top panel of Fig. \ref{fig:CLUSMHDprofiles} at roughly $1~\rm{Mpc}$.

Recently \citet{niMHD11} used a constant dissipation level of $\eta_m=6\times10^{27}\rm{cm}^2\rm{s}^{-1}$ to produce the observed magnetic field in their cosmological simulation. They also derived a value of $\eta_m=2\times10^{27}\rm{cm}^2\rm{s}^{-1}$ from turbulence arguments. Their $\eta_m$ simply corresponds to $\alpha^B/2$. We find an average value of $\alpha^B = 3.35\times 10^{-4}$. Converting this to cgs units, we obtain $\eta_m = 2.14\times 10^{27} \rm{cm}^2\rm{s}^{-1}$ which is similar to their values.

The lower plot of Fig. \ref{fig:CLUSMHDprofiles} shows the radial profile for the average $\nabla\cdot \textbf{B}$ error calculated by eq. (14). The error remains at the level of a few percent throughout the cluster volume and is acceptable for all levels of applied magnetic dissipation.

\begin{figure}
 \includegraphics[width=\hsize,height=6.00cm]{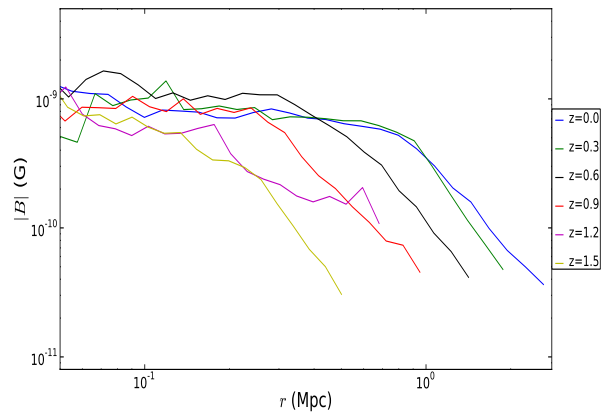}
 \caption{Radial profiles for the magnetic field for the redshift range of $z=1.5$ to $z=0.0$ of simulation C1. The magnetic field in the cluster has already roughly reached it maximum value and correct profile by the time the final cluster is forming at z=1.5. The in-fall of small structures causes the value to change slightly.}
\label{fig:BevoC1}
\end{figure}

The evolution of the magnetic field from $z=1.5$ to the current epoch in simulation C1 is shown in Fig. \ref{fig:BevoC1}. At $z=1.5$ there are two major protoclusters and these then merge before $z=1.2$. The peak field in the cluster remains roughly constant from $z=1.5$ to the final value found in the simulation. The effect of merging and in-falling material can be seen as the radial field profile evolves and the change in field profile at $z=1.2$ is  due to the major merger of the two large protoclusters. The field then relaxes back via dissipation to its final shape.

There is a very slight increase in the peak field at $z=0.6$. This is when the last big in-fall of material occurs for the cluster. This shows that in the absence any additional source terms the magnetic field is amplified by the merging of protoclusters and by minor mergers at later epochs.

\begin{figure}
 \includegraphics[width=\hsize,height=6.00cm]{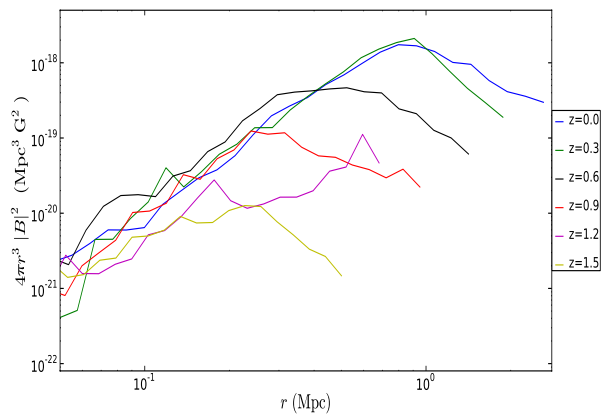}
 \caption{Radial profiles for the strength of the magnetic field for the redshift range of $z=1.5$ to $z=0.0$ of simulation C1. As the cluster evolves the strength of the magnetic field increases. The power peaks at increasing radius as the cluster grows.}
\label{fig:BpowC1}
\end{figure}

Fig. \ref{fig:BpowC1} shows the evolution of the $overall$ magnetic field strength over the redshift range of $z=1.5$ to $z=0.0$. The overall magnetic field strength grows as the cluster assembles itself. This is due to the in-fall of material on to the cluster. This accretion causes amplification of the magnetic field, increasing its strength. The peak changes with redshift due to the increase in size of the cluster. As the cluster grows, the largest possible coherence length of the magnetic field will increase accordingly and so the peak of the strength will tend to larger radii. In this simulation the strength profile at $z=1.2$ is very different to the others due to the effects of the major merger disturbing it.

\begin{figure}
  \includegraphics[width=\hsize,height=6.00cm]{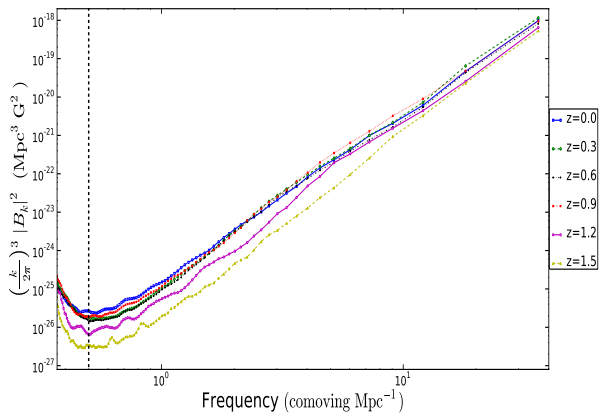}
 \caption{The power of the Fourier transform of the magnetic field as a function of frequency for simulation C1, between the redshift of $z=1.5$ to $z=0.0$. The plot shows a clear increase in the low frequency power of the field as the cluster evolves, while the higher frequency power increases very little. The dotted line represents a cut-off where window effects dominate the spectrum. }
\label{fig:C1FFTs}
\end{figure}

The Fourier transform of the magnetic field was calculated over the same redshift range as Fig. 21. The edge of the cluster was defined by its virial radius and this set the box size for each transform. We adopt the comoving scale, allowing the different redshift profiles to be directly compared. The Fourier power spectra of the magnetic fields are shown in Fig. \ref{fig:C1FFTs}. The general trend shows an increase in the power of the magnetic field as the cluster evolves. At high frequencies the power increases very little as the cluster grows, but at lower frequencies the field increases between $z=1.5$ and $z=0.0$ implying that the field becomes more coherent as the cluster evolves. The major merger is clearly visible as the low frequency power spectrum increases very rapidly as the two protoclusters merge between $z=1.5$ and $z=1.2$. (Note that any frequency lower than the dotted line is dominated by the window function and should be ignored.)

\section{Summary}
We have introduced the MHD component to the N-body/SPH code GCD+. We discussed the addition of the equations of ideal MHD, the choice of instability correction and the addition of dissipative terms to treat discontinuities in the magnetic field. We implement schemes to remove the tensile instability, suggested by \citet{ic01}, and for artificial magnetic dissipation, following \citet{cp04a}. The code's ability to vary this dissipation in the simulation and allow each particle to evolve it own dissipation constant is presented. We put the code through a set of standard 1D shocktube tests, the fast rotor test, the Orszag-Tang vortex test and the MHD Point like explosion test. The numerical parameters were varied for all tests and the best compromise between noise reduction and minimised smoothing was found. The code with the best compromised parameter set performs very well in these tests and agrees with the reference solutions provided by the ATHENA mesh code, where they are available. The code shows no sign of the tensile instability and the magnetic dissipation scheme produces very little smoothing, while allowing the code to accurately capture the features. We then applied the code to a cosmological simulation for the formation of the Santa Barbara galaxy cluster. The code produces the expected hydrodynamic parameters of the cluster. The magnetohydrodynamic parameters are also well captured and show a similar level of magnetic field to the literatures with similar resolution \citep{cp09}. We demonstrate that no minimum limit for the parameter of the dissipation of magnetic field, $\alpha^B_{min}=0.0$, is necessary to minimise the artificial dissipation for the cluster scale magnetic field. This requirement is significantly lower than the previous SPMHD implementations, except for \citet{niMHD11}. Our extensive test simulations in Section 3 demonstrate that $\alpha^B_{min}=0.0$ still leads to satisfactory results. Encouraged by the success of our new MHD code, GCMHD+, we will apply it to higher resolution cosmological simulations, and study how magnetic fields developed in the evolving universe.

\section*{Acknowledgments}

The authors acknowledge the support of the UK's Science \& Technology Facilities Council (STFC Grant ST/H00260X/1). The calculations for this paper were per-formed on Cray XT4 at Center for Computational Astrophysics, CfCA, of National Astronomical Observatory of Japan and the DiRAC Facility jointly funded by STFC and the Large Facilities Capital Fund of BIS. The authors acknowledge support of the STFC funded Miracle Consortium (part of the DiRAC facility) in providing access to the UCL Legion High Performance Computing Facility. The authors additionally acknowledge the support of UCL's Research Computing team with the use of the Legion facility. We also thank James M. Stone for making ATHENA, which we used to obtain reference solutions to many of the tests, publicly available. We also thank the referee for their suggestions and constructive comments.

\bibliographystyle{mn}
\bibliography{ms.bbl}

\appendix
\section{Code parameters}

\begin{figure}
 \includegraphics[width=\hsize,height=5.70cm]{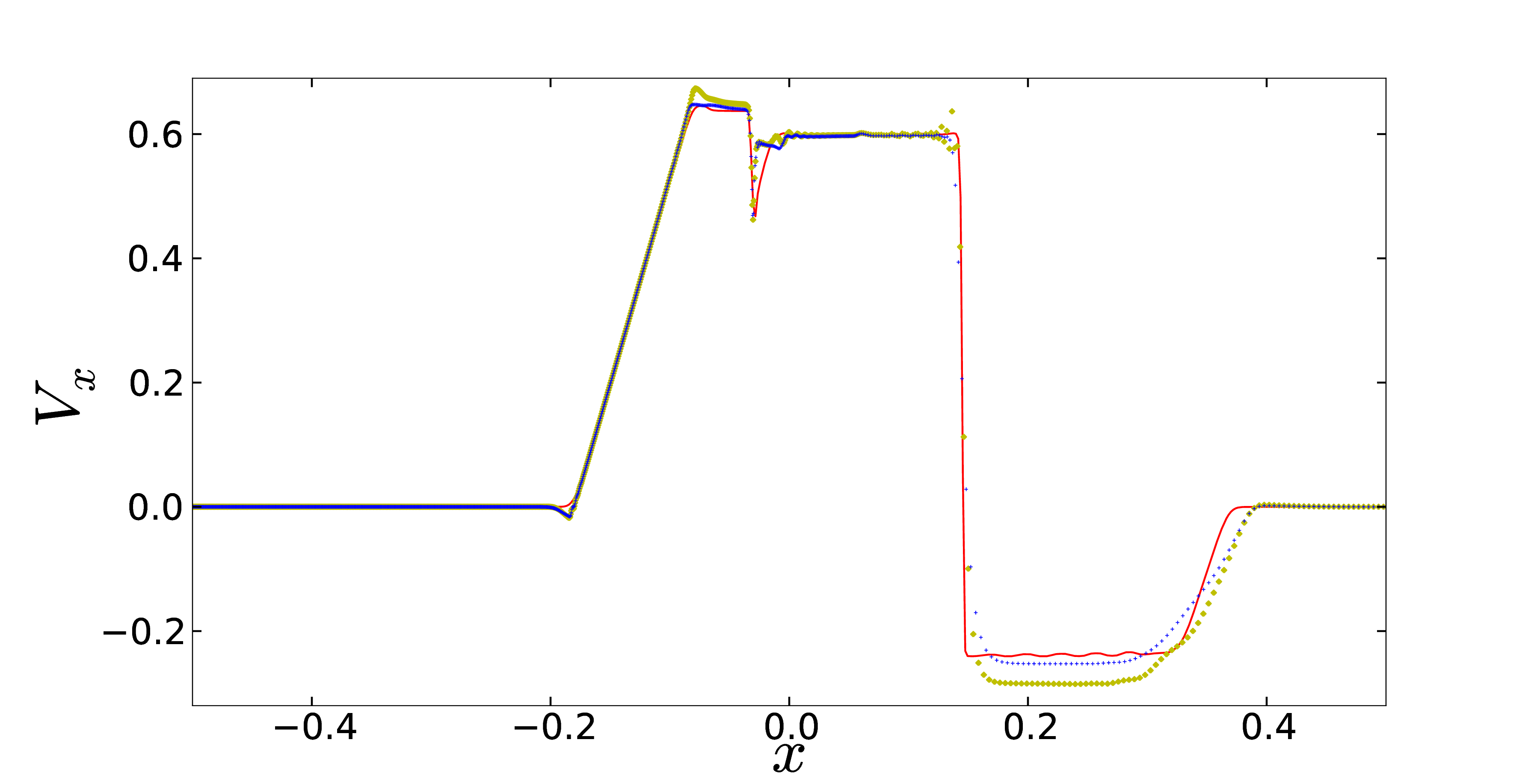}
 \caption{Plot of $v_x$ for shocktube 5A. The reference solution, provided by ATHENA, is shown in red. Two solutions produced by GCMHD+ with (yellow diamonds) and without (blue crosses) the Balsara switch are shown for $v_x$. The solution in the negative velocity region is improved by removing the Balsara switch.}
\label{fig:bal}
\end{figure}

\begin{figure}
 \includegraphics[width=\hsize,height=5.70cm]{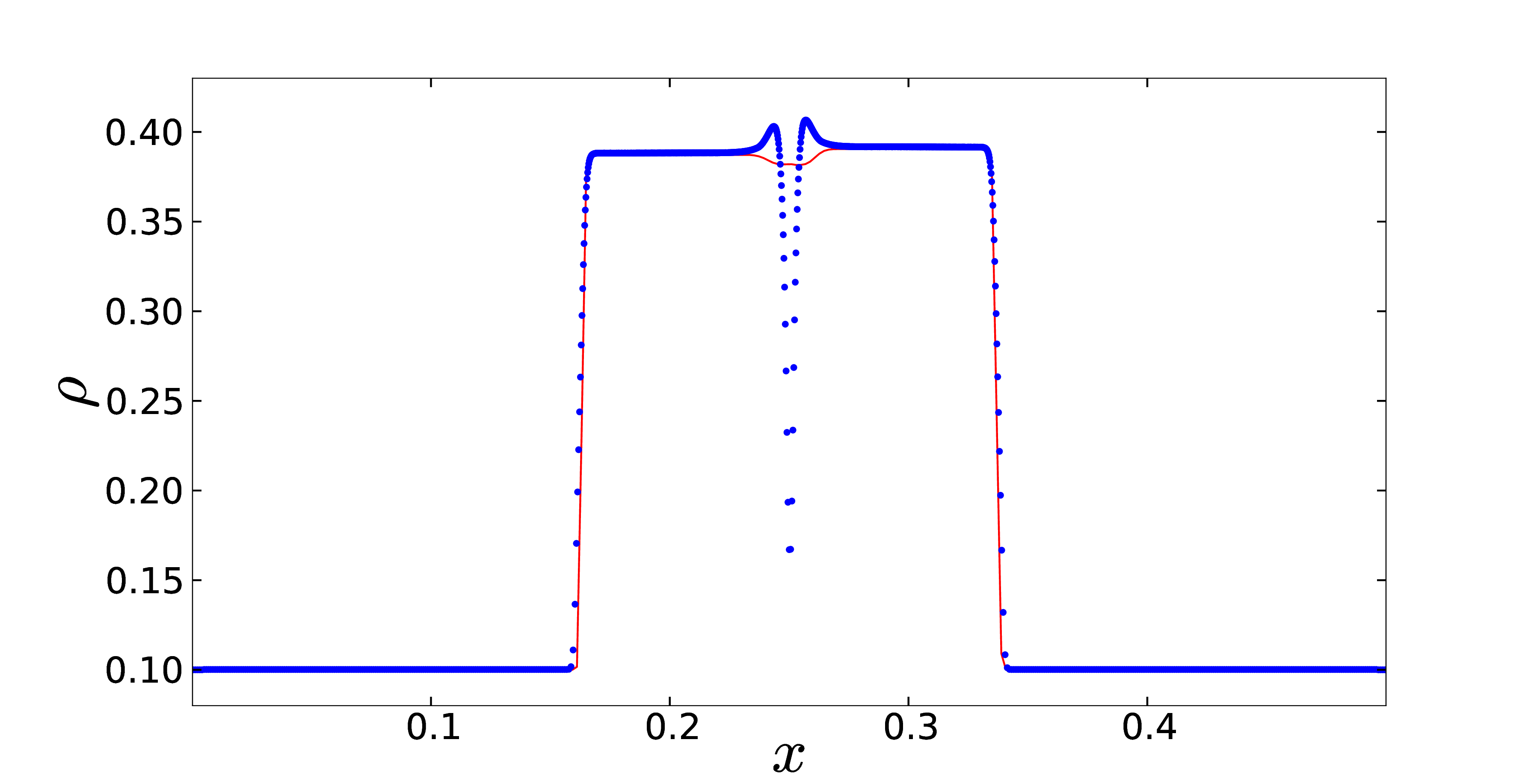}
 \caption{Plot of the density for shocktube 3A. The ATHENA solution is shown by the red solid line, the code's solution without A.C. is shown by the blue points. This is significantly worse when compared to Fig. 5.}
\label{fig:ap3A}
\end{figure}

\begin{figure}
 \includegraphics[width=\hsize,height=5.70cm]{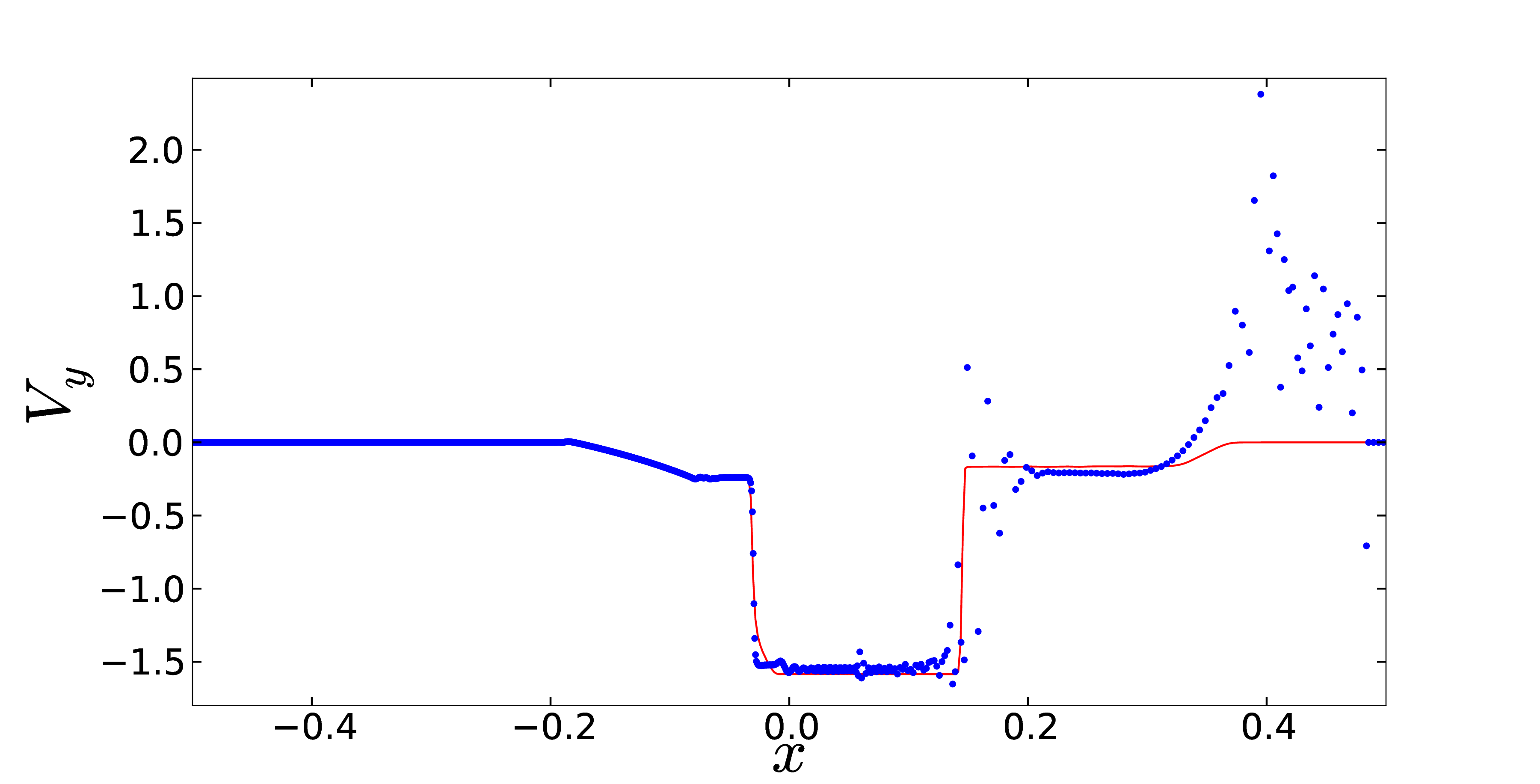}
 \caption{Plot of $v_y$ for shocktube 5A. The reference solution, provided by ATHENA, is shown in red. The solution produced by GCMHD+ with $\hat{\beta}=0.2$ is shown by the blue circles. }
\label{fig:ap5A1}
\end{figure}

\begin{figure}
 \includegraphics[width=\hsize,height=5.70cm]{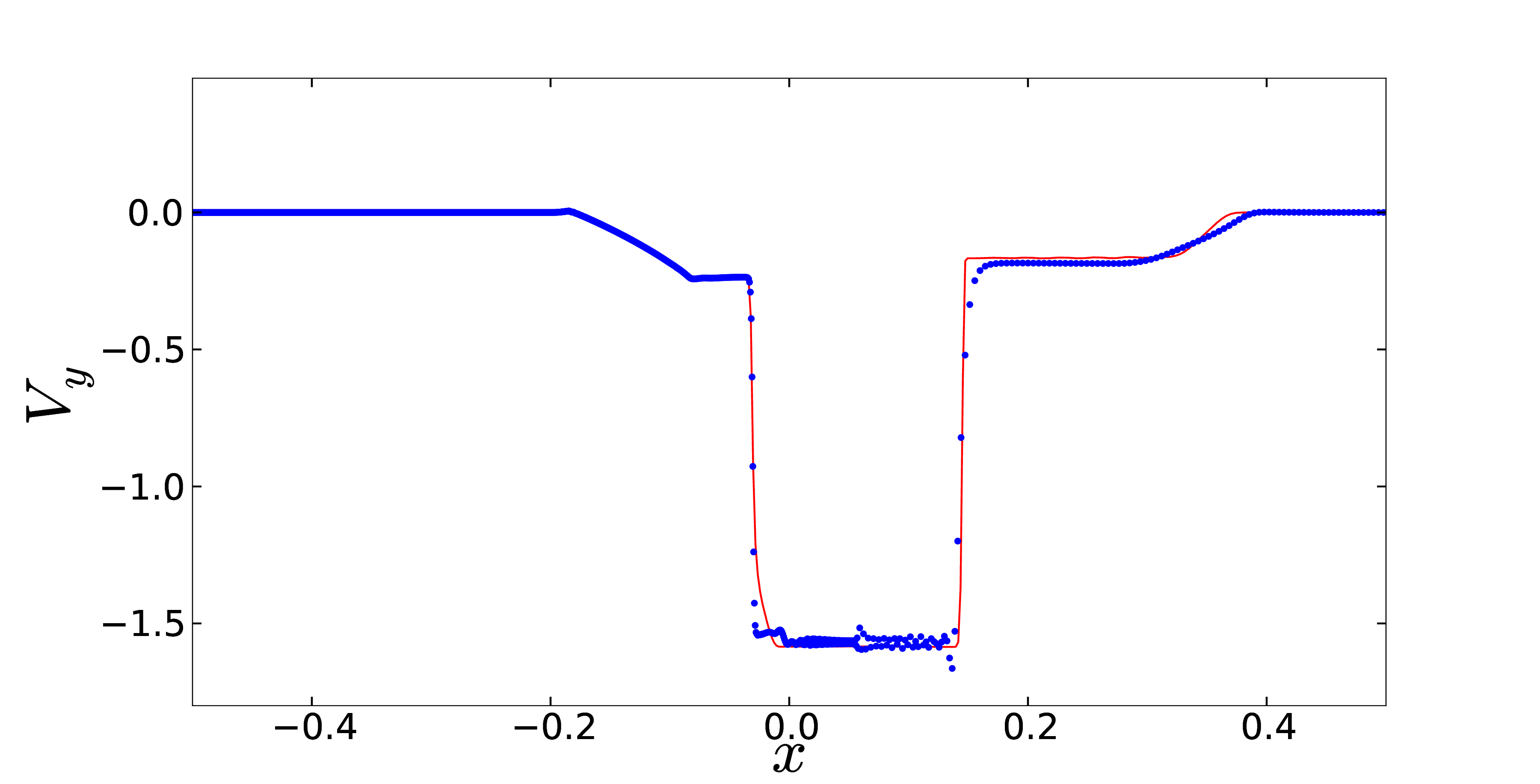}
 \caption{Plot of $v_y$ for shocktube 5A. The reference solution, provided by ATHENA, is shown in red. The solution produced by GCMHD+ with $\hat{\beta}=0.5$ is shown by the blue circles.}
\label{fig:ap5A2}
\end{figure}

\begin{figure}
 \includegraphics[width=\hsize,height=5.70cm]{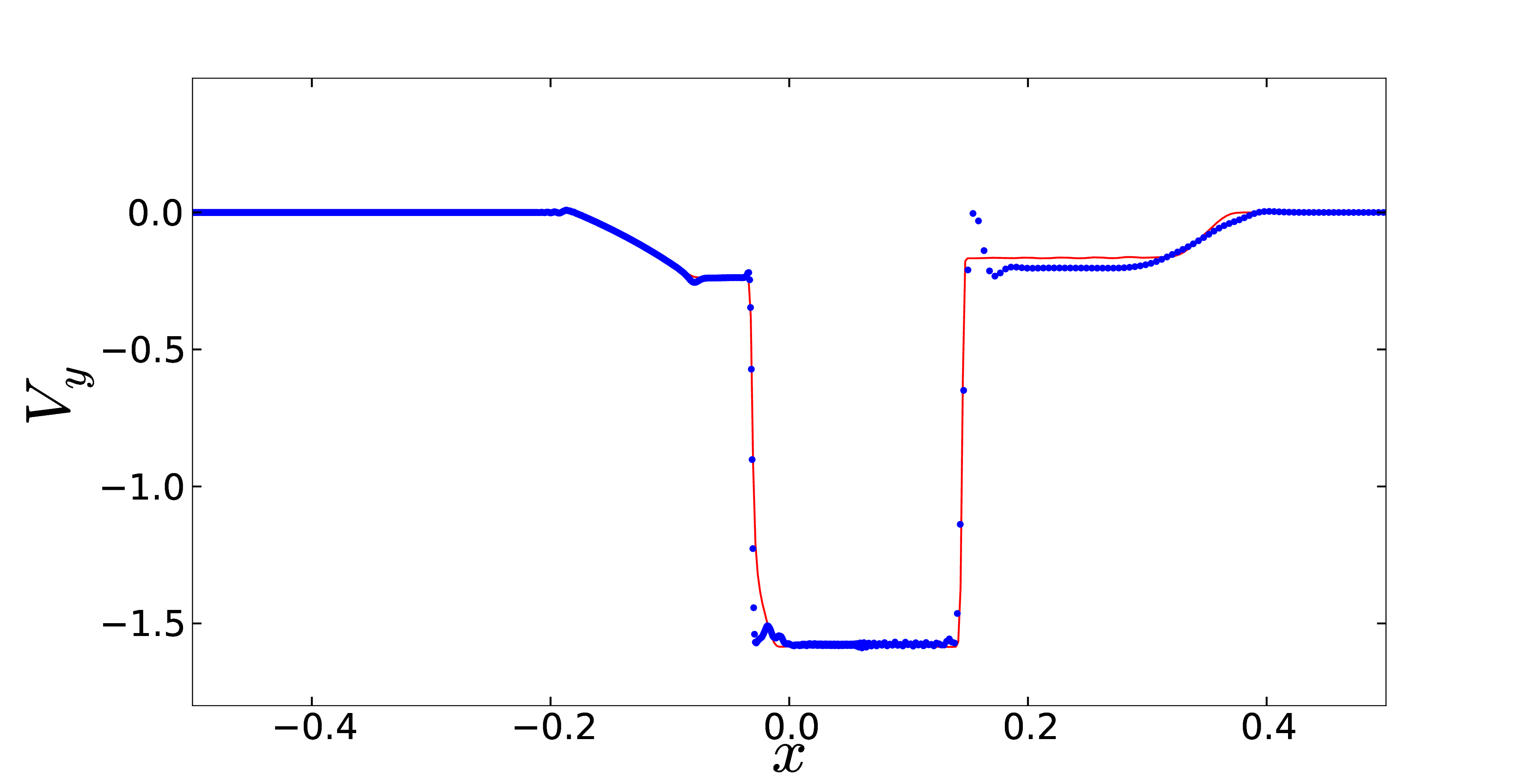}
 \caption{Plot of $v_y$ for shocktube 5A. The reference solution, provided by ATHENA, is shown in red. The solution produced by GCMHD+ with $\hat{\beta}=1.0$ is shown by the blue circles.}
\label{fig:ap5A3}
\end{figure}

Sections 3 and 4 show test results for the best compromised parameter set of GCMHD+. Changing these parameters will change the solutions produced by the code for each test. Here we present the effect of changing these parameters. 

In GCMHD+ the Balsara switch has been removed from the implementation. When using the switch we found that the velocity solutions produced for shocktube 5A were incorrect. Fig. \ref{fig:bal} shows the $v_x$ solution produced by GCMHD+ with and without the Balsara switch. When the switch is included the velocity in negative velocity region is incorrectly captured. When the switch is removed the velocity solution produced by the code shows improved agreement with the reference solution provided by ATHENA. For this reason the switch was removed.

In Section 3.1 we showed the results for the 1D shocktube test 3A. The density profile produced by the code shows evidence of a wall heating error at $x=0.25$. A wall heating error can be reduced by applying A.C. at the point where it occurs. GCMHD+ uses the same time varying A.C. as GCD+ \citep{gcd11}, which reduces to zero where it is not required. Applying a minimum value of A.C. will produce unnecessary smoothing in areas where it is not needed.

Fig. \ref{fig:ap3A} shows the density profile produced by the code without it's scheme for A.C.. The wall heating error becomes significantly worse when the A.C. scheme is turned off. The wall heating problem shown in test 3A can only be solved by introducing a new switch for the A.C. which increases the strength of the dissipation when this type of shock front is detected.

In Section 3.5 we showed how the $\nabla\cdot \textbf{B}$ error varied as a function of the parameters $\alpha^B_{min}$ and $\hat{\beta}$ and optimised the values of these parameters. The accuracy of the solution produced by the code changed when the parameters varied. This is easily seen in shocktube test 5A, as shown in Fig. \ref{fig:betapara}. The $\nabla\cdot \textbf{B}$ error reduces as the parameter $\hat{\beta}$ is reduced. However, Figs. \ref{fig:ap5A1}-\ref{fig:ap5A3} show the effect of varying the parameter $\hat{\beta}$ on the solution produced for $v_y$. With $\hat{\beta}=0.2$ the tensile instability is clearly not suppressed in the low resolution region, right of $x=0.15$. The solution produced with $\hat{\beta}=1.0$ shows a deviation from the ATHENA solution at $x=0.15$, where the velocity overshoots the reference solution. In order to produce a $v_x$ profile which agrees with the reference solution and has the least error a value of $\hat{\beta}=0.5$ was chosen.

\section{Late time Orszag-Tang vortex}

\begin{figure*}
 \includegraphics[width=5.00cm,height=5.0cm]{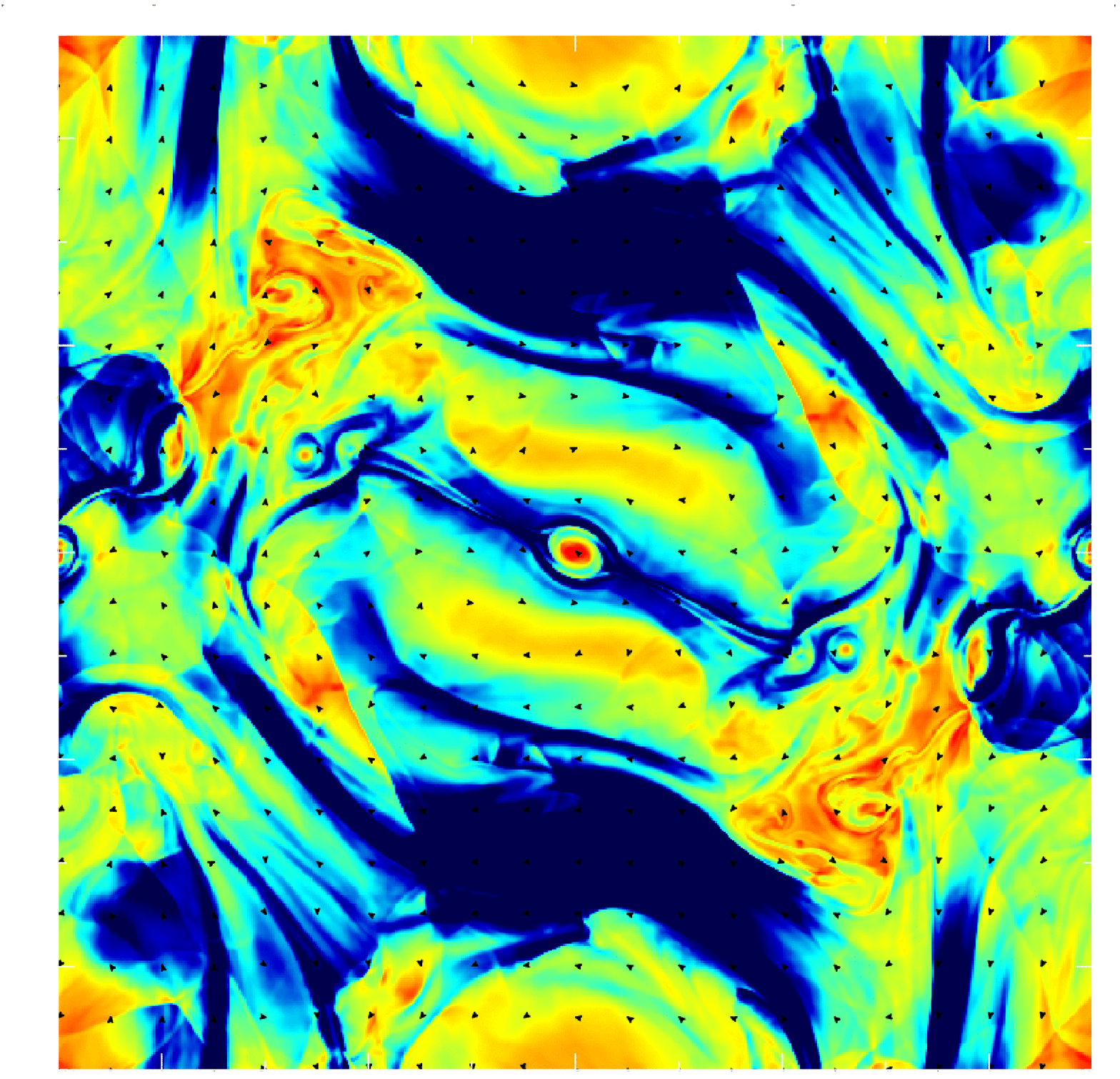}
 \includegraphics[width=5.00cm,height=5.0cm]{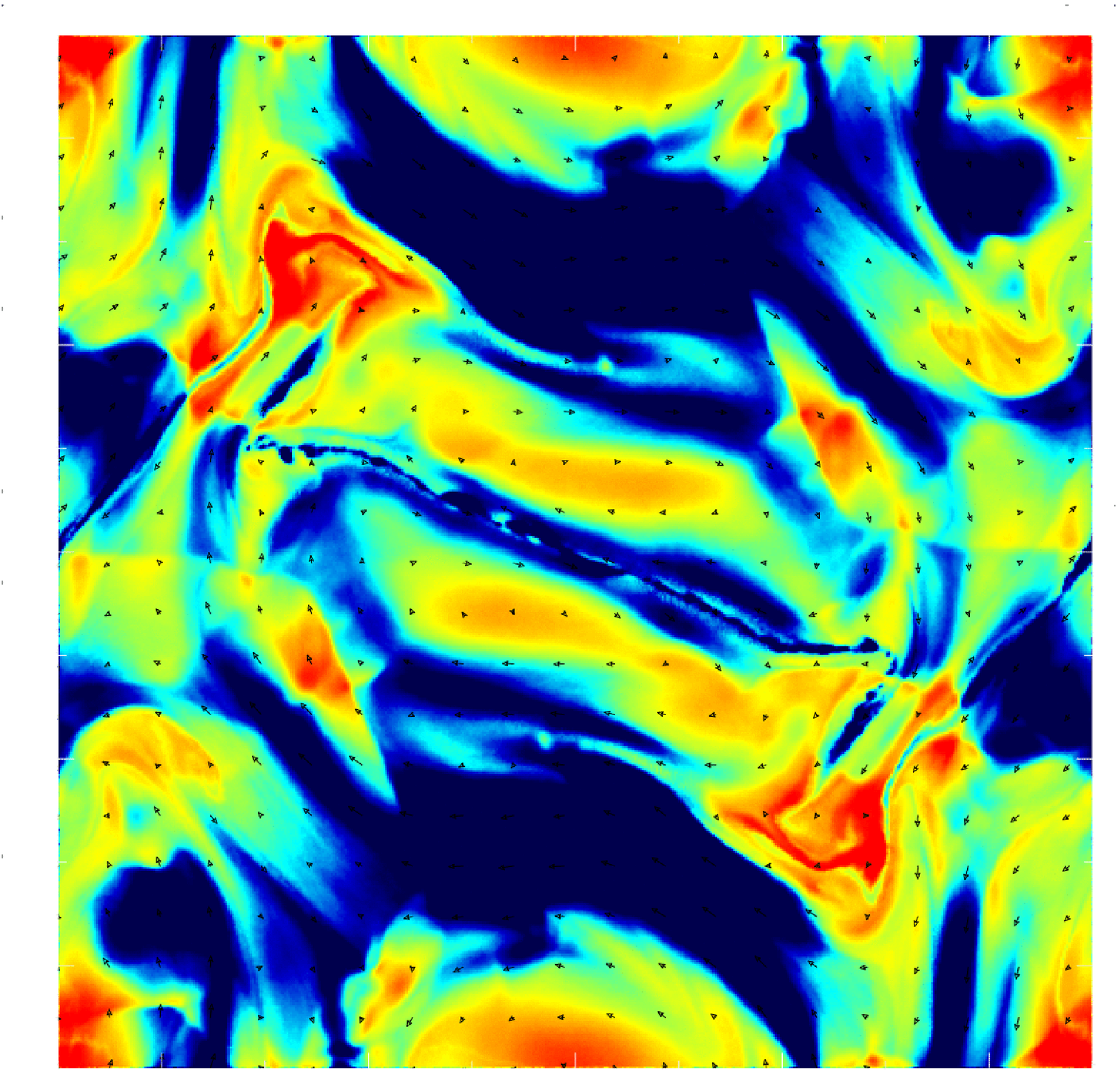}
 \includegraphics[width=5.00cm,height=5.0cm]{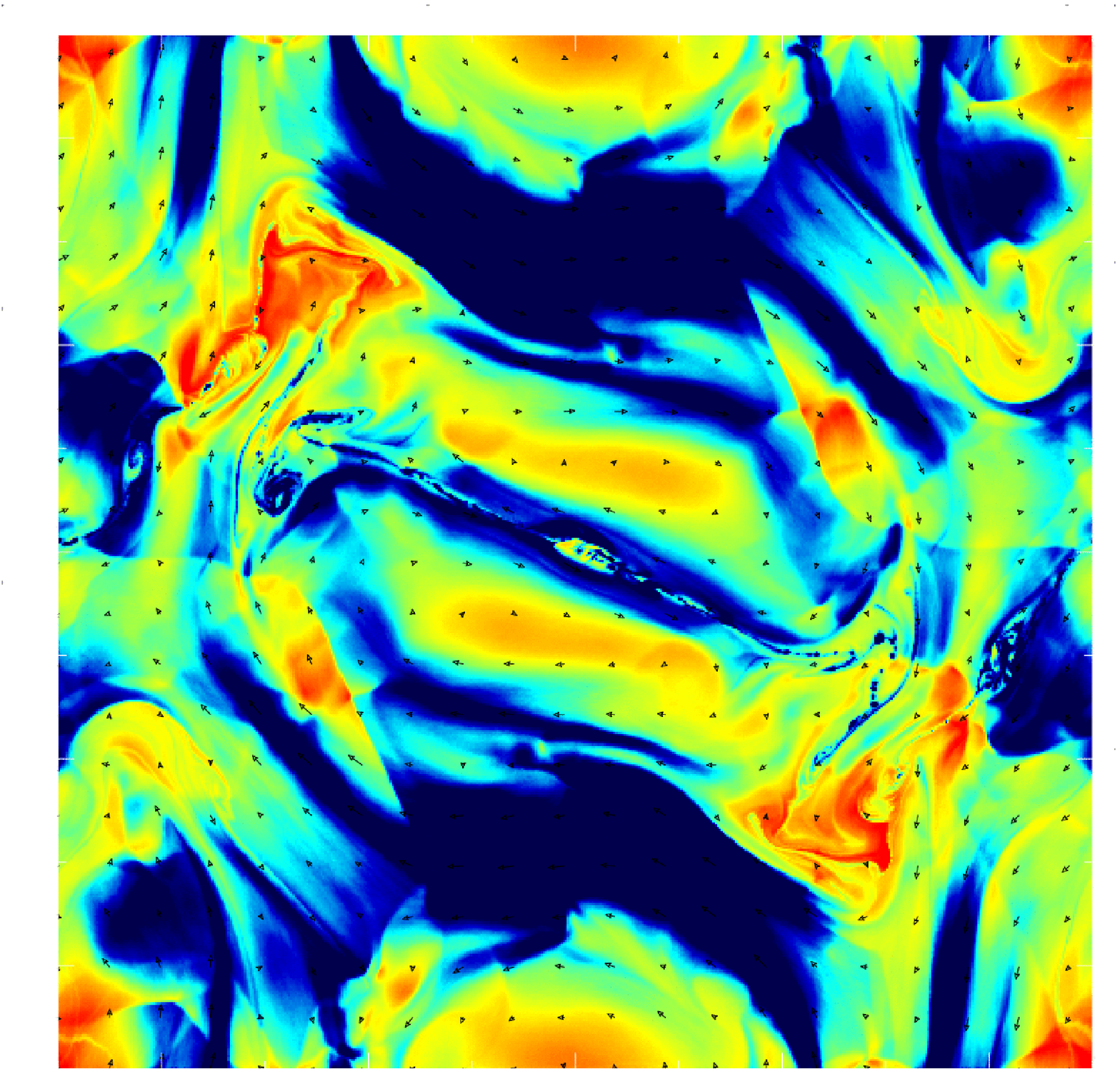}
 \includegraphics[width=5.00cm,height=5.0cm]{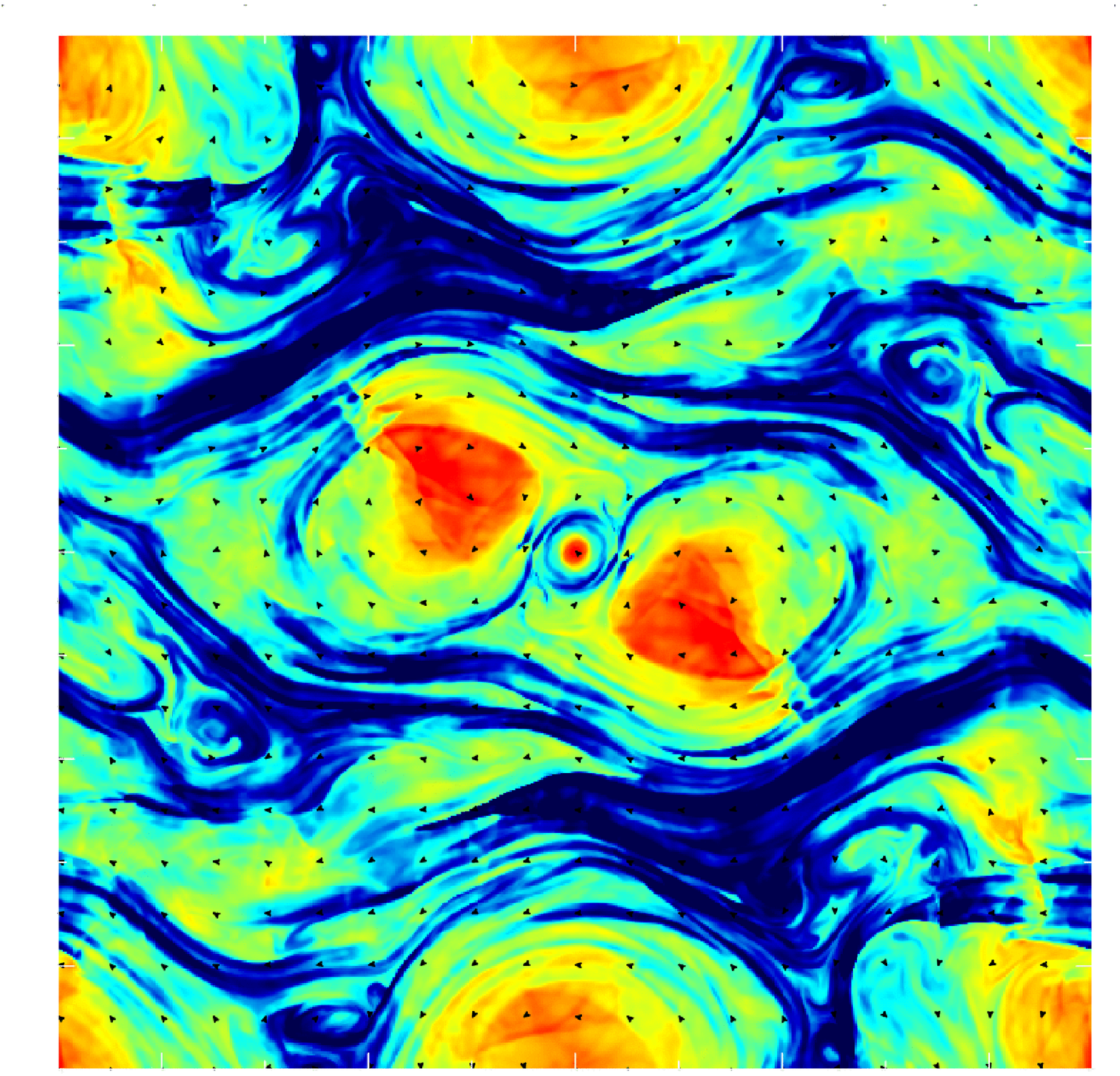}
 \includegraphics[width=5.00cm,height=5.0cm]{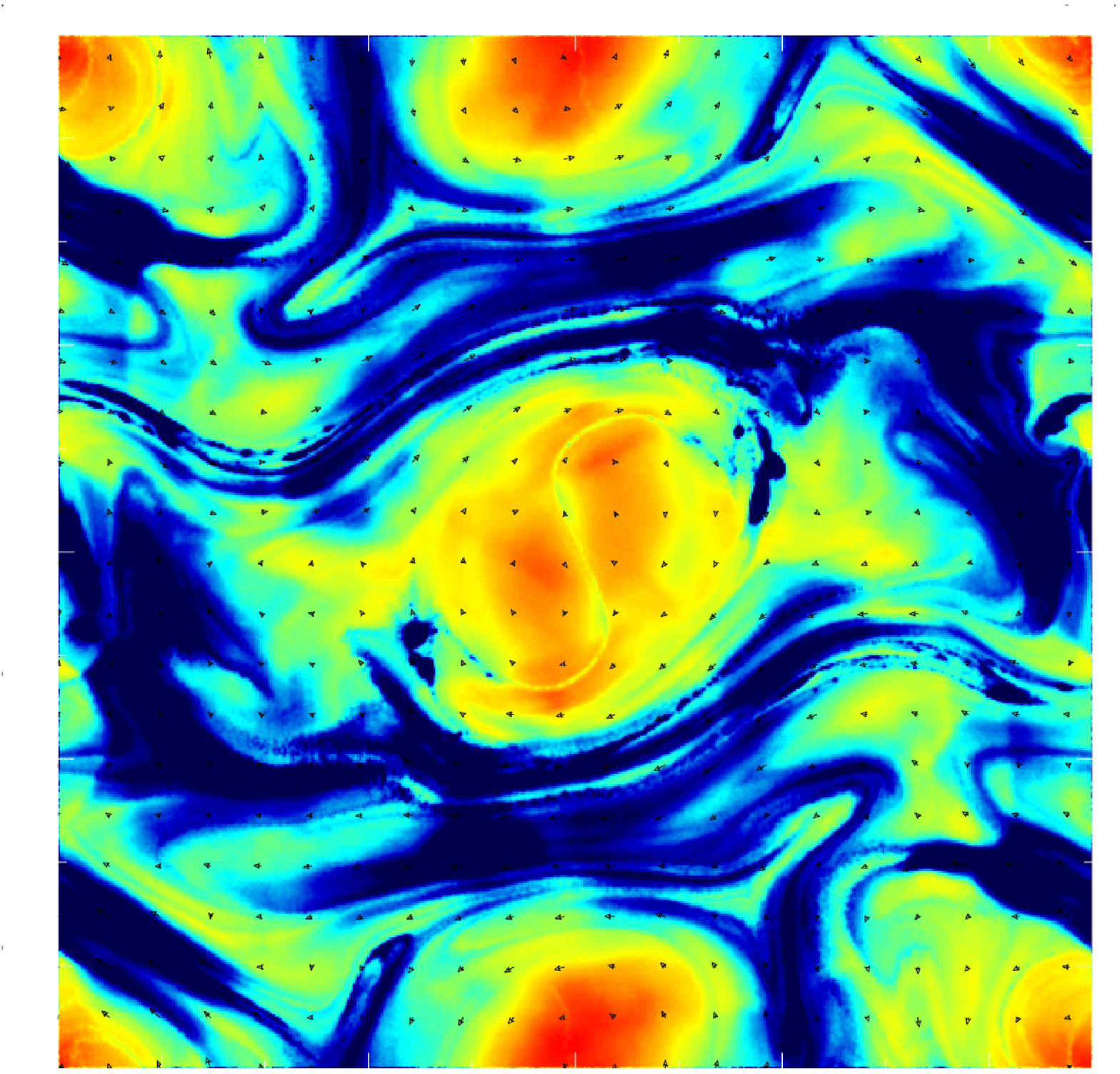}
 \includegraphics[width=5.00cm,height=5.0cm]{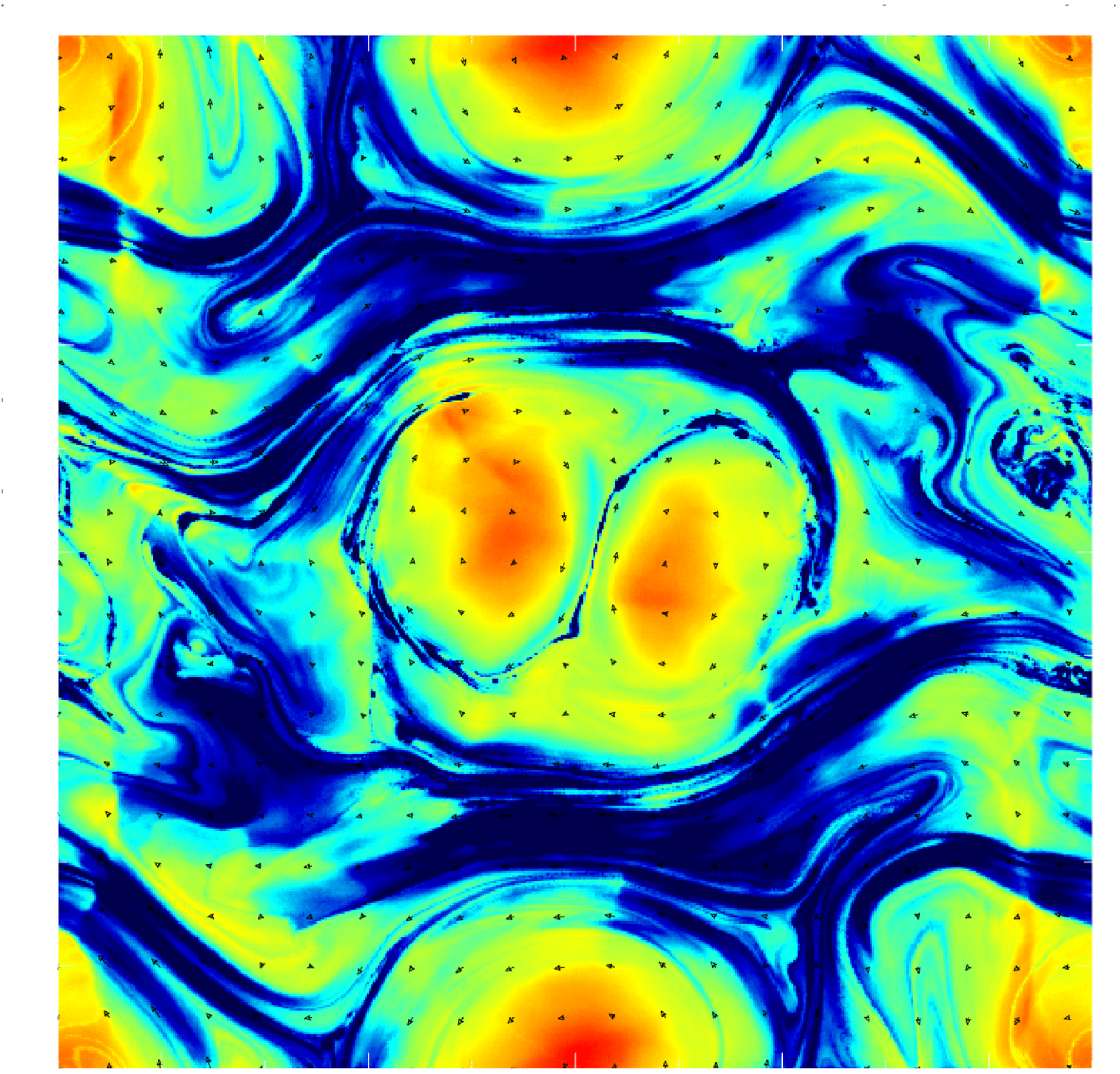}
 \caption{Plot of the Orszag-Tang vortex at times $t=1.0$ (upper row) and $t=2.5$ (lower row) for ATHENA (left column), GCMHD+ [$600 \times 692$] (middle column) and GCMHD+ [$1800 \times 2076$] (right column). The arrows show the strength and direction of the magnetic field. The code fails to capture the central density feature even at the higher resolution and both resolutions are not perfectly symmetric. }
\label{fig:apOT10}
\end{figure*}

In Section 3.3 the results for the Orszag-Tang vortex at $t=0.5$ were shown. At later times the test develops turbulence which then decays away. This development and subsequent decay of turbulence is very challenging for SPMHD codes to capture. We ran the Orszag-Tang vortex test again with a resolution of $600 \times 692$ and $1800 \times 2076$ and compared with the result produced by ATHENA with a resolution of $600 \times 600$ cells.

The result is shown in Fig. \ref{fig:apOT10}. The code correctly captures the majority of the features present in the density field. However, the code fails to capture the density peak at the center of the simulation. As a result the central density features seen in the ATHENA simulations are not fully reproduced in the GCMHD+ simulations.

\section{Kelvin-Helmholtz instability}

\begin{figure*}
 \includegraphics[width=5.00cm,height=5.0cm]{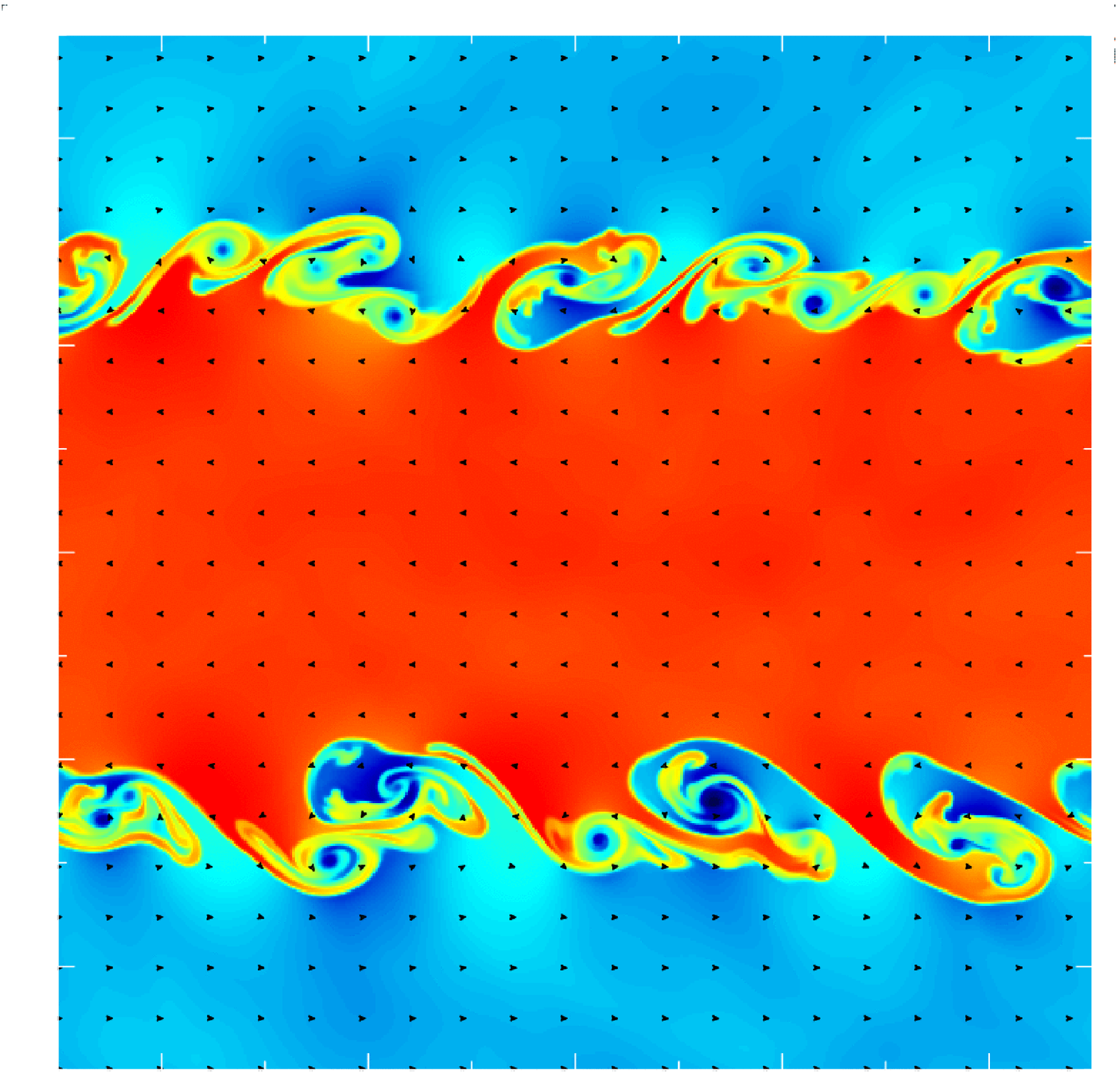}
 \includegraphics[width=5.00cm,height=5.0cm]{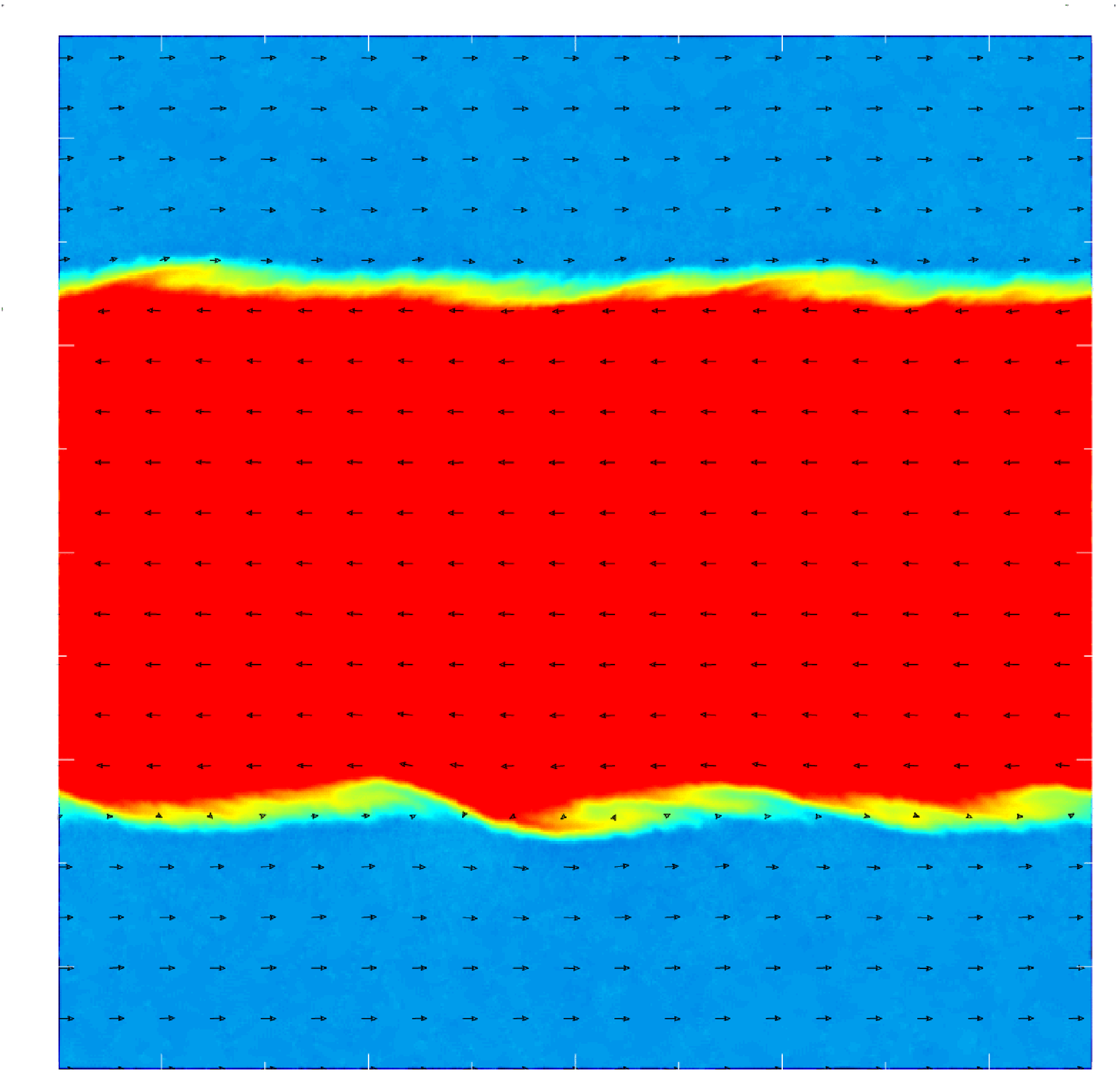}
 \includegraphics[width=5.00cm,height=5.0cm]{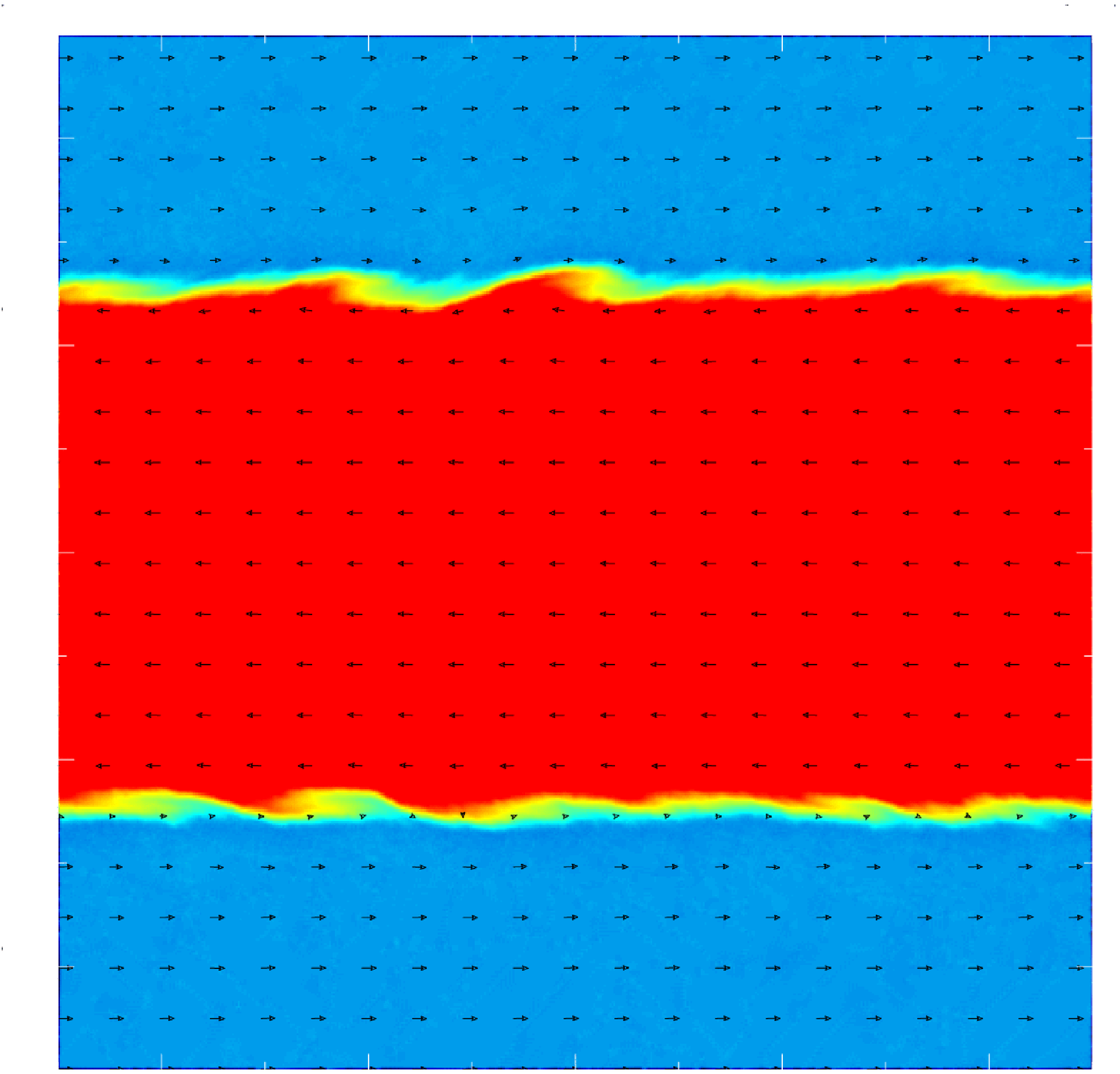}
 \includegraphics[width=5.00cm,height=5.0cm]{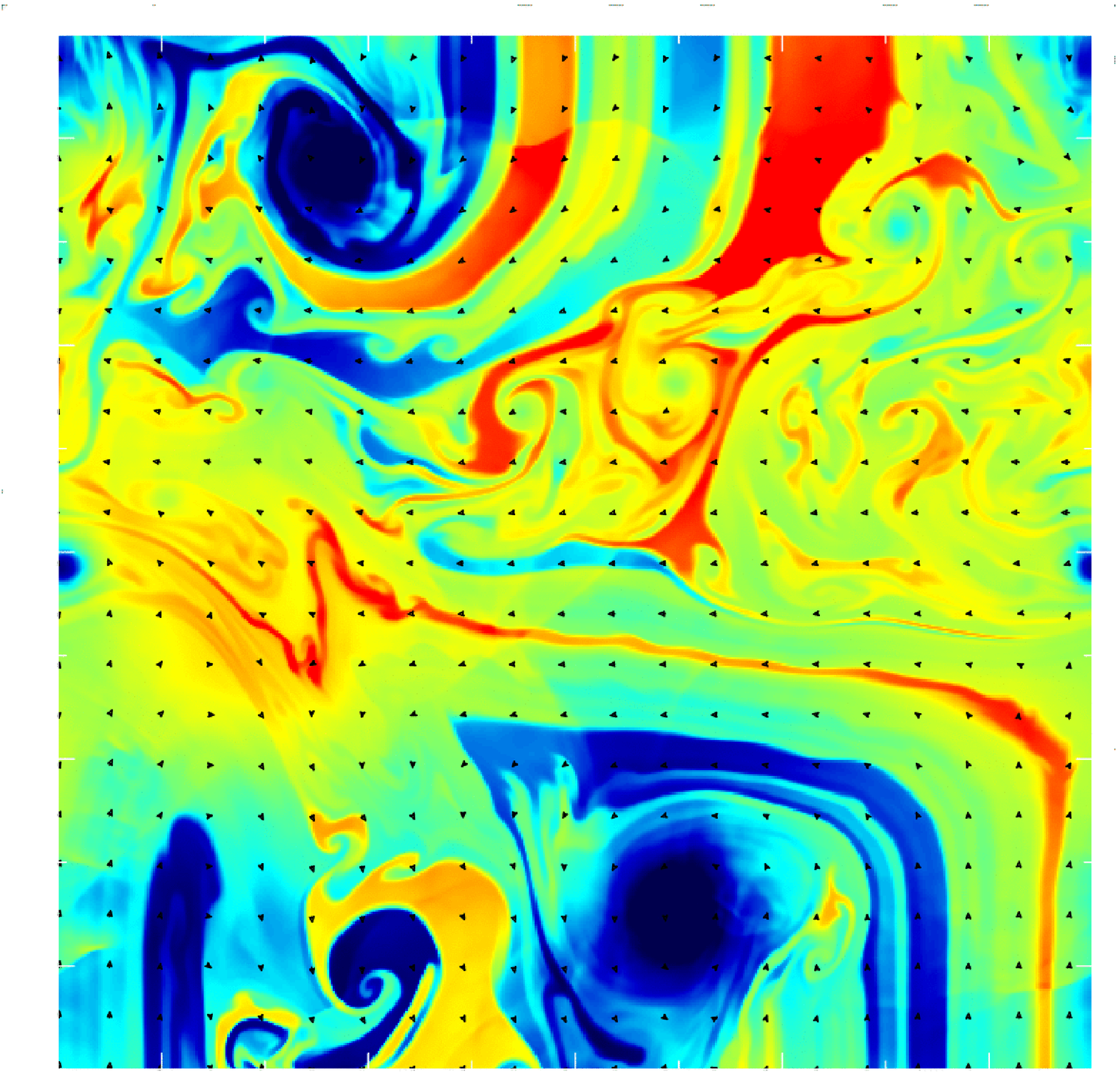}
 \includegraphics[width=5.00cm,height=5.0cm]{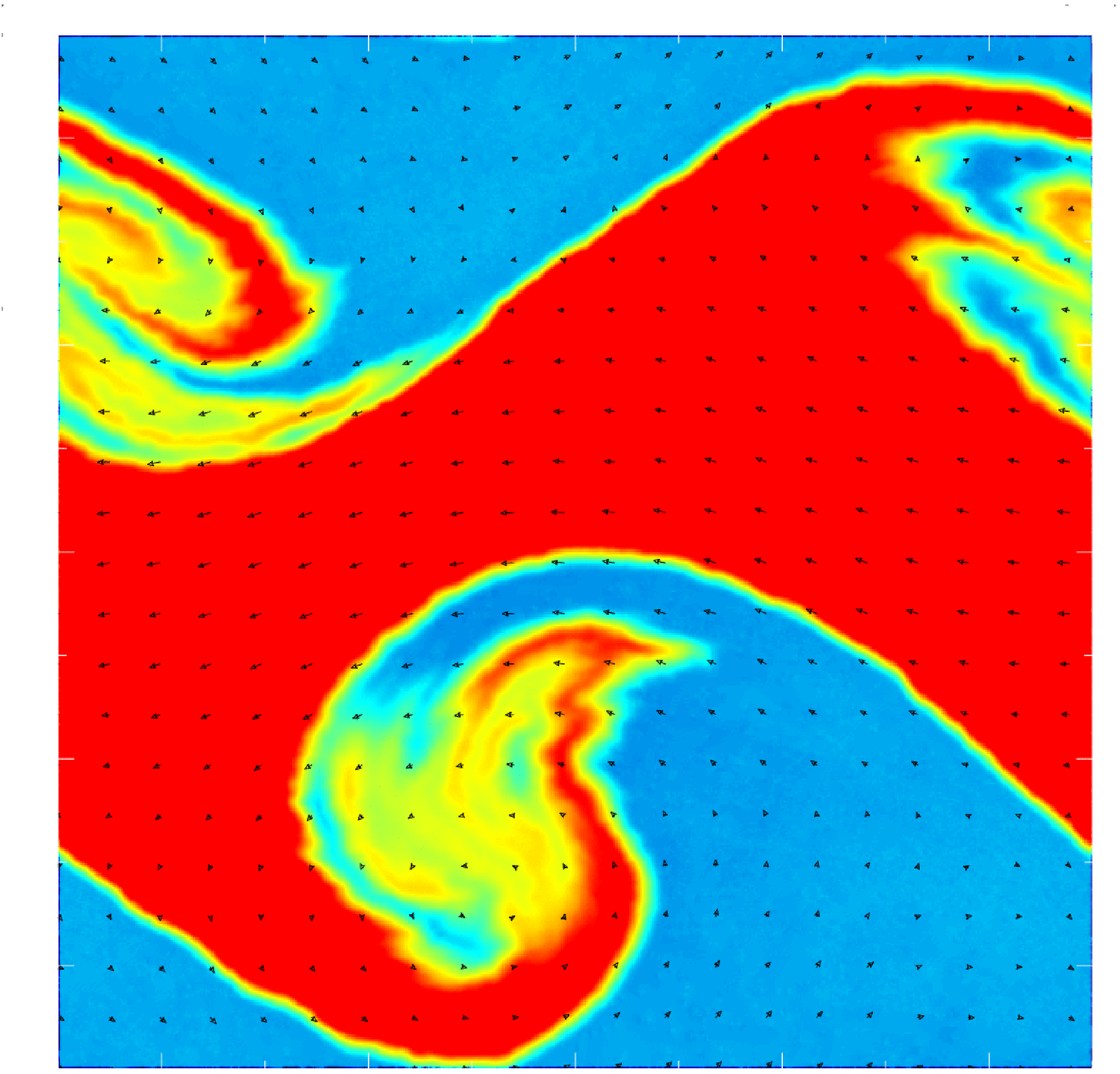}
 \includegraphics[width=5.00cm,height=5.0cm]{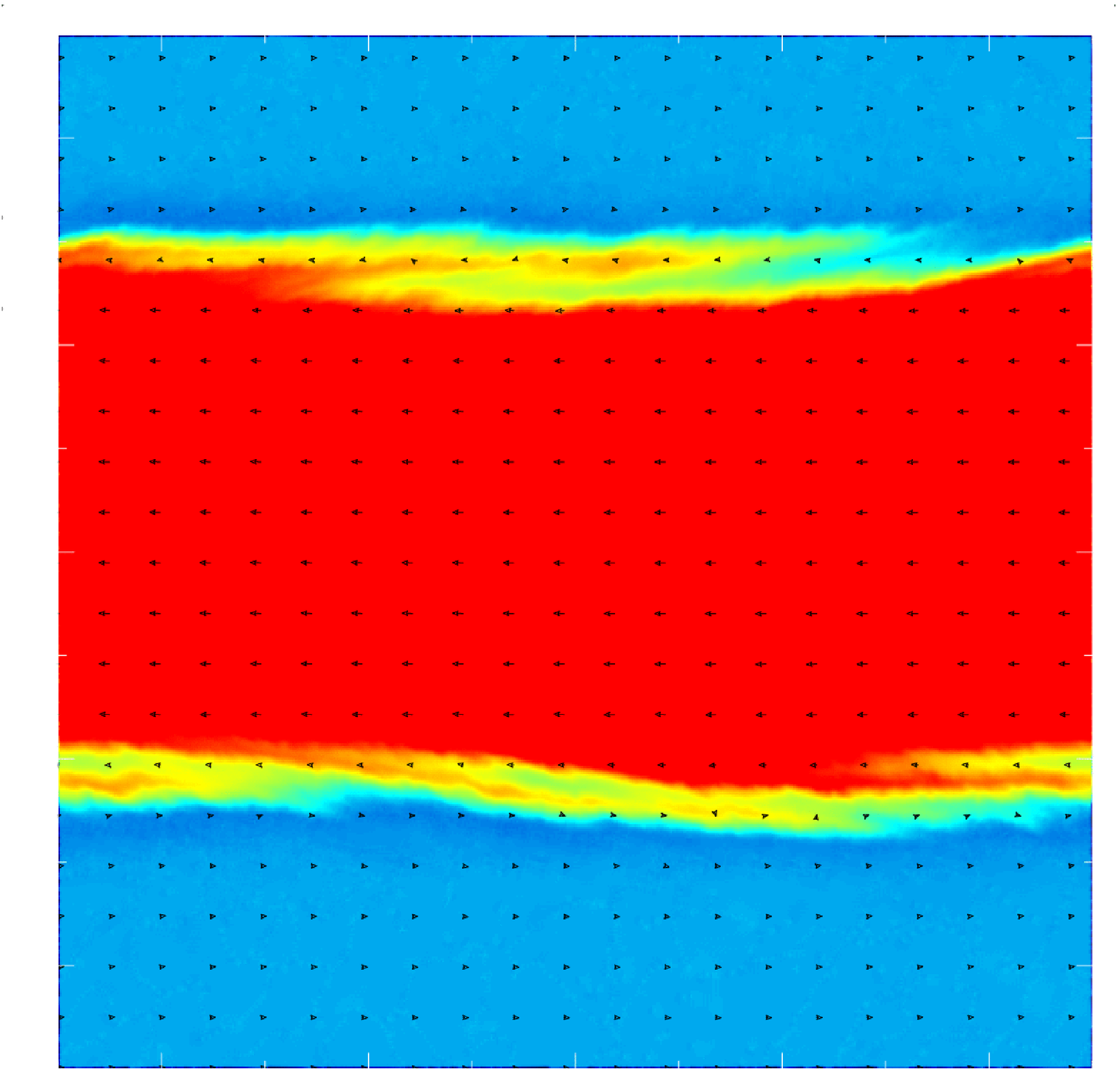}
 \caption{KHI test results for ATHENA (left), GCMHD+ [$\alpha^{AV}_{min}=0.1$] (middle) and GCMHD+ [$\alpha^{AV}_{min}=0.6$] (right) at $t=1.0$ (upper row) and $t=5.0$ (lower row). The arrows show the velocity field. The superior ability of the grid code to resolve the KHI is clear when the ATHENA result is compared to the SPH results.}
\label{fig:khi}
\end{figure*}

\begin{figure*}
 \includegraphics[width=5.00cm,height=5.0cm]{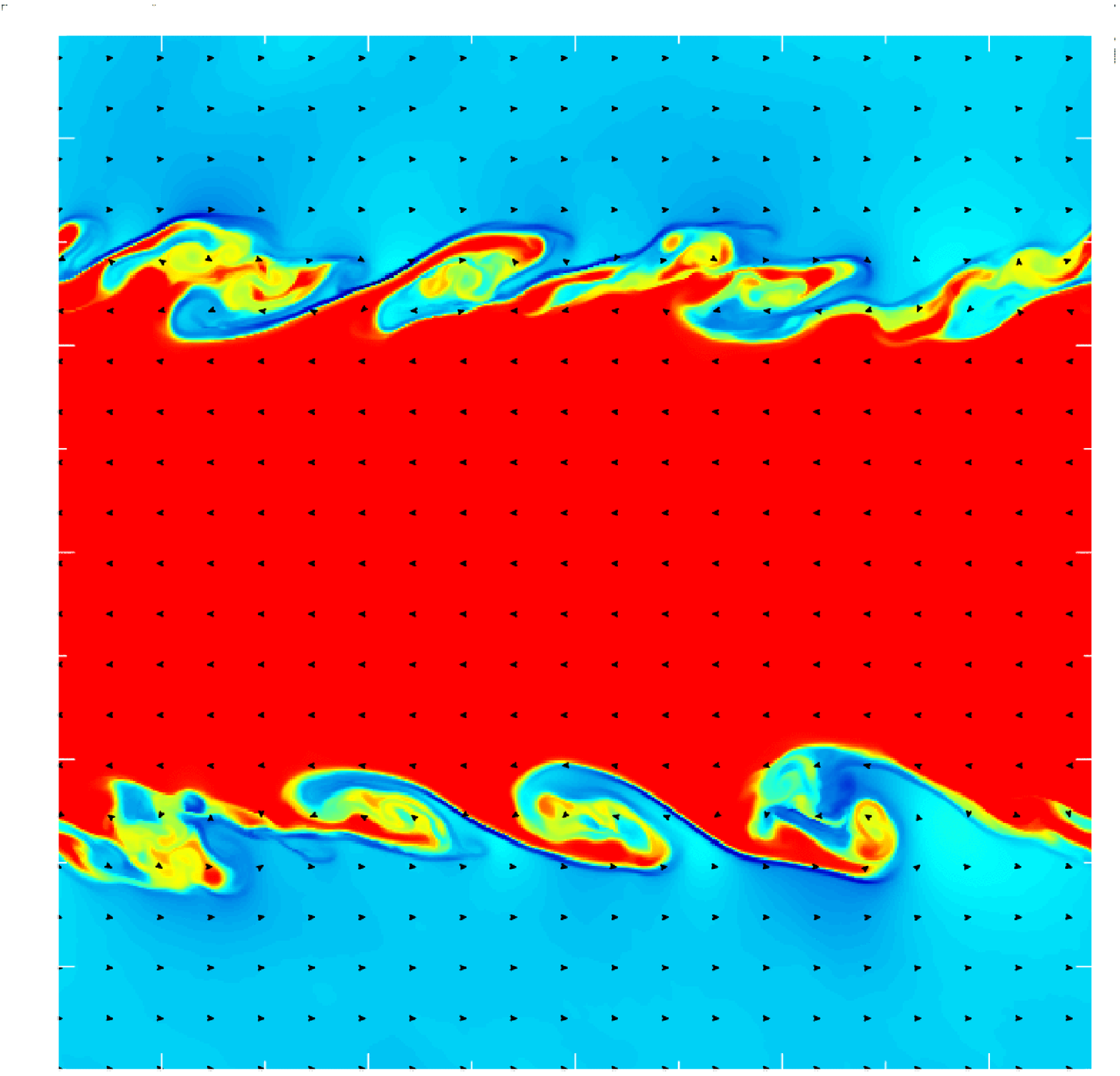}
 \includegraphics[width=5.00cm,height=5.0cm]{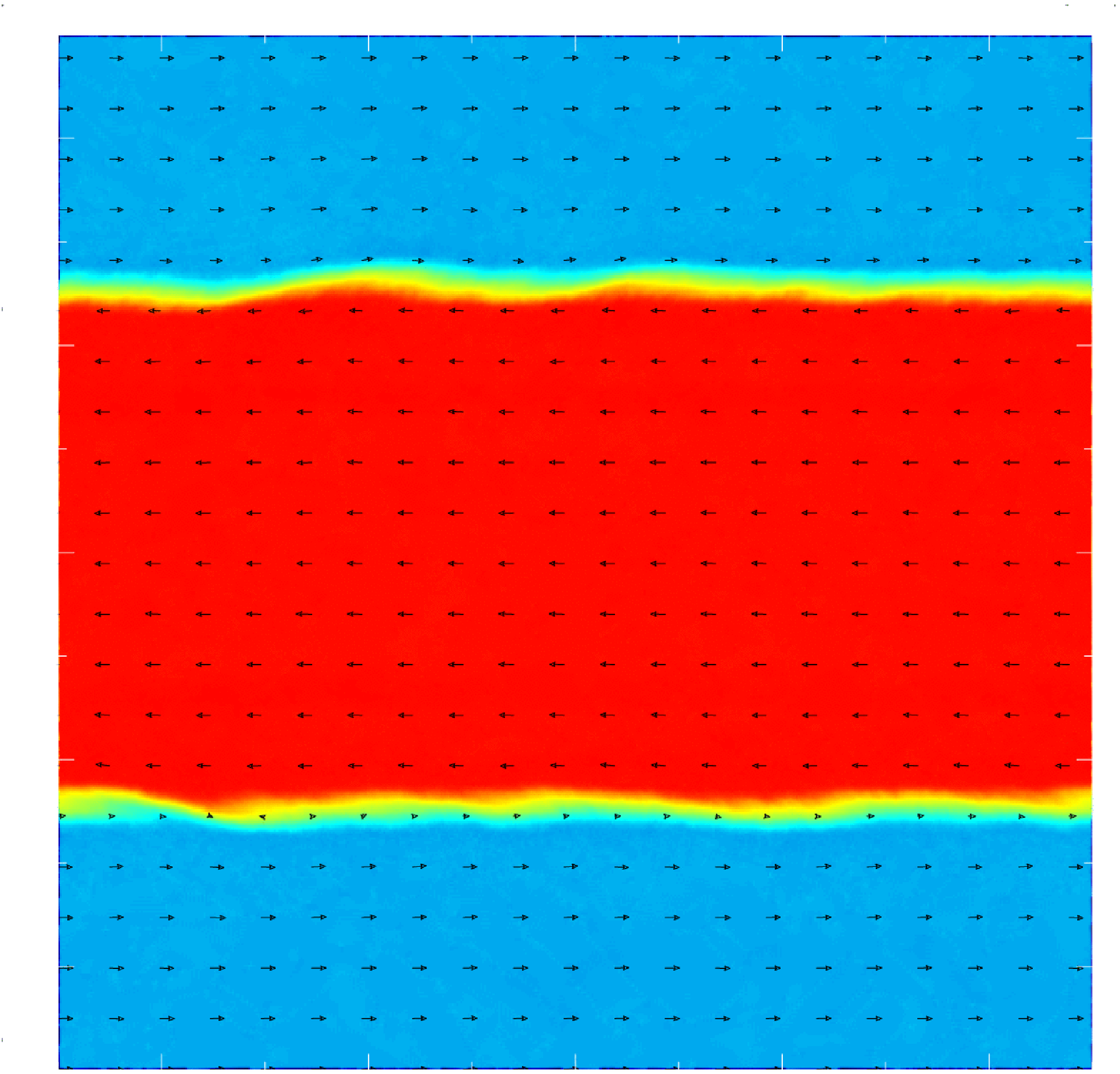}
 \includegraphics[width=5.00cm,height=5.0cm]{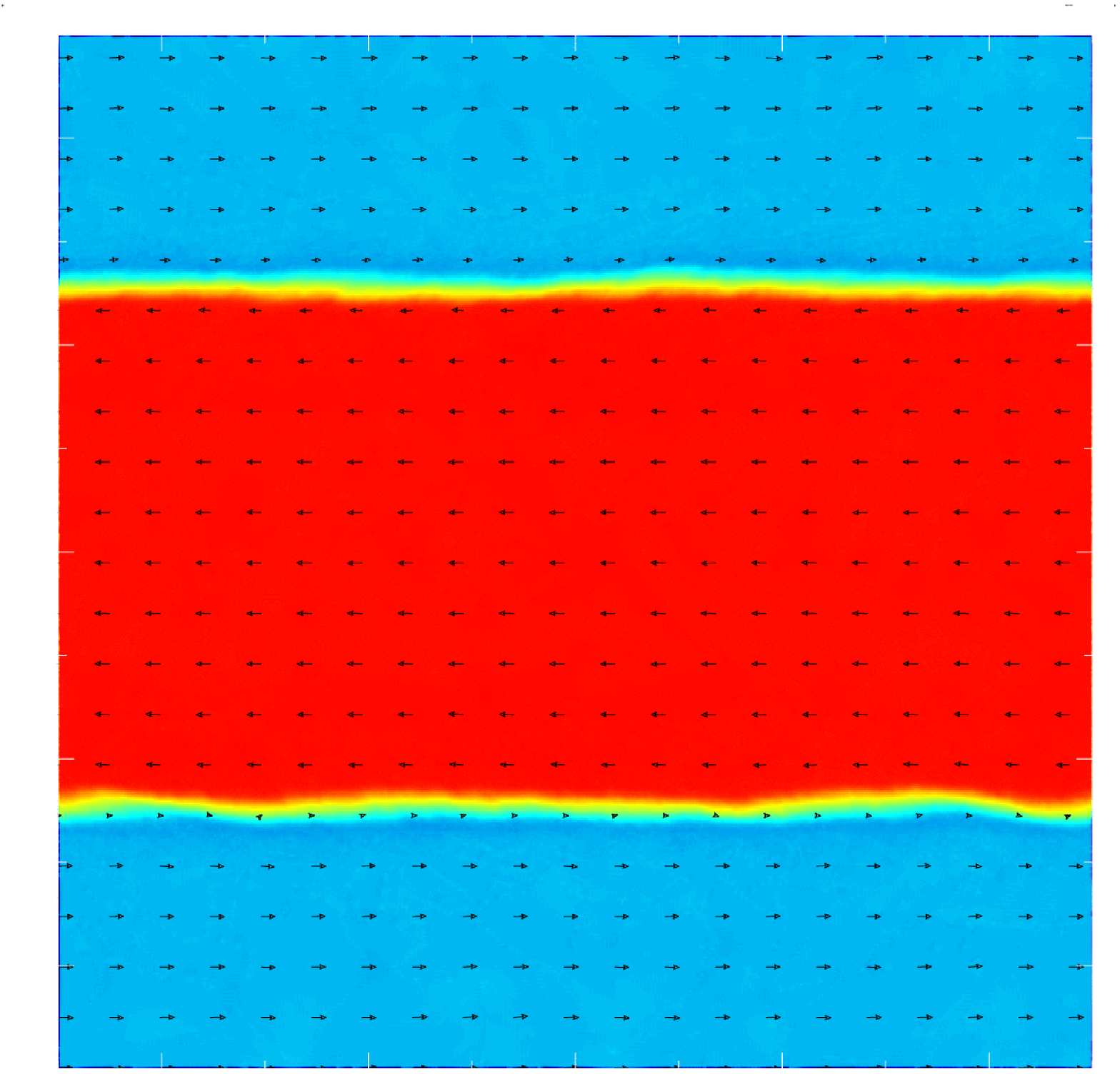}
 \includegraphics[width=5.00cm,height=5.0cm]{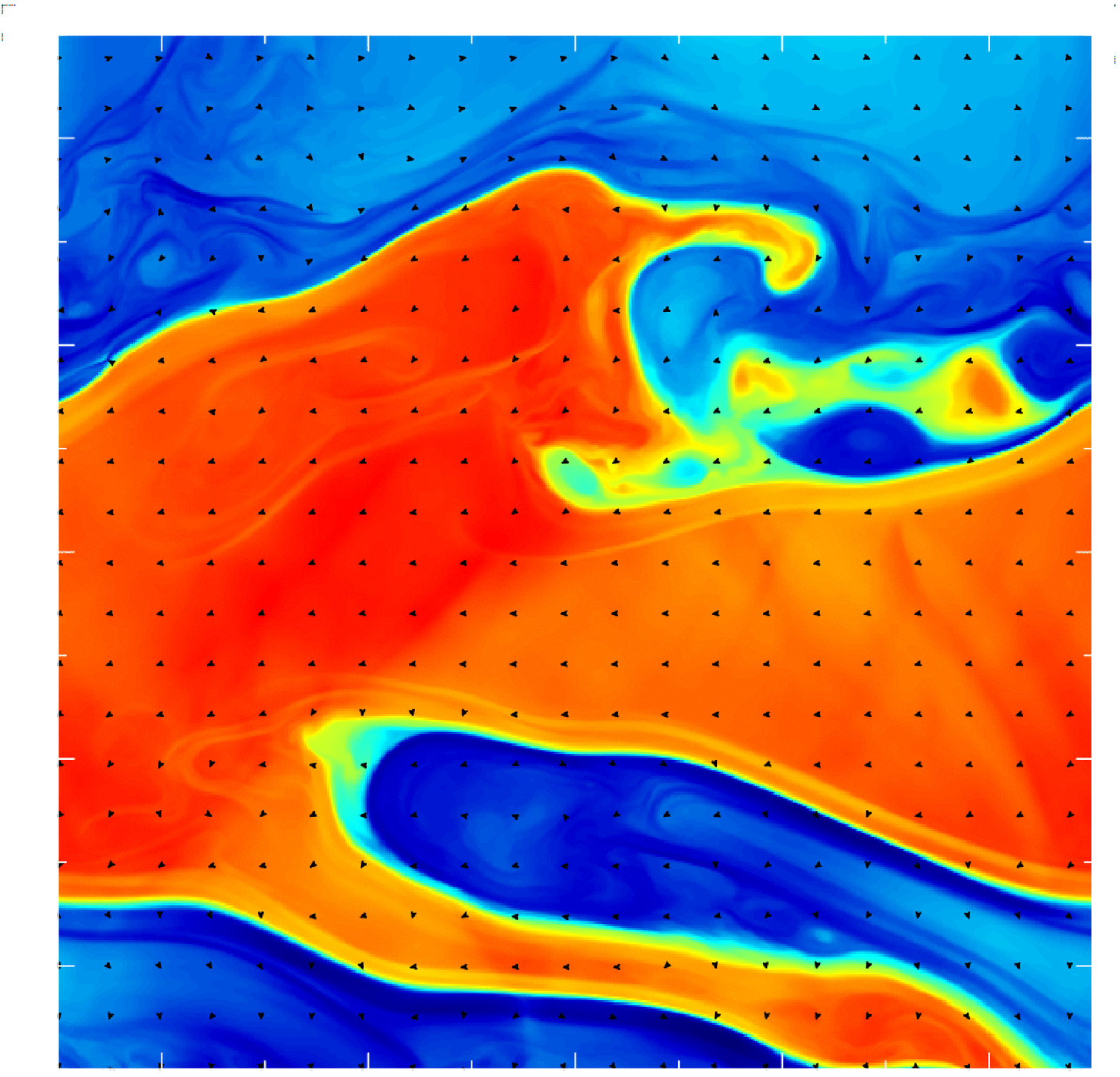}
 \includegraphics[width=5.00cm,height=5.0cm]{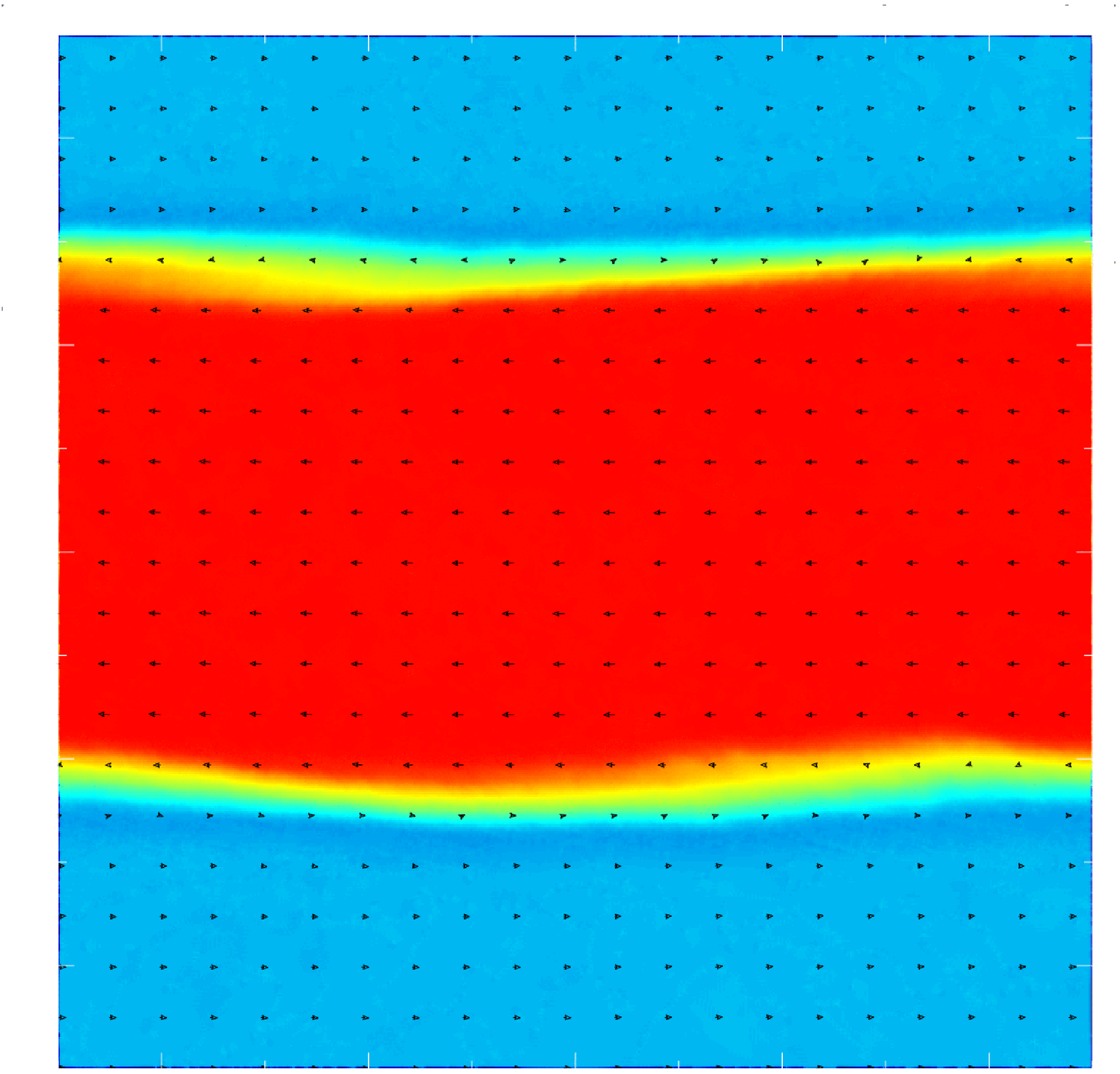}
 \includegraphics[width=5.00cm,height=5.0cm]{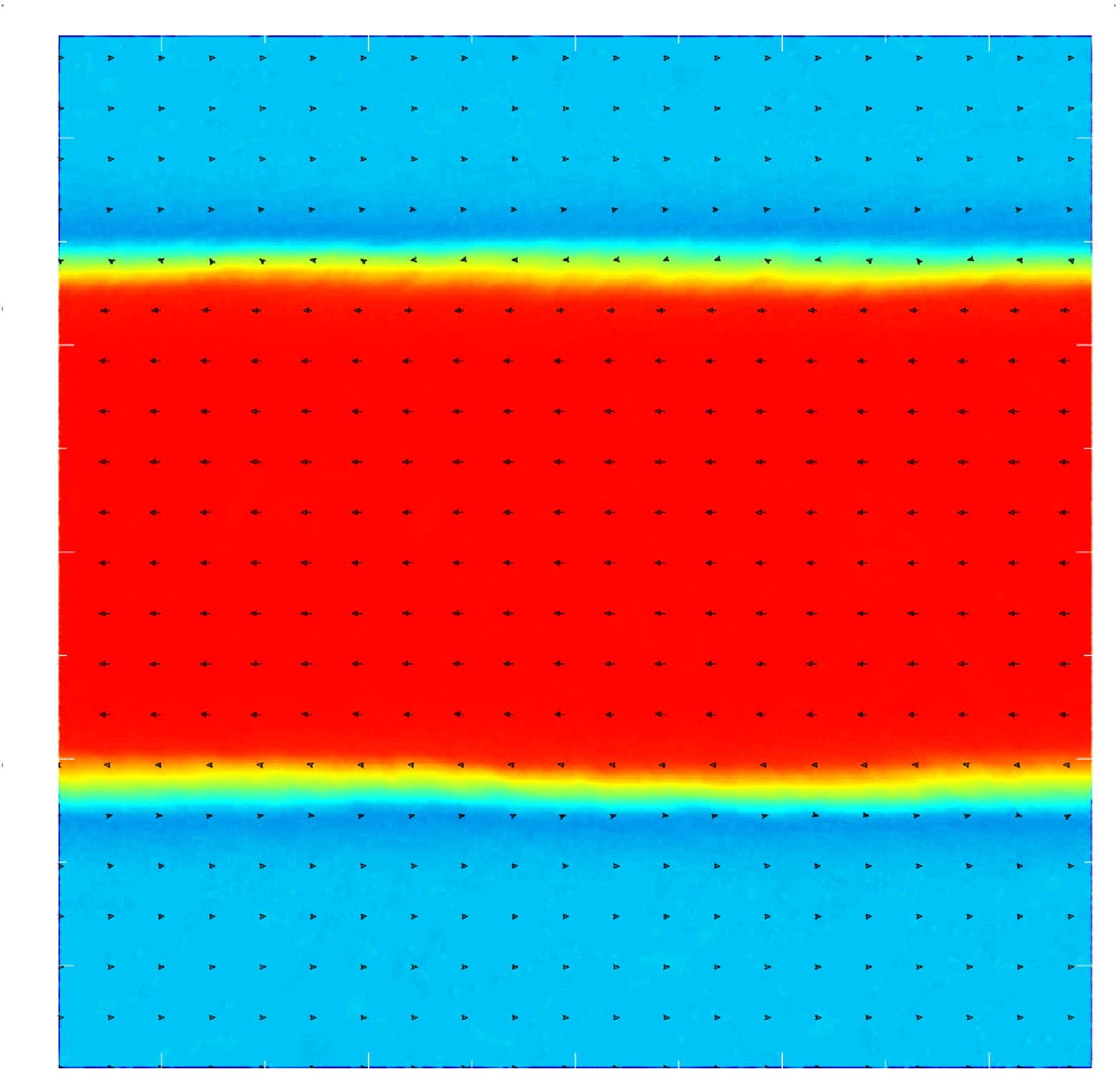}
 \caption{MHD KHI test results for ATHENA (left), GCMHD+ [$\alpha^{AV}_{min}=0.1$] (middle) and GCMHD+ [$\alpha^{AV}_{min}=0.6$] (right) at $t=1.0$ (upper row) and $t=5.0$ (lower row). The presence of a magnetic field in the $x$ direction prevents the growth of the instability in the GCMHD+ solutions.}
\label{fig:mkhi}
\end{figure*}

SPH codes in general struggle to resolve the Kelvin-Helmholtz instability (KHI). \citet{gcd09} demonstrated that the instability could be resolved when implemented with A.C.. We consider a periodic two dimensional box from  $x=-0.5$ to $0.5$ and from $y=-0.5$ to $0.5$. Equal mass particles are set out on a square lattice with $724$ particles along the x-axis in the high density region, between $y=-0.25$ and $y=0.25$, and $512$ particles along the x-axis in the low density region. The high density region has a density of $\rho_h=1.0$ and a velocity $v_x=-0.5$. The low density region has a density of $\rho_l=0.5$ and a velocity $v_x=0.5$. The two regions are in pressure equilibrium with $P_h=P_l=2.5$. The instability is seeded by adding random perturbations to the $x$ and $y$ components of the velocity with an amplitude of $0.01$.

Fig. \ref{fig:khi} shows the result of the KHI test for the ATHENA reference code, GCMHD+ with $\alpha^{AV}_{min}=0.1$ and GCMHD+ with $\alpha^{AV}_{min}=0.6$. Two different GCMHD+ solutions are shown to show the effect of increasing the minimum applied artificial viscosity. The instability develops more quickly in the grid method solution of ATHENA compared to the SPH solutions. This is clearly seen at both time steps displayed. The increase in minimum A.V. of GCMHD+ reduces the development of the instability considerably when compared to the lower GCMHD+ implementation of $\alpha^{AV}_{min}=0.1$.

We then performed an MHD KHI test. The same set up as above was used with the addition of a magnetic field in the $x$ direction. A uniform and homogeneous magnetic field of strength $B_x=0.129$ was applied to all particles in the simulation. Fig \ref{fig:mkhi} shows the results of this test for ATHENA, GCMHD+ ($\alpha^{AV}_{min}=0.1$) and GCMHD+ ($\alpha^{AV}_{min}=0.6$). The magnetic field stabilises the instability and prevents it from fully developing. The ATHENA result again shows greater development of the instability when compared to the SPMHD solutions, but the effect of the magnetic field is clear when the results are compared to Fig. \ref{fig:khi}. The solutions produced by the code show very minimal development of the instability. Further work is required to ensure that GCMHD+ is capable of resolving Kelvin-Helmholtz instabilities.

\label{lastpage}

\end{document}